\documentclass[traditabstract]{aa}              

\usepackage{graphicx}
\usepackage{txfonts}
\usepackage{natbib}
\usepackage{enumitem}
\usepackage{lscape}

\begin{document}

\title{Measures of galaxy dust and gas mass with \textit{Herschel} photometry and prospects for ALMA
\thanks{\textit{Herschel} is an ESA space observatory with science instruments provided by 
European-led Principal Investigator consortia and with important participation from NASA.}}

\titlerunning{Measures of galaxy dust and gas mass}

\author{S. Berta\inst{1}
        \and
        D. Lutz\inst{1}
        \and
        R. Genzel\inst{1}   
        \and
        N.M. F\"{o}rster-Schreiber\inst{1}
        \and
        L.J. Tacconi\inst{1}
        }

\offprints{Stefano Berta, \email{berta@mpe.mpg.de}}

\institute{Max-Planck-Institut f\"{u}r extraterrestrische Physik (MPE),
Giessenbachstr., D-85748 Garching, Germany.
}

\date{Received: November 15th, 2015; accepted December 8th, 2015}

\abstract{
Combining the deepest \textit{Herschel} extragalactic surveys (PEP, GOODS-H, HerMES),
and Monte Carlo mock catalogs, we explore the robustness of dust mass estimates 
based on modeling of broadband spectral energy distributions (SEDs) with 
two popular approaches: Draine \& Li (2007; DL07) and a modified blackbody (MBB).
We analyze the cause, drivers, and trends of uncertainties and systematics in thorough 
detail. 

As long as the observed SED extends to at least 160-200 $\mu$m in the rest frame,
$M_{\rm dust}$ can be recovered with a $>3\sigma$ significance and without the occurrence 
of systematics. An average offset of a factor $\sim$1.5 exists between DL07- and MBB-based 
dust masses, based on consistent dust properties. The performance of DL07 modeling turns out to be more robust than 
that of MBB since relative errors on M$_{\rm dust}$ are more mildly dependent 
on the maximum covered rest-frame wavelength and are less scattered.

At the depth of the deepest \textit{Herschel} surveys (in the GOODS-S field),
it is possible to retrieve dust masses with a signal-to-noise ratio, S/N$\ge$3 for galaxies on the 
main sequence of star formation (MS) down to $M^\ast\sim10^{10}$ $[$M$_\odot]$
up to $z\sim 1$. At higher redshift ($z\le2$), the same result is only achieved 
 for objects at the tip of the MS or for those objects lying above the tip owing to sensitivity 
and wavelength coverage limitations.

Molecular gas masses, obtained by converting $M_{\rm dust}$ through 
the metallicity-dependent gas-to-dust ratio $\delta_{\rm GDR}$, are consistent with 
those based on the scaling of depletion time, $\tau_{\rm dep}$, and on CO
sub-mm spectroscopy. Focusing on CO-detected galaxies at $z>1$, the $\delta_{\rm GDR}$
dependence on metallicity is consistent with the local relation, provided that 
a sufficient SED coverage is available.

Once we established that Herschel-only and sub-mm-only estimates of
dust masses  can be affected by large uncertainties and possibly 
systematics in some cases, we combined far-IR \textit{Herschel} data and sub-mm ALMA expected fluxes
to study the advantages of a full SED coverage. The uncertainty on $M_{\rm dust}$ 
reduces to $<30$\% for more than 85\% of \textit{Herschel} galaxies, thus potentially 
facilitating a fast statistical study of $M_{\rm dust,gas}$ for large samples, 
at least up to $z\sim2$.
}


\maketitle


\section{Introduction}\label{sect:intro}

Tightly connected, dust and gas are the key ingredients 
of star formation, which is governed by their complex, mutual interplay as in a cyclic dance. 
Stars form in cold, dense molecular clouds 
and dust works as catalyst in transforming atomic hydrogen into molecular 
hydrogen \citep[e.g.,][]{wolfire1995}. Gas is again expelled from stars during their lifetime 
in the form of winds and finally at the supernova (SN) stage. Dust is believed to 
be mainly produced in the envelopes of asymptotic giant branch (AGB) stars 
and at the end of the life of massive stars during the explosive SN
phase \citep[see][for a short summary]{dunne2011}. Supernovae shocks, on the other hand, destroy dust 
grains, which can form again in the interstellar medium (ISM) by an accretion process.
Dust absorbs the ultraviolet (UV) emission of young stars, allowing gas to cool and condense to 
form new stars. 

The dust and gas content of galaxies are linked to each other through metallicity, 
as shown in the local Universe \citep{leroy2011,draine2007b}: the gas-to-dust mass ratio
in the  ISM increases as a function of metallicity. 
Dust has therefore been often adopted as a proxy to gas in absence 
of time-expensive, sub-millimeter spectroscopic observations \citep[e.g.,][]{magdis2012,santini2014,scoville2014,genzel2015} at high redshift 
also, assuming that local relations hold. 
The energy absorbed by dust at short wavelengths is re-emitted in the infrared (IR)
and sub-millimeter (sub-mm) regimes, where the thermal emission of grains dominates the spectral 
energy distribution (SED) of galaxies ($\sim$8-1000 $\mu$m). 
Dust emission is frequently used to trace the ongoing 
rate of star formation (SFR; e.g., Nordon et al. \citeyear{nordon2010,nordon2012}, Elbaz et al. 
\citeyear{elbaz2011}, Kennicutt \citeyear{kennicutt1998}).

It is thus no wonder that the interest in the study of galaxy dust and gas properties 
has grown in the past two decades and that an always increasing number of studies 
based on \textit{Herschel} data are dedicated to the dust emission of galaxies at all redshifts. 
The \textit{Herschel} satellite \citep{pilbratt2010} provides far-infrared (FIR) 
photometric observations of local and distant galaxies thanks to its 
two onboard cameras, the \textit{Photodetector Array Camera and Spectrometer} 
(PACS, 70-160 $\mu$m, Poglitsch et al. \citeyear{poglitsch2010}) 
and the \textit{Spectral and Photometric Imaging REceiver} (SPIRE, 250-500 $\mu$m, Griffin et al. \citeyear{griffin2010}).
The availability of a good SED coverage from 24 to 500 $\mu$m, obtained by combining \textit{Herschel}
and {\em Spitzer} observations, allows us to apply different types of galaxy emission models 
with the aim of deriving the main properties of dust, such as 
its mass, temperature, and, to some degree,  composition. 
Different families of models, with different levels of complexity, can be broadly recognized: 
{\em a)} full radiation transfer treatments \citep[e.g.,][]{silva1998,efstathiou2000,piovan2006,siebenmorgen2007} 
requiring geometric assumptions, very detailed observed SEDs, and large amounts of computational power; {\em b)} evolutionary and
mixed stellar population synthesis, including FIR emission \citep[e.g.,][]{dacunha2008}; {\em c)} 
physically motivated FIR dust emission models \citep[e.g.,][]{dl07,galliano2011}; {\em d)} simple or multiple modified blackbody (MBB) 
SED fitting; {\em e)} template families following locally calibrated relations \citep[e.g., the $L$-$T$ relation,][]{chary2001,dale2002};  and
{\em f)} semiempirical template fitting \citep[e.g.,][]{polletta2007,wuyts2011a,berta2013a}.

Focusing on dust emission, of particular relevance and very popular are 
types {\em c)} and {\em d)}, i.e., SED fitting with physically motivated models, comprising 
several dust components or MBB fitting under the assumption 
that dust emissivity can be described as a simple frequency power law and dust is in thermal equilibrium.

\citet{dunne2011} applied MBB SED fitting to $z<0.5$ H-ATLAS \citep{eales2010} star-forming
galaxies selected at 250 $\mu$m and studied the evolution of the dust mass function 
over the last 5 billion years, showing that the dust/stellar mass ratio was 
three to four times larger at $z=0.4-0.5$ than today. At these redshifts, PACS and SPIRE data cover 
the FIR peak of the SED and part of its Rayleigh-Jeans tail (RJ), thus leading
to reliable $M_{\rm dust}$ estimates. 

\citet{cortese2012} combine \textit{Herschel} photometry and 
multiwavelength data to build detailed SEDs of 
nearby galaxies from the \textit{Herschel} Reference Survey \citep[HRS;][]{boselli2010}
and fit them with MBB and \citet[][DL07]{dl07} models. In combination with 
radio data, these authors study the dust-to-{\sc Hi}  and dust-to-stellar mass ratios
of their targets, finding that the former increases as a function of stellar mass, 
while the latter tends to decrease as a function of stellar mass.
\citet{ciesla2014} apply DL07 modeling to HRS SEDs 
and have built a new set of reference, local SED templates. 

\citet{dale2012} and \citet{bianchi2013} apply DL07 and MBB 
fitting to the FIR-submm SEDs of KINGFISH \citep{kennicutt2011} 
local galaxies. The former authors show evidence of sub-mm (500 $\mu$m) 
flux excess in dwarf galaxies with respect to the expectation based on the extrapolation 
from {\em Spitzer} data. They also find a factor $\sim$2 difference 
between DL07- and MBB-based dust masses, while \citet{bianchi2013} 
shows that this difference is avoided if using consistent 
assumptions for the two approaches.

\citet{remyruyer2013} combine KINGFISH and other samples 
of nearby galaxies observed by {\textit Herschel}, 
including direct measurements of molecular gas mass, to study 
the gas/dust vs. metallicity relation over a 2 dex metallicity range. 
Dust masses are in this case estimated with \citet{galliano2011} 
models.
\citet{hunt2014} come back to the KINGFISH sample and present the 
surface brightness profiles of these nearby galaxies. These authors  again
study the spatially resolved SEDs and  resolved properties 
of dust with DL07 and MBB modeling.

\citet{eales2012} foresee 
the possibility of applying a single conversion factor 
to derive gas masses from sub-mm fluxes, based on 
a Milky Way (MW) observation and recalibrated 
on HRS nearby galaxies with CO and {\sc Hi} observations. 
They point out that 
this method suffers from the limitation that their 
sample is not necessarily representative of 
all galaxy populations and metallicity dependencies play an important 
role both for dust- and CO-based masses.

Successfully applied to local galaxies, 
the SED fitting techniques and scaling relations mentioned above have also 
been adopted by several authorsto study 
the IR emission of distant galaxies. 
Broadly speaking, two main approaches can be identified:
using a single sub-mm continuum observation to  scale it to dust or gas mass using known relations \citep[e.g.,][]{eales2012,scoville2014};
or characterizing the galaxy SED including the FIR peak \citep[e.g.][]{dale2012,galametz2013}.
Nevertheless, when dealing with high-$z$ galaxies, the use of these techniques is 
not as straightforward as in the nearby 
Universe. It is necessary not 
only to keep in mind that the properties of dust and gas might be 
evolving as a function of time, but also to face the limitations 
of the available data.

In the simplest case of all, \citet{eales2012} discuss the 
limitations of SED coverage at high redshift, and show that small
dust temperature gradients \citep[see also][]{hunt2014} within 
a galaxy can lead to incorrect measurements when using 
a MBB fit to the global SED. \citet{galametz2014} point out the 
effects of limited angular resolution on the derivation 
of model parameters.

\citet{scoville2014} apply a similar concept to 
the RJ side of the SED of local and high-$z$   
galaxies. They suggest that gas masses can be measured 
using single-band sub-mm observations, adopting 
dust emissivities calibrated on {\em Planck} observations of the MW. 
They apply this method to early ALMA observations 
of four small samples of galaxies at four different 
cosmic epochs. Using simple MBB simulations, 
\citet{genzel2015} discuss the systematic effects induced by this method
and show that an approach involving two sub-mm bands and a proper 
MBB fit would be preferable. 

\citet{magdis2011} and \citet{magnelli2012a}
study the SEDs of star-forming and 
sub-mm galaxies that benefit from {\em Spitzer}, \textit{Herschel}, 
and sub-mm photometry. Modeling of MBB and DL07  is applied to their SEDs 
with the aim of studying the dust properties of these objects and 
comparing these to CO observations. In this case, \citet{magdis2011} 
combine dust and CO observations to obtain a measurement 
of the $\alpha_{CO}$ CO-to-gas conversion factor. Their 
method is then further refined by \citet{magdis2012} 
on a larger sample of sources via stacking on FIR and sub-mm maps.

Stacking is also employed by \citet{santini2014}, who bin 
24 $\mu$m-selected galaxies in a grid of $z$-$M^\ast$-SFR and 
study the evolution of dust/stellar 
mass ratio and star formation efficiency (SFE) as a function of redshift
\citep[see also][]{genzel2015}.
These authors find no evidence that the $M_{\rm dust}$ 
evolves with redshift at a given $M^\ast$ or SFR. 
They find that the SFE at $z\sim2$ was a factor $\sim$ 5 higher than at $z\sim0$.
\citet{bethermin2014} also include sub-mm data in their stacking analysis 
and thus improve on the application of DL07 models to the average SEDs 
of distant galaxies.

Despite the wide use of SED fitting to derive dust and gas properties
of distant galaxies, the underlying assumptions and implications 
of instrumental limitations are rarely highlighted nor discussed. 
In this work, we aim to show how real-world limitations affect 
dust and gas mass determinations based on DL07 and MBB SED fitting. 
To achieve our goal, we use the deepest PACS data available to date, namely the 
GOODS fields observations belonging to 
the \textit{PACS Evolutionary Probe} \citep[PEP,][]{lutz2011} and the 
GOODS-\textit{Herschel} \citep{elbaz2011} surveys combined with SPIRE 
photometry from the \textit{Herschel} Multi-tiered Extragalactic Survey 
\citep[HerMES;][]{oliver2012}. The analysis consists  of 
measurements carried out on real sources (individual galaxies and stacked photometry) 
and Monte Carlo simulations. 

The paper is organized as follows. 
Sections \ref{sect:gas_mass} and \ref{sect:sed_fitting}
present the methods adopted to derive gas and dust masses, including
the details of DL07 and MBB SED fitting. 
In Sect. \ref{sect:data} the different collections of data 
used in the analysis are described.
Monte Carlo simulations and their results are presented in Sections \ref{sect:simu_Mstar_SFR_z}
and \ref{sect:restframe_simu} in two flavors: analyzing $M_{\rm dust}$ uncertainties and systematics 
in the $z$-$M^\ast$-SFR space and as a function of rest-frame spectral coverage.
Real-world limitations and prospects for ALMA observations are discussed in Sect. \ref{sect:discussion}.
Finally, Sect. \ref{sect:summary} summarizes our findings.

Throughout this manuscript, we adopt a $\Lambda$-CDM cosmology with 
$\left( h_0,\ \Omega_m,\ \Omega_\Lambda\right)=\left( 0.71,\ 0.27,\ 0.73\right)$; a \citet{chabrier2003} 
initial mass function (IMF); and the \citet{pettini2004} metallicity scale.
Moreover, as described in the next Sections, we make use of the parameterization of 
the $M^\ast$-$Z$ relation by \citet{genzel2015} and of the definition 
of reference (main) sequence of star formation (MS) by \citet{whitaker2014}.
By the term gas mass, we mean  the molecular gas component of galaxies 
unless otherwise specified.


\section{Deriving gas mass}\label{sect:gas_mass}

Molecular gas is generally quantified using 
measurements of the intensity of rotational transitions of the CO 
molecule and adopting a CO-to-H$_2$ conversion factor \citep[see][for a review]{bolatto2013}.
Local samples include several hundred targets 
\citep[e.g.,][]{saintonge2011a}, but
CO detections of star-forming galaxies at intermediate- or high-redshift
are still limited to modest statistics and mostly to luminous galaxies
\citep[][and references therein]{carilli2013}, although they increasingly reach 
so-called main-sequence objects \citep{tacconi2013}.
\citet{genzel2015} 
collected CO line emission measurements for 484 star-forming galaxies at
$z=0$-$3$, including more than 200 $z\le0.05$ galaxies from the COLDGASS \citep[][]{saintonge2011a} 
survey and more than 80 at $z=0.5$-$1.5$, $z=2.0$-$2.5$ from the PHIBSS and PHIBSS2 \citep[][and in prep.]{tacconi2013}
surveys.

Alternative, faster ways to gain gas masses for distant galaxies
involve the use of locally calibrated relations linking gas mass to dust mass
through gas/dust ratio scaling as a function of metallicity or to SFR through 
the scaling of depletion times. In this work, we extensively employ the former
approach, based on the gas/dust ratio, and compare the results to CO- and to $t_{\rm depl}$-based results.

\subsection{Scaling of depletion times}\label{sect:ks}

Based on current CO observations of local and $z\sim1$ objects
\citep[e.g.,][]{saintonge2011a,saintonge2012,tacconi2013}, 
and integrating the relation found by \citet{schmidt1959} and \citet{kennicutt1998b} 
between volume (surface) density of star formation and of molecular gas mass, 
the relation between the total $M_{\rm gas}$ and SFR of MS galaxies can be described as a simple scaling with 
a depletion timescale mildly dependent on redshift 
\begin{equation}\label{eq:ks}
M_{\rm gas}/\textrm{SFR}=\tau_{\rm dep}\textrm{,}
\end{equation}
referred to as the integrated Kennicutt-Schmidt relation (KS)
in what follows. 
Based on CO observations of $z\sim 1.2$ and 2.2 galaxies, compared to 
local galaxies, \citet{tacconi2013} derived a dependence of 
$\tau_{\rm dep}$ on redshift of 
\begin{equation}
\tau_{\rm dep}=1.5\, 10^9\times(1+z)^{-1}\ [\textrm{Gyr}]\textrm{.}
\end{equation}
Based on their collection of $\sim$ 500 CO measurements and on dust mass 
derivations, \citet{genzel2015} derived a three-parameter expression of 
$\tau_{\rm dep}$, depending on stellar mass, $M^\ast$, SFR, and redshift. 
We use the latest update of this relation (R. Genzel, priv. comm.):
%
\begin{eqnarray}
\log\left(\tau_{\rm dep}\right) &=& 0.13 -0.37\times \log\left(1+z\right) \nonumber\\
&&-0.43\times \log\left(\textrm{sSFR}/\textrm{sSFR}_{MS}\right) \nonumber\\
&&+0.10\times \left(\log\left(M^\ast\right)-10.5\right)\textrm{,} \label{eq:genzel2015_tau_dep}
\end{eqnarray}
adopting the \citet{whitaker2014} definition of MS and the Milky Way value
of the $\alpha\left(\textrm{CO}\right)$ conversion factor between CO luminosity
and molecular gas mass.

Here, star formation rates include both IR and UV contributions.
The SFR$_{\rm IR, UV}$ are computed with the \citet{kennicutt1998} calibrations, 
scaled to the \citet{chabrier2003} IMF. Infrared luminosities between 8 and 1000 $\mu$m, 
$L(\textrm{IR})$, are derived by integrating the results of far-IR SED fitting 
(see Section \ref{sect:dl07}). An alternative $L(\textrm{IR})$
estimate was obtained by 
fitting SEDs with the \citet{berta2013a} templates library, leading to equivalent results.

\subsection{The dust method}\label{sect:mgas_mdust}

As described by \citet{magdis2011,magdis2012} and \citet{eales2012}, 
and as widely found in the literature, it is possible to derive
gas masses of (distant) galaxies by relying on measurements of dust mass, 
$M_{\rm dust}$, to be transformed into $M_{\rm gas}$, once the gas-to-dust 
ratio (GDR or $\delta_{\rm GDR}$) of the galaxy is known, i.e.,

\begin{equation}
M_{\rm gas} = M_{\rm dust} \times \delta_{\rm GDR}
.\end{equation} 

A number of assumptions can be adopted to estimate the GDR, from 
simply adopting the Milky Way (MW) value \citep[which turns out to work well for local 
Virgo cluster galaxies;][]{eales2012}, to adopting observed relations, such as the 
\citet{leroy2011} local dependence of GDR on metallicity. 
Here we use a revisited form of the \citet{leroy2011} relation, 
consistently recalibrated to the \citet[][PP04]{pettini2004} metallicity scale 
by \citet{magdis2012} as follows:
\begin{equation}\label{eq:magdis_GDR}
\log\left(\delta_{\rm GDR}\right)=\left(10.54\pm1.0\right)-\left(0.99\pm0.12\right)\times\left(12+\log\left(\textrm{O}/\textrm{H}\right)\right)\textrm{,}
\end{equation}
with a scatter of 0.15 dex. 
In absence of direct spectroscopic measurements, galaxy metallicity can 
be derived from the known stellar mass-metallicity relation and its 
redshift dependence. Recently, \citet{genzel2015} combined the $M^\ast$-$Z$ 
parameterizations of \citet{erb2006}, \citet{maiolino2008}, \citet{zahid2014}, 
and \citet{wuyts2014} into a single relation
\begin{equation}\label{eq:mass_metall}
12 + \log\left(\textrm{O}/\textrm{H}\right)_{\rm PP04} = a-0.087 \times\left(\log M^\ast -b\right)^2
,\end{equation}
where $a=8.74$ and 
$b=10.4+4.46\times \log\left(1+z\right)-1.78\times\left[\log\left(1+z\right)\right]^2$.
In what follows, we adopt this relation, which provides metallicities in 
the PP04 scale. Alternative parameterizations or the so-called
$M^\ast$-$Z$-SFR fundamental relation \citep{mannucci2010,mannucci2011} give similar results, for $z\le2$ massive star-forming galaxies
with $Z\sim Z_\odot$, which dominate the samples discussed 
below,- modulo the adopted metallicity scale.

We stress that it is very important to use a self consistent set of relations, 
i.e., the $\delta_{\rm GDR}$ vs. $Z$ scaling and the $M^\ast$ vs. $Z$ relation must 
be calibrated to the same metallicity scale to produce meaningful results.


\section{Deriving dust mass through far-IR SED fitting}\label{sect:sed_fitting}

Dust mass can be estimated by fitting the far-IR SED
of galaxies in several ways, e.g., by means of a MBB function  or with 
dust models such as those prepared by \citet[][DL07]{dl07}. 
In this Section we provide more details about these two approaches.

\subsection{Modified blackbody}\label{sect:mbb}

The FIR SED of galaxies can be simplistically reproduced with a single-temperature
modified blackbody, assuming that emission comes from a single-temperature $\delta$
distribution of dust. In this case, the rest-frame SED is modeled as \citep[e.g.,][]{blain2002}
\begin{equation}\label{eq:MBB}
L_\nu=\Omega\epsilon_\nu B_\nu(T_{\rm dust})
,\end{equation}
where $\Omega$ is the solid angle of emission, 
$\epsilon_\nu$ is the emissivity coefficient, 
and $B_\nu(T_{\rm dust})$ is the Planck function
\begin{equation}\label{eq:planck}
B_\nu(T_{\rm dust}) = \frac{2h}{c^2} \frac{\nu^3}{\exp\left(\frac{h\nu}{k_{\rm B}T_{\rm dust}}\right)-1}
\end{equation}
in units of $[$erg s$^{-1}$ Hz$^{-1}$ m$^{-2}$ sr$^{-1}]$.

For a uniform medium (e.g., a dust cloud in our case) of optical depth, $\tau_{\nu}$, radiative transfer implies that 
radiation is reduced by a factor $\exp\left(-\tau_{\nu}\right)$ \citep[e.g.,][]{rybicki1979}. 
Therefore, we can also write $\epsilon_\nu= 1-exp\left(-\tau_{\nu}\right)$ \citep[e.g.,][]{benford1999,omont2001}.
We allow $\nu_0$ to be the frequency where $\tau=\tau_0=1$. The optical depth at the generic frequency $\nu$ can be approximated as a power law 
\begin{equation}\label{eq:beta}
\tau_\nu=\tau_0 \left(\nu / \nu_{0}\right)^{\beta}\textrm{.}
\end{equation}
Therefore, 
\begin{equation}
\epsilon_{\nu}=1-exp\left(-\tau_{\nu}\right)=1-exp\left[-\tau_0 (\nu / \nu_{0})^{\beta}\right]\textrm{.}
\end{equation}

On the other hand, the frequency-dependent optical depth can be written as \citep[e.g.,][]{beelen2006}
\begin{equation}
\tau_\nu = \kappa_\nu \int_s \rho\, ds^\prime\textrm{,}
\end{equation}
where $\kappa_\nu$ is the mass absorption coefficient of dust at rest frequency $\nu$ and $\rho$ is 
the total mass density. If the emission is optically thin (i.e., $\tau\ll 1$), 
we can expand $\epsilon_\nu$ in its Taylor series expression\footnote{using $e^x=1+x+x^2/2!+x^3/3!+...$} and 
thus obtain
\begin{equation}
\epsilon_\nu\sim\kappa_\nu M_{\rm dust}\textrm{.}
\end{equation}
Therefore, the emergent luminosity from a given dust mass is given by
\begin{equation}\label{eq:mbb_full_expr}
L_\nu = 4\pi M_{\rm dust} \kappa_\nu B_\nu(T_{\rm dust})\textrm{,}
\end{equation}
where we have assumed that emission is isotropic over a spherical surface and 
$\kappa_\nu=\kappa_0\left(\nu/\nu_0\right)^\beta$. For a galaxy FIR dust emission, $\beta$ is in the range $1.5-2.0$
\citep[e.g.,][and references therein]{magnelli2012a}.
Here $M_{\rm dust}$ is in units of $[$kg$]$, $\kappa_\nu$ is given 
in $[$m$^2$ kg$^{-1}]$, and therefore $L_\nu$ turns out to have units of $[$erg s$^{-1}$ Hz$^{-1}]$.

Making the dependence of frequency explicit, Eq. \ref{eq:mbb_full_expr} can be rewritten as
\begin{equation}\label{eq:mbb_simpl}
L_\nu \propto \frac{\nu^{3+\beta}}{exp\left(\frac{h\nu}{k_{\rm B} T_{\rm dust}}\right)-1}\textrm{,}
\end{equation}
which is the usual simplified form of the modified blackbody function.

Inverting Eq. \ref{eq:mbb_full_expr}, it is possible to obtain an estimate of 
dust mass, once the rest-frame luminosity of a galaxy at a given FIR wavelength is known, a
MBB fit to the data is possible, and a value of $\kappa_\nu$ (or $\kappa_0$ and $\nu_0$) for dust is 
adopted, as
\begin{equation}\label{eq:mdust_MBB}
M_{\rm dust} = \frac{L_\nu}{4\pi \kappa_\nu B_\nu\left(T_{\rm dust}\right)}\textrm{.}
\end{equation}
This holds when working at rest-frame frequencies and needs to be transformed into 
an observed frame in the usual way. 
Finally, combined with the knowledge of the gas-to-dust ratio, $\delta_{\rm GDR}$, 
Eq. \ref{eq:mdust_MBB} leads to an estimate of gas mass.

\subsection{The use of $\kappa_\nu$}\label{sect:kappa}

To compute dust masses through a modified blackbody fit, 
one needs to assume a value of $\kappa_\nu$ at a given wavelength, possibly within
the rest-frame wavelength range covered by actual photometric data or by a value of 
$\kappa_0$ at the reference frequency $\nu_0$, where $\tau_\nu=\tau_0=1$.

One viable approach is to adopt the set of $\kappa_\nu$ computed by \citet[][see their Table 6]{ld01},
and either use the value appropriate for the given rest-frame wavelength from their Table (or interpolated) or apply Eq. \ref{eq:beta}.

It is common \citep[e.g.,][]{magdis2012,magnelli2012b} to follow the second approach, 
using the values of $\kappa_\nu$ tabulated by \citet{ld01} and applying $\kappa_\nu=\kappa_0\left(\nu/\nu_0\right)^\beta$.
In this procedure, it is important to properly apply $k$ correction to all terms in 
Eq. \ref{eq:mdust_MBB} and use the correct values of $\kappa_0$, $\nu_0$, and $\beta$.
In fact, Eq. \ref{eq:beta} assumes that the slope $\beta$ used while 
fitting the observed SEDs (see Eq. \ref{eq:mbb_simpl}) is the same $\beta$ adopted in 
the computation of $\kappa_\nu$. 

Often, $\beta$ is let free to vary while fitting, if enough datapoints are available, 
or if there are not enough available, a fixed value (e.g., $\beta=1.5$) is adopted.
On the other hand, the value of $\beta$ adopted by \citet{dl07} is $\sim2$ \citep[see][]{cortese2012}
and the values of $\kappa_\nu$ in Table 6 of \citet{ld01} (in its electronic version with the old 2001 computation and for the MW dust mixture with $q_{\rm PAH}=0.47$) turn out to show
$\beta_{\rm LD01, a}=2.07$ between $\lambda=100$ and 600 $\mu$m, and 
$\beta_{\rm LD01, b}=1.68$ at $\lambda>600$ $\mu$m.

The direct consequence of the mismatch between the adopted value of $\beta$ and the 
$\beta\sim 2$ describing the actual \citet{ld01} $\kappa_\nu$ is that one
should apply an additional correction factor when applying Eq. \ref{eq:beta} (i.e., 
Eq. 6 in Magdis' paper and Eq. 4 in Magnelli's paper) or there will be 
a dependence of $M_{\rm dust}$ on the value of $\nu_0$ used.
This dependence on frequency has a power of roughly $-0.5$ when adopting $\beta=1.5$.

\citet{bianchi2013} points out this effect and also shows that using the $\kappa_\nu$
values tabulated by \citet{draine2003}, i.e., those actually used in the \citet{dl07} 
models, is 
more appropriate and leads to consistent results when comparing MBB-based and DL07-based 
dust masses. Also, Bianchi points out that adopting a different value of $\beta\ne2.08$
leads to a dependence of model normalizations and, hence, of $M_{\rm dust}$, on 
the reference wavelength adopted when computing $\kappa_\nu$ starting from $\kappa_0$.
Finally, \citet{bianchi2013} shows that after accounting for the needed 
corrections on $\kappa_\nu$, a mild dependence on blackbody shapes is still left because 
that the best-fit dust temperature also depends on $\beta$ \citep[see also][]{bianchi1999}.

In practice, these subtle differences in the treatment of $\kappa_\nu$ 
and $\beta$ have led to MBB-based dust masses changing by a factor 
3-5 for the same SED and the same basic reference dust opacity
(see also Sect. \ref{sect:compa_dl07_mbb}).

\begin{figure*}[!ht]
\centering
\includegraphics[width=0.32\textwidth]{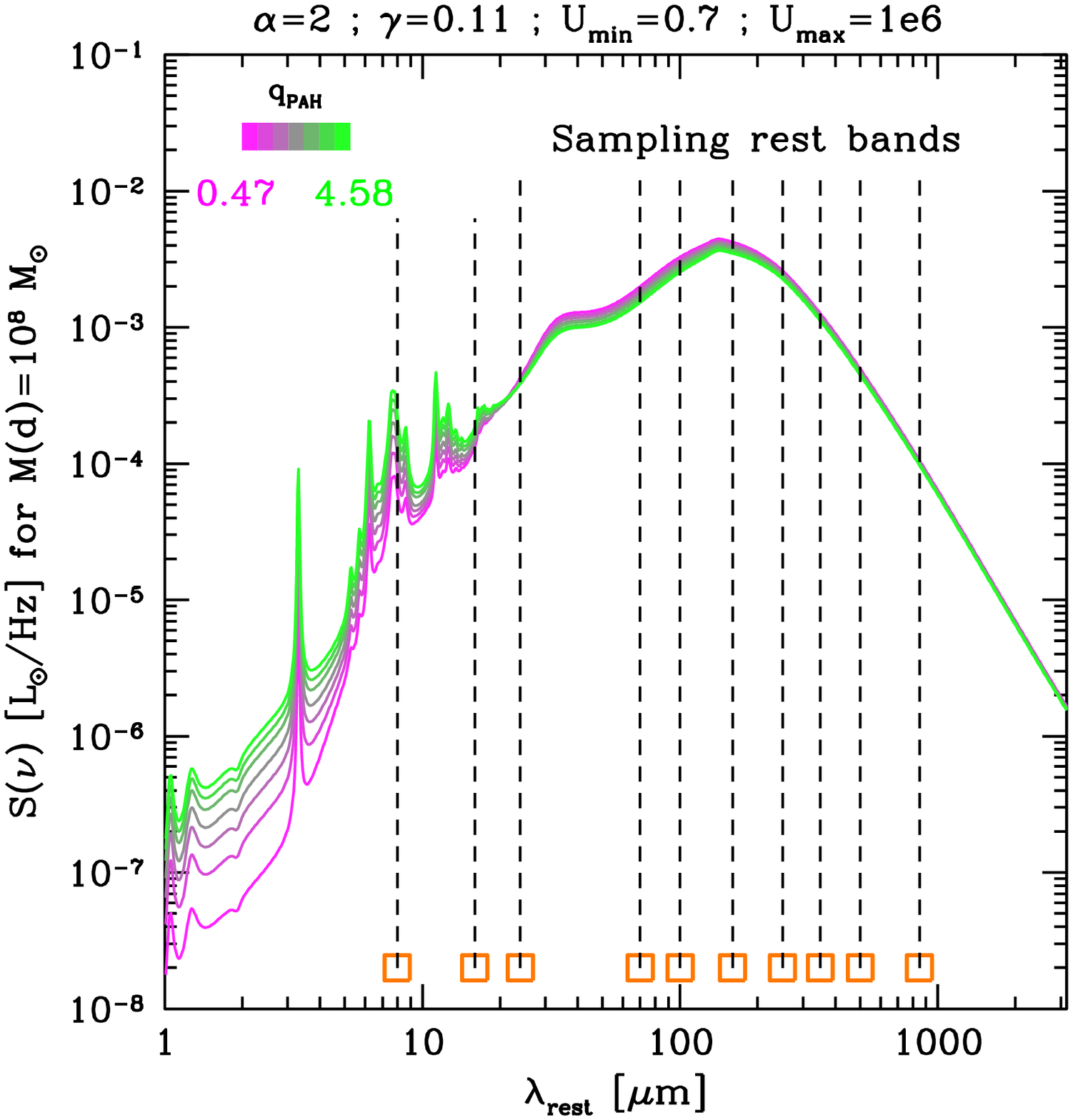}
\includegraphics[width=0.32\textwidth]{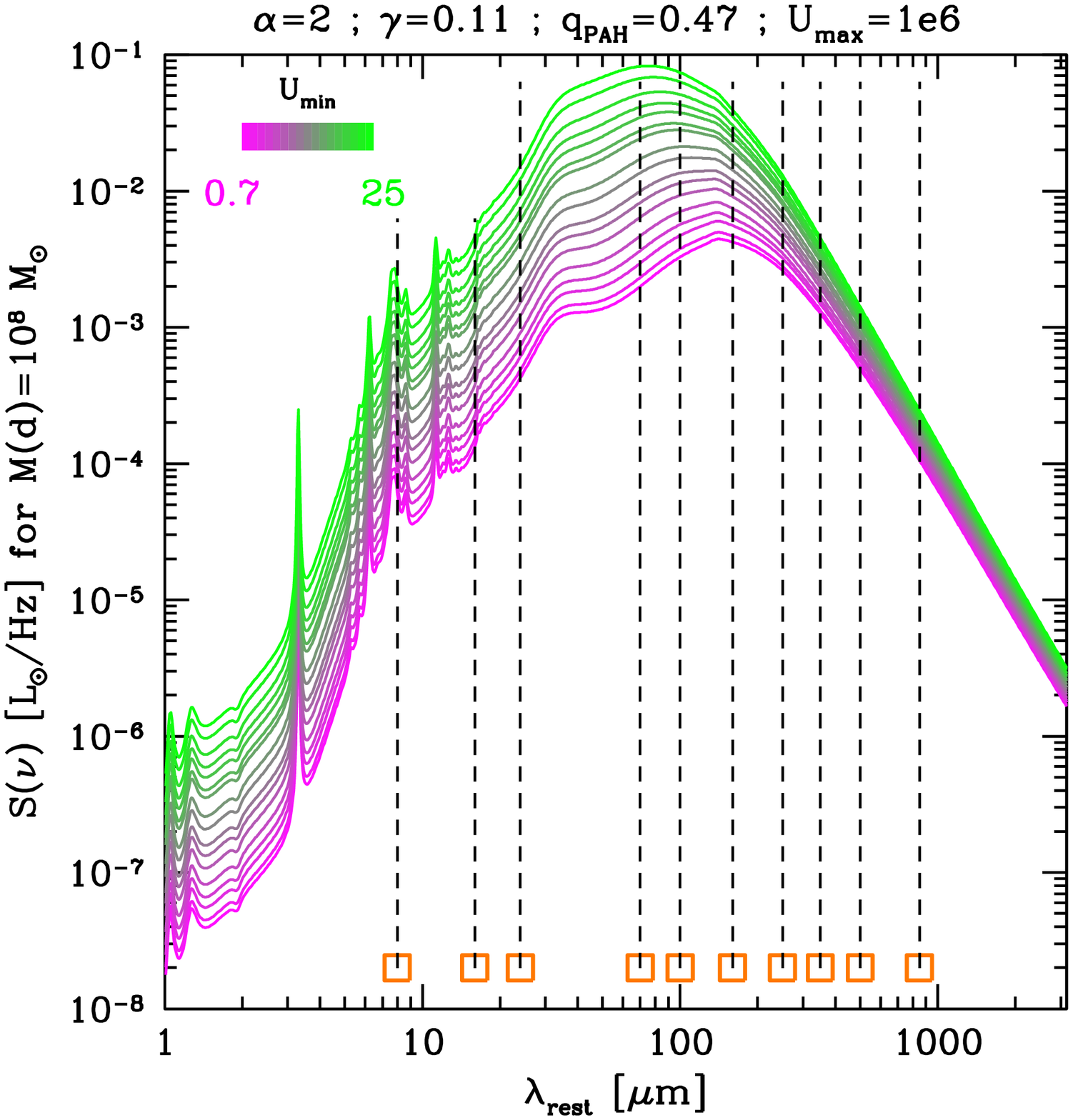}
\includegraphics[width=0.32\textwidth]{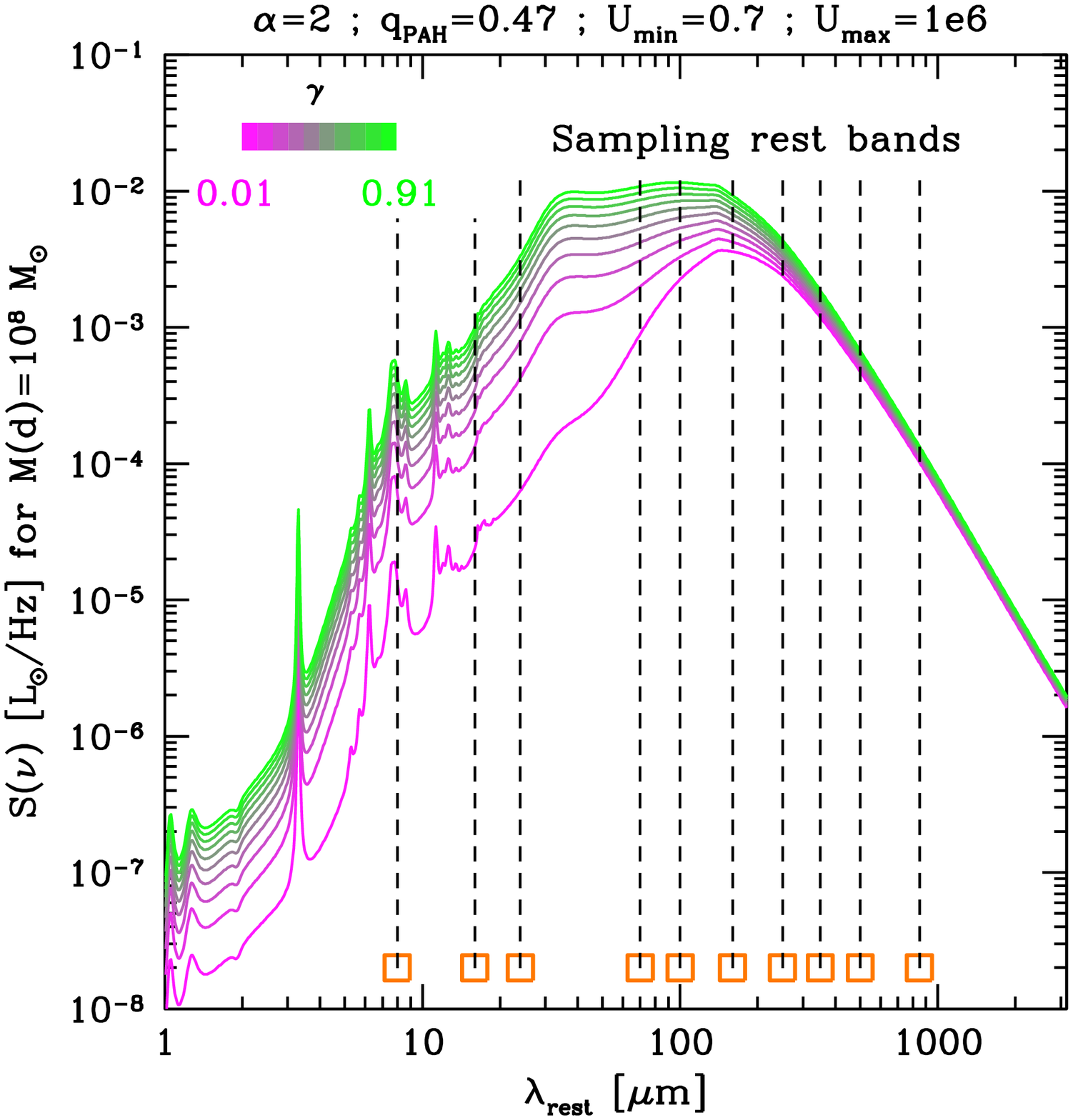}
\caption{Effect of varying $q_{\rm pah}$ (left), $U_{\rm min}$ (center), $\gamma$ (right), 
in \citet{dl07} models, independently. On top of each panel, the values of the frozen parameters are 
quoted. The orange squares at the bottom indicate the position of the chosen rest-frame 
bands used to generate artificial catalogs.}
\label{fig:dl07}
\end{figure*}

\subsection{Draine \& Li (2007) models}\label{sect:dl07}

A more sophisticated and physically motivated approach to FIR SED fitting and derivation 
of dust masses is to adopt the \citet[][DL07]{dl07} dust models, which are an upgrade of those originally
developed by \citet{dl01}, \citet{ld01}, and \citet{weingartner2001}.

In short, these models describe interstellar dust as a mixture of carbonaceous and amorphous silicate 
grains, whose size distributions are chosen to mimic the observed extinction law in the MW, 
Large Magellanic Cloud (LMC), or Small Magellanic Cloud (SMC) bar region.
As described by \citet{dl07}, carbonaceous grains have the properties
of polycyclic aromatic hydrocarbons (PAH) molecules and ions when the effective radius is $a<$ 5.0 $[$nm$]$, the
properties of graphite spheres when  $a\gg$ 10 $[$nm$]$, and optical properties
intermediate between those of PAH particles and graphite particles for $5 < a < 10$ $[$nm$]$.
The ionization fraction $x_{\rm ion(a)}$ of the PAH particles is assumed to be
the average for the diffuse ISM.
The properties of grains are parameterized by the PAH index $q_{\rm PAH}$, which is defined as the fraction 
of dust mass in the form of PAH molecules.

The dust distribution is divided in two components: the diffuse interstellar medium (ISM), usually 
constituting the majority of the dust budget and dust enclosed in photodissociation 
regions (PDRs). The former is heated by a radiation field of constant intensity $U_{\rm min}$.
The latter, representing a fraction $\gamma$ of the total amount of dust, is exposed to
starlight with intensities ranging from $U_{\rm min}$ to $U_{\rm max}$. 
Although PDRs usually provide a small fraction of the total dust mass, they 
can contribute to the majority of dust emission in mid-IR SEDs. 
The mass $dM$ of dust exposed to starlight intensities in the range $[U,U+dU]$ is 
given by the following power-law distribution:
\begin{equation}
dM_{\rm dust} = \textrm{const} \times U^{-2} \, dU  \ \ \ \ \ \  \textrm{for } U_{\rm min} < U < U_{\rm max}\textrm{.}
\end{equation}
The diffuse ISM is modeled by setting $U_{\rm max} = U_{\rm min}$.

Summing the diffuse and PDR components, we obtain
\begin{equation}
\frac{dM_{\rm dust}}{dU}=\left(1-\gamma\right)\delta\left(U-U_{\rm min}\right)+\gamma M_{\rm dust}\frac{\alpha-1}{U_{\rm min}^{1-\alpha}-U_{\rm max}^{1-\alpha}} U^{-\alpha}\textrm{,}
\end{equation}
for $\alpha\ne 1$; here $\delta$ is the delta function representing the diffuse interstellar radiation field 
of intensity $U=U_{\rm min}=U_{\rm max}$.

Using the numerical methods described in \citet{dl01}, \citet{ld01}, and \citet{dl07}, 
temperature distribution functions are computed for all particles
small enough for quantized heating to be important; large grains are
treated as having steady-state temperatures determined by starlight
heating and radiative cooling equilibrium. With temperature distributions and 
dust absorption cross sections, the time-averaged IR emission for a given grain size and type is computed, and 
finally the power radiated per unit frequency and unit mass for the given 
dust mix exposed to the radiation field $U$, is computed summing over all grain types and sizes.
Integrating between $U_{\rm min}$ and $U_{\rm max}$, this gives 
the specific power per unit mass $p_\nu(q_{\rm PAH},U_{\rm min}, U_{\rm max}, \alpha)$ \citep[see Eq. 9 in][]{draine2007b}.

\citet{dl07} thus propose to fit the emission of galaxies through a linear 
combination of emission from diffuse ISM dust and PDRs emission. 
The emission spectrum, expressed as emissivity per hydrogen nucleon, $j_\nu=\left(\frac{M_{\rm dust}}{M_{\rm H}}\right)m_{\rm H}\frac{p_\nu}{4\pi}$, can then 
be simply written as 
\begin{eqnarray}
j_\nu\left(q_{\rm PAH},U_{\rm min},U_{\rm max},\alpha,\gamma\right) &=& (1-\gamma)j_\nu\left[U_{\rm min},U_{\rm min}\right] \nonumber\\
&& + \gamma j_\nu\left[U_{\rm min},U_{\rm max},\alpha\right]\textrm{,}
\end{eqnarray}
in units of $[$erg s$^{-1}$ Hz$^{-1}$ sr$^{-1}$ H$^{-1}]$; $j_\nu$ is the quantity contained in the 
model files available online\footnote{http://www.astro.princeton.edu/$\sim$draine/dust/irem.html}.

As shown by \citet{draine2007b}, the total dust luminosity is given by
\begin{equation}\label{eq:ldust}
L_{\rm dust}=\langle U\rangle P_0 M_{\rm dust}
,\end{equation}
where the dust-weighted mean starlight intensity, or mean radiation field, the scale factor is
\begin{equation}\label{eq:U_avg}
\langle U\rangle = \left[\left(1-\gamma\right)U_{\rm min}+\frac{\gamma\ln \left(U_{\rm max}/U_{\rm min}\right)}{U_{\rm min}^{-1}-U_{\rm max}^{-1}}\right]\textrm{,}
\end{equation}
if $\alpha=2$ as in our assumptions and $P_0$ is the power absorbed per unit dust mass 
in a radiation field $U=1$.

In principle, the model includes six free parameters: 
$q_{\rm PAH}$; $U_{\rm min}$; $U_{\rm max}$; $\alpha$; $\gamma$; 
$M_{\rm dust}$. Dust mass is basically the model's normalization.
In practice, studying local galaxies in the \textit{Spitzer} Nearby Galaxy Survey (SINGS) and fitting their FIR SEDs, \citet{draine2007b} 
demonstrate that some of the parameters can be limited to a restricted range of values or even fixed to a single value.
The overall fit is not very sensitive to the dust model adopted (MW, LMC, SMC), and can be limited to 
MW models alone, implying only seven values of $q_{\rm PAH}$. Also, the actual values of $\alpha$ and $U_{\rm max}$ 
do not significantly influence the performance of models, and fixing them 
to $\alpha=2$ and $U_{\rm max}=10^6$ is a reasonable choice to successfully reproduce 
SINGS galaxies. Finally, since small values of $U_{\rm min}$ correspond to dust temperatures below 15 K
that cannot be constrained by FIR photometry alone, in the absence of rest-frame sub-mm data, they suggest 
 limiting U$_{\rm min}$ to the range $0.7 \le U_{\rm min} \le 25$. A direct consequence is a possible 
underestimate of dust mass, if a significant amount is stored in a very cold component, but \citet{draine2007b} 
conclude that omitting sub-mm data increases the scatter on $M_{\rm dust}$ by 50\%, but does not 
induce any significant systematic offset.
Most observational papers in the literature 
\citep[e.g.,][]{magdis2012,magnelli2012a,santini2014} adopt
this optimized choice of parameters, relying on a setup that 
has only been thoroughly verified for local galaxies near solar
metallicity \citep{draine2007b}.

Figure \ref{fig:dl07} shows how varying each parameter ($q_{\rm pah}$, $U_{\rm min}$, $\gamma$) independently, and fixing 
the others, induces modifications on the shape of the modeled SED. We also indicte also 
a set of IR rest-frame bands, which are available in surveys of nearby or distant galaxies. 

The choice of the actual PAH abundance, 
parameterized by $q_{\rm PAH}$ and  limited to MW models, mostly influences the short-wavelength 
SED (1-20 $\mu$m), and is not making a difference at $\lambda>20$ $\mu$m. 
Therefore FIR surveys of galaxies are not very sensitive to this parameter 
and cannot put tight constraints on its value.

The parameter $U_{\rm min}$ regulates the range of starlight intensities that are heating dust. 
Since the amount of dust exposed to starlight intensities in between $U$ and $U+dU$ 
is modeled as a power law $\propto U^{-\alpha}$, the smaller $U_{\rm min}$, 
the larger is the amount of dust subject to low-energy radiation. As a consequence, the 
dust component dominating the SED is colder for smaller values of $U_{\rm min}$ and 
the FIR peak shifts to longer wavelengths; at the same time, at a given wavelength 
and for a given total dust mass, the model intensity decreases.
Thus, we expect degeneracies and possibly some amount of systematics between 
$M_{\rm dust}$ and $U_{\rm min}$.

The fraction $\gamma$ of dust locked in PDRs influences the repartition of the emitted energy between the 
$U_{\rm max}=U_{\rm min}$ component (ISM) and the rest. 
When fitting the SEDs of local SINGS galaxies, \citet[][see their Table 4]{draine2007b}
found that the SEDs of vast majority of galaxies were successfully reproduced with $\gamma\le0.15$ with only 
a couple of cases extending to $\gamma\sim0.3-0.4$.

Once the FIR SED is fitted with these models, its dust mass can be computed as
\begin{equation}
M_{\rm dust}=\left(\frac{M_{\rm dust}}{M_{\rm H}}\right) m_{\rm H} \frac{L_\nu}{4\pi j_\nu}\textrm{,}
\end{equation}
where the right-hand term is evaluated at a given rest-frame frequency $\nu$ of choice, which is at best covered by observations; $L_\nu$ is the rest-frame luminosity at frequency $\nu$; 
$m_{\rm H}$ is the mass of the hydrogen nucleon; and $M_{\rm dust}/M_{\rm H}$ is the 
dust-hydrogen ratio characteristic of the adopted dust model (and tabulated together with $q_{\rm PAH}$).
Making sure to use the right units for all terms, we obtain $m_{\rm H}=1.67262178\, 10^{-27}$ $[$kg$]$, 
$L_\nu$ in units of $[$erg s$^{-1}$ Hz$^{-1}]$, and therefore dust mass turns out to have units of $[$kg$]$.


\section{Available data}\label{sect:data}

The PEP survey \citep{lutz2011} has covered 
the most popular blank fields with the PACS \citep{poglitsch2010} instrument 
aboard \textit{Herschel} \citep{pilbratt2010}
at 100 and 160 $\mu$m. In addition, the same fields have been observed by 
the HerMES survey \citep{oliver2012}
with SPIRE \citep{griffin2010} at 250, 350, 500 $\mu$m. Furthermore, the 
GOODS-\textit{Herschel} \citep{elbaz2011} survey has provided deeper PACS coverage of 
the GOODS fields, reaching the confusion limit at 100 and 160 $\mu$m.
Finally, the PEP team has combined all the available PACS data in the GOODS-N
and GOODS-S fields, including science verification (SV) and ECDFS observations, 
producing the deepest FIR maps obtained so far \citep{magnelli2013}.

Here we make use
of PEP first data release 
(DR1\footnote{http://www.mpe.mpg.de/ir/Research/PEP/public\_data\_releases},
Lutz et al. \citeyear{lutz2011}),
the HerMES DR2 and DR3\footnote{hedam.lam.fr/HerMES/index/download} 
\citep{oliver2012,roseboom2010,roseboom2012}, and combined PEP + GOODS-H data 
\citep[included in PEP DR1,][]{magnelli2013}.

\subsection{Individual sources}\label{sect:data_individual}

The deepest PACS fields, GOODS-N and GOODS-S benefit from an 
extensive coverage at all wavelengths. We use the PACS 70, 100, 160 $\mu$m
data presented by \citet{magnelli2013}; the SPIRE HerMES data \citep{oliver2012,roseboom2010}; 
the multiwavelength catalogs built by \citet{berta2011} and 
by the MUSIC team \citep{grazian2006}; the collection of spectroscopic
redshift by \citet{barger2008}, \citet{balestra2010}, and \citet{berta2011}; and 
the photometric redshifts by \citet{berta2011} and \citet{wuyts2011a}. 
We refer to the dedicated publications for more details on each dataset and to \citet{berta2011,berta2013a} 
for an overview.

We select objects with at least a 3$\sigma$ detection in the PACS 160 $\mu$m band, 
which for the redshift range of interest turns out to be the 
best single-band proxy of star formation rate 
\citep{elbaz2011,nordon2013}.

To this generic sample of $\sim$1400 160 $\mu$m sources, we add the list 
of 61 sub-mm galaxies (SMGs) compiled by \citet{magnelli2012a} 
and distributed in the main PEP fields. 
For these sources, a rich multiwavelength dataset is available, 
ranging from the optical to the sub-mm as compiled by \citet{magnelli2012a}.

\begin{table}
\centering
\begin{tabular}{c c c}
\hline
\hline
$z$     & $\log\left(\textrm{sSFR}\right)$ & $\log\left( M^\ast\right)$\\
\hline
0.0-0.1 & -10.5 to -9.3 & 10.0 to 11.5 \\
0.1-0.3 & -10.5 to -9.0 & 10.0 to 11.5 \\
0.3-0.6 & -10.0 to -8.5 & 10.0 to 11.5 \\
0.6-1.0 & -9.8 to -8.0  & 10.0 to 12.0 \\
1.0-1.5 & -9.5 to -8.0  & 10.0 to 12.0 \\
1.5-2.5 & -9.2 to -7.8  & 10.0 to 12.0 \\
\hline
\end{tabular}
\caption{Regions of the  $M^\ast$-SFR-$z$ space 
where good stacked SEDs are available (see Figs. 
4 and 5 in Magnelli et al. \citeyear{magnelli2014}).}
\label{tab:good_ranges}
\end{table}

\subsection{Stacked photometry}\label{sect:data_stacks}

\citet{magnelli2014} studied the dust temperature, $T_{\rm dust}$, 
in star-forming galaxies 
as a function of their position in the $z$-$M^\ast$-SFR space.

These authors used the $M^\ast$, SFR, $z$ estimates by \citet{wuyts2011a,wuyts2011b}
in the GOODS-N, GOODS-S, and COSMOS fields. Stellar masses are based on optical-NIR SED fitting 
carried out adopting \citet{bc03} templates. Star formation rates are based on a ladder 
of indicators (SED fitting, mid-IR photometry, far-IR photometry), calibrated on 
\textit{Herschel} observations of $\sim 7000$ galaxies in PEP fields.
Redshifts are the combination of a collection of spectroscopic measurements and 
photometric estimates (see above).

\citet{magnelli2014} produced a grid in the $M^\ast$-SFR-$z$ space, binned a 
$K_s$-selected sample accordingly, and finally produced an average SED 
(70, 100, 160, 250, 350, 500 $\mu$m) for each bin by means of  stacking 
on \textit{Herschel} maps.

We make use of the stacked photometry by \citet{magnelli2014}, 
plus similar \textit{Spitzer}/MIPS 24 $\mu$m data, taking care 
to limit our analysis to those regions of the  $M^\ast$-SFR-$z$ space, 
where good stacked SEDs are available (see Figs. 
4 and 5 in Magnelli et al. \citeyear{magnelli2014}). Table \ref{tab:good_ranges}
summarizes the $z$, $M^\ast$, sSFR ranges of interest.

\subsection{CO-detected sources}\label{sect:data_CO}

A direct measurement of $M_{\rm gas}$ comes from the intensity of spectral lines 
of molecular gas tracers, such as CO, modulo the conversion factor from the 
given molecular species to total gas.
Here we collect the far-IR and sub-mm photometry of galaxies that were 
observed in CO-line spectroscopy to  compare
CO-based and dust-based $M_{\rm gas}$ estimates directly.

The following samples of CO-detected objects are taken into account:
\begin{itemize}
\item PHIBSS $z\sim1$ galaxies \citep{tacconi2013};
\item other star-forming galaxies by \citet{magnelli2012b} and \citet{daddi2010a};
\item lensed galaxies \citep{saintonge2013};
and\item sub-mm galaxies (SMGs) by \citet{bothwell2013}.
\end{itemize}

Appendix \ref{sect:app_CO_data} describes the publicly available data for each
case.
In summary, the number of CO-detected objects that have enough FIR photometry
to allow a DL07 SED fitting and that have a $M^\ast$ estimate is: 23 sources in EGS from the PHIBSS catalog, 5 BzK galaxies by \citet{daddi2010a}, six sources by \citet{magnelli2013},
ten lensed galaxies by \citet{saintonge2013}, 
and seven SMGs by \citet{bothwell2013}; this is a total of 51 objects.

Stellar masses of these CO-detected sources come from the 
respective publications dealing with each sample. 
They were all computed with \citet{chabrier2003} IMF and BC03 
models or are consistent with these assumptions. 
For SMGs, the adopted stellar masses are similar to the scale of 
\citet{hainline2011}.
The adopted cosmological parameters are all in line with those 
assumed in this work.


\section{Accuracy of  $M_{\rm dust}$ and $T_{\rm dust}$, as allowed by \textit{Herschel} surveys}\label{sect:simu_Mstar_SFR_z}

\begin{figure*}[!ht]
\centering
\includegraphics[width=0.37\textwidth]{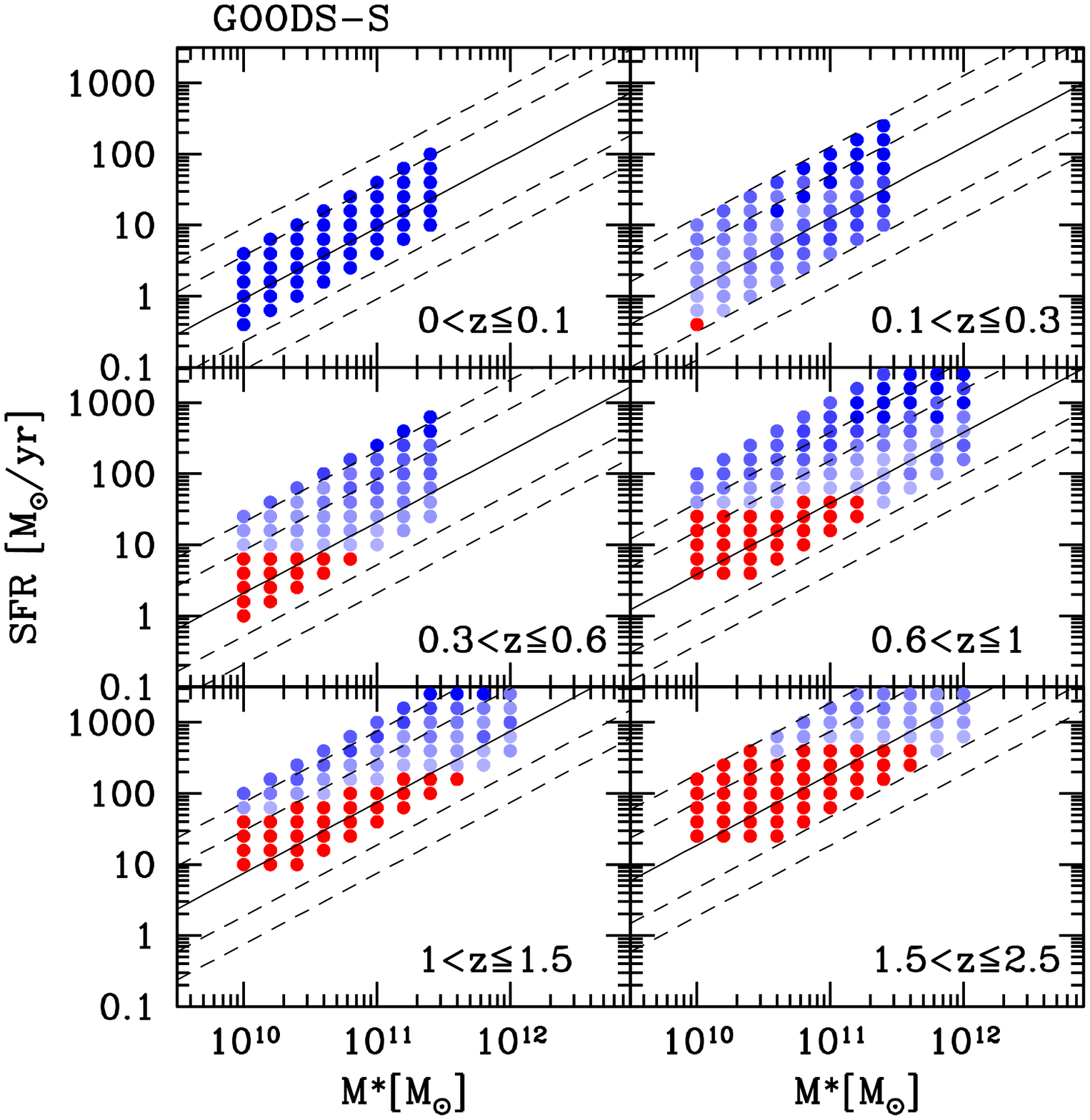}
\includegraphics[width=0.37\textwidth]{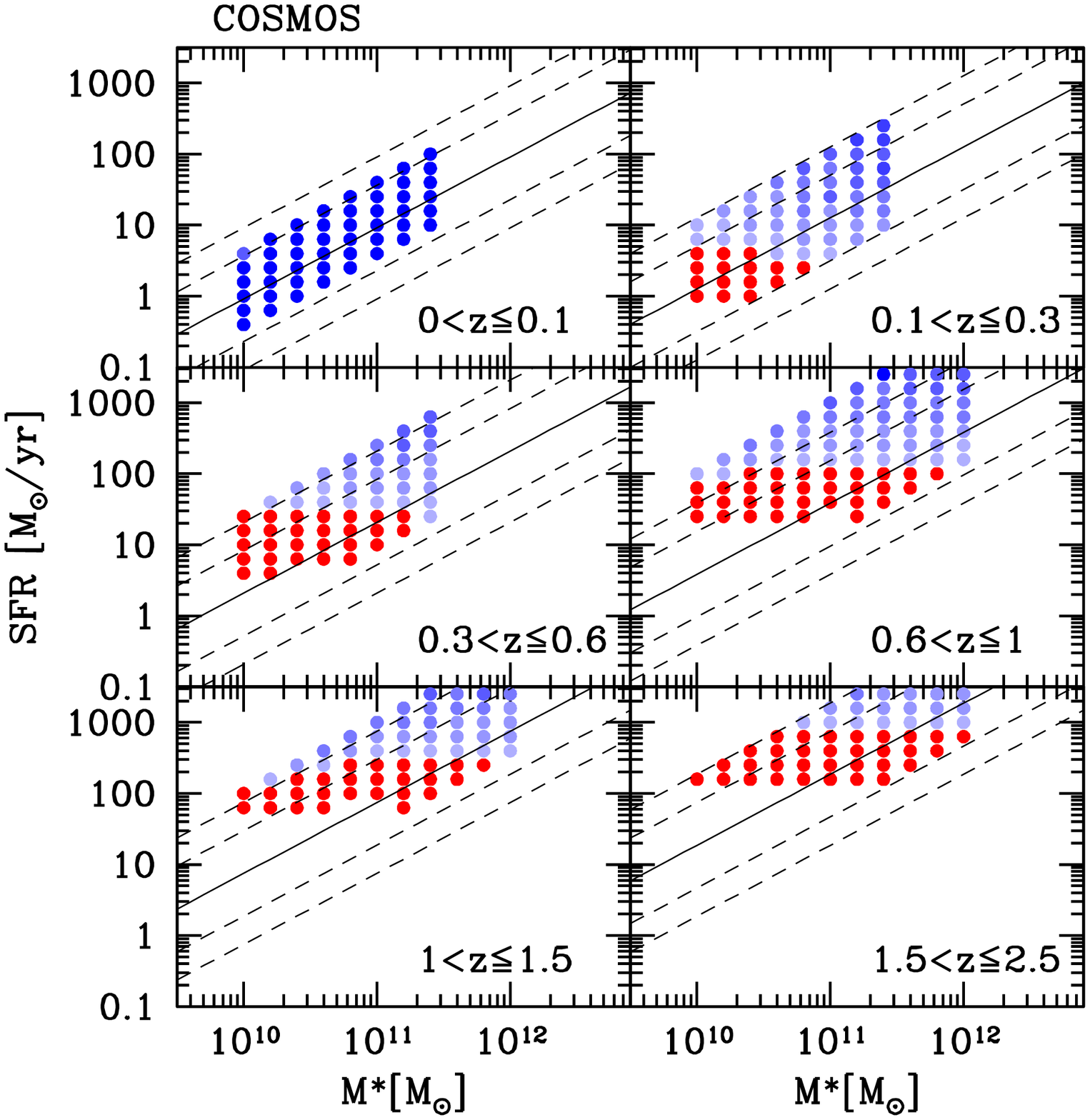}
\includegraphics[height=0.37\textwidth]{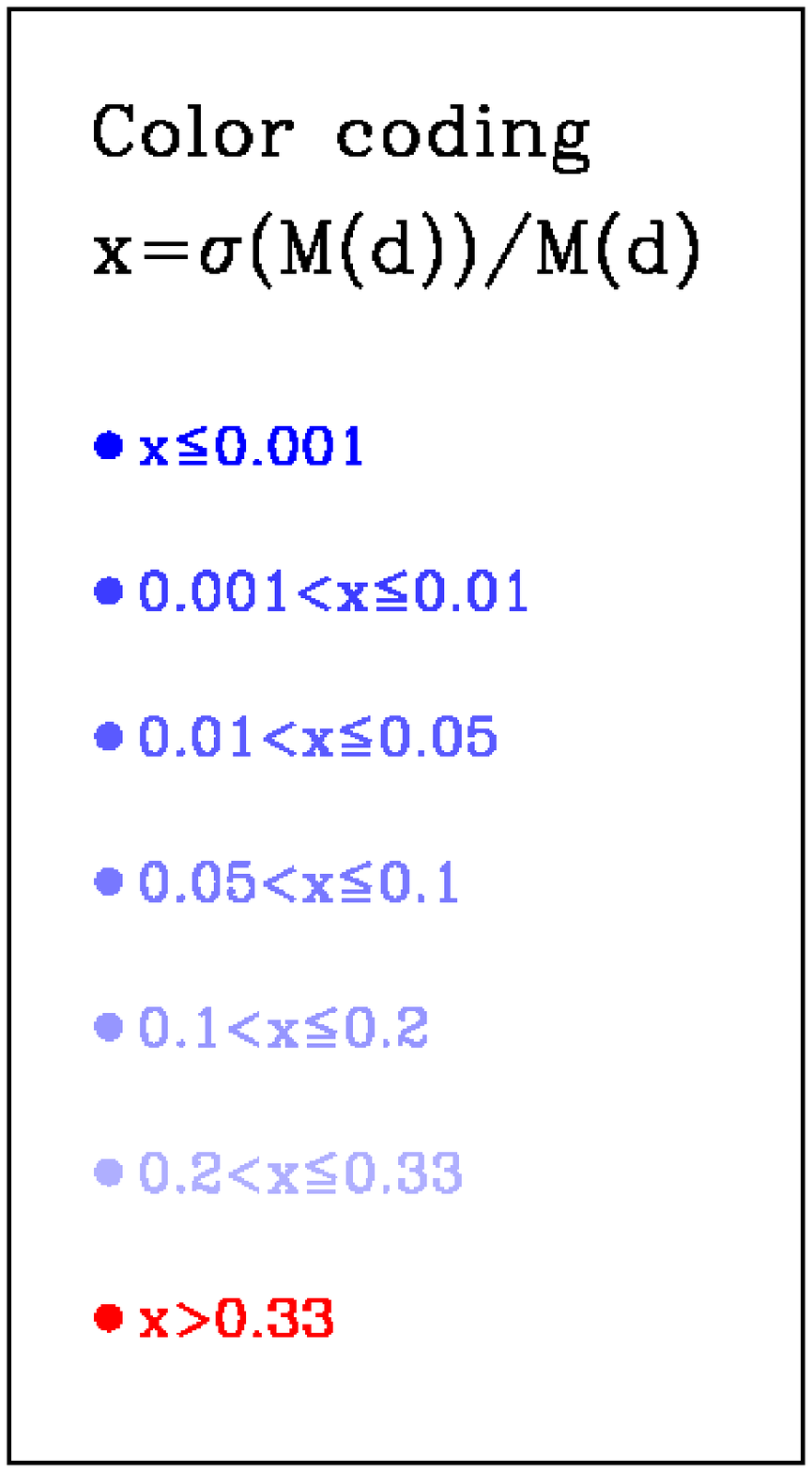}
\caption{Relative uncertainty of dust masses (defined as $\sigma(M_{\rm dust})/M_{\rm dust}$),
based on DL07 SED modeling, as a function of position in the $z$-$M^\ast$-SFR space.
The two diagrams belong to two independent simulations, obtained noise levels of individual 
detections in GOODS-S and COSMOS, applied 
to artificial photometry based on the SEDs by \citet{magnelli2014}.
For reference, black lines denote the position of the main sequence (MS, solid line) of star formation \citep{elbaz2011} and 
$\pm4$, $\pm10$ MS levels (dashed).}
\label{fig:mstar_sfr_2b_NEW}
\end{figure*}

\begin{table}[!t]
\centering
\begin{tabular}{c c c}
\hline
\hline
Band    & GOODS-S       & COSMOS \\
\hline
24 $\mu$m       & 20.0 $\mu$Jy  & 45 $\mu$Jy    \\
70 $\mu$m       & 1.0 mJy       & --            \\
100 $\mu$m      & 0.54 mJy      & 5.0 mJy       \\
160 $\mu$m      & 1.3 mJy       & 10.2 mJy      \\
250 $\mu$m      & 11.7 mJy      & 12.4 mJy      \\
350 $\mu$m      & 14.0 mJy      & 14.4 mJy      \\
500 $\mu$m      & 15.9 mJy      & 16.6 mJy      \\
\hline
\end{tabular}
\caption{\textit{Herschel} PACS and SPIRE 3$\sigma$ depths in the GOODS-S and COSMOS
fields, adopted in Monte Carlo simulations. SPIRE noise includes confusion \citep[see][]{nguyen2010}.}
\label{tab:depths}
\end{table}

\citet{magnelli2014}  provide a tool to 
derive $T_{\rm dust}$ of a galaxy once its redshifts, $M^\ast$ and SFR are known, 
as well as to obtain its expected SED based on the \cite{dale2002} template library.
\citet{genzel2015} re-analyze the data by \citet{magnelli2014} and produce 
new scalings linking $T_{\rm dust}$ and $M_{\rm gas}$ to $M^\ast$, sSFR, and $z$.

In this Section, we would like to study how the observational limitations 
of \textit{Herschel} photometry affect the derivation of  $M_{\rm gas}$ and $T_{\rm dust}$. 
To this aim, we adopt the signal-to-noise ratios of two case studies:
the deepest field of the PEP survey, i.e., GOODS-S; and the widest, but 
$\sim$ 8 times shallower, field of PEP, i.e., COSMOS. In what follows, we produce a list of artificial 
sources, characterized by $T_{\rm dust}$, $M_{\rm gas}$, and SEDs given by the 
relations by \citet{magnelli2014} and \citet{genzel2015}, and fit them with MBB and DL07 
models. We limit the analysis to the parameter ranges listed in Table \ref{tab:good_ranges}.
Our simulation is structured as follows:
\begin{description}
\item[1)] We adopt the $z$-$M^\ast$-sSFR grid by \citet{magnelli2014} and 
we limit the analysis to the range of parameters over which the relation by \citet{magnelli2014} holds
(see Sect. \ref{sect:data_stacks}). 
\item[2)] The recipe by \citet{magnelli2014} produces a value of $T_{\rm dust}$ given the position 
in the $z$-$M^\ast$-sSFR grid along with the typical far-IR 
SEDs of a galaxy in that position. These SEDs are 
based on (and limited to) the \citet{dale2002} templates library.
\item[3)] These SEDs are convolved with photometric filters: MIPS 24 $\mu$m; PACS 70, 100, and 160 $\mu$m; and SPIRE 250, 350, and 500 $\mu$m.
\item[4a)] In the case of the DL07 simulation, the convolved SEDs are fit with the DL07 models, adopting a $\le1$\% photometric uncertainty in all bands. 
In this way, a so-called input catalog is produced, defining the DL07 parameters and a reference $M_{\rm dust}$ 
to be associated with each $z$-$M^\ast$-SFR bin.
\item[4b)] In the case of MBB simulation, \citet{genzel2015} provide scaling relations to define 
the input value of $T_{\rm dust}$ and $M_{\rm gas}$ (to be transformed 
into $M_{\rm dust}$ adopting a gas/dust mass ratio)
at any position in the $M^\ast$-sSFR-$z$ space. These relations are calibrated on \citet{magnelli2014}
data, which hold for $\beta=1.5$. Thus, using the relation by \citet{magnelli2014} leads to similar results.
In case a value $\beta=2.0$ was adopted, temperatures need to be 
increased by 4 K \citep{magnelli2014} and dust masses corrected as discussed in Sect. \ref{sect:kappa}.
\item[5)] Real noise levels are associated with each band \citep[see][]{magnelli2013,berta2013a}. We use the PEP/GOODS-H/HerMES
noise values for individual detections of two different fields: GOODS-S (PEP plus GOODS-H and HerMES depths; see Magnelli et al. \citeyear{magnelli2013};
Oliver et al. \citeyear{oliver2012}) 
and COSMOS (PEP and HerMES depths; see Lutz et al. \citeyear{lutz2011};
Oliver et al. \citeyear{oliver2012}). Table \ref{tab:depths} lists the adopted depths.
 Two independent 
simulations are run: one for each set of depths. Only bands with S/N$\ge3$ are taken into account.
\item[6)] We fit the noisified catalog with DL07 or MBB models. In the DL07 case, photometric points
at $\lambda\ge8$ $\mu$m (rest frame) are considered; in the MBB case, only bands at $\lambda>50$ $\mu$m 
(rest frame) are used.
\item[7)] Solutions are
found both through $\chi^2$ minimization and through 1000 Monte Carlo (MC) realizations for each 
entry in the synthetic catalog and are obtained by letting the photometry vary within the ``observed'' noise.
In the two cases, uncertainties are computed based on $\Delta\chi^2$ or as
the dispersion of all MC realizations, respectively. The two approaches lead to comparable results. 
The analysis and figures presented here are based on the MC approach.
\end{description}

\subsection{Results of DL07: Relative uncertainties on $M_{\rm dust}$}\label{sect:NEW_rel_unc}

We first verify down to what precision our procedure is able to determine 
dust masses as a function of position in the redshift, $M^\ast$, SFR space, 
and at the depths reached by \textit{Herschel} for individual detections in the 
GOODS-S and COSMOS fields.

\begin{figure}[!ht]
\centering
\rotatebox{-90}{
\includegraphics[height=0.45\textwidth]{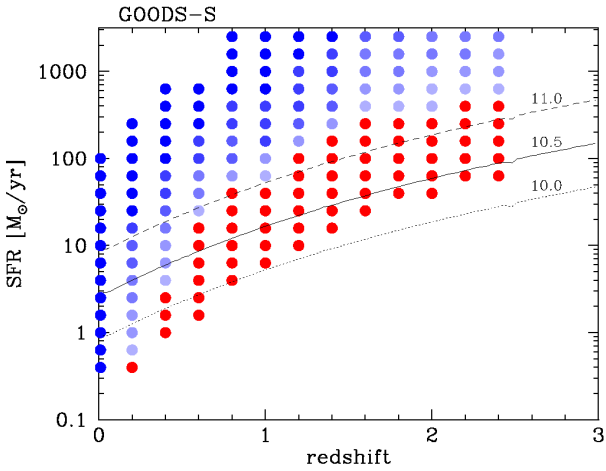}
}
\caption{Relative uncertainty of dust masses (defined as $\sigma(M_{\rm dust})/M_{\rm dust}$)
at the GOODS-S depth,
based on DL07 SED modeling, as a function of redshift and SFR. Color coding is
as in Fig. \ref{fig:mstar_sfr_2b_NEW}. The dotted, solid, and dashed lines indicate the 
position of the MS of star formation at different values of stellar mass 
\citep{elbaz2011}. These values were obtained  
from the dependence of the sSFR$_{\rm MS}$ on redshift by adopting 
$\log\left(M^\ast/\textrm{M}_\odot\right)$=10.0, 10.5, and 11.0, respectively.}
\label{fig:sfr_z_2b}
\end{figure}

Figure \ref{fig:mstar_sfr_2b_NEW} shows the $M^\ast$-SFR plane 
in different redshift slices, color coding each bin on the basis of 
its average $\sigma(M_{\rm dust})/M_{\rm dust}$ value. 
Bins indicated in red have an average $M_{\rm dust}/\sigma(M_{\rm dust})$ 
smaller than 3.
 It was not possible to run SED fits for bins missing at the low SFR side 
because too few photometric bands are available.
Figure \ref{fig:sfr_z_2b} collapses this diagram along the $M^\ast$ axis 
for the GOODS-S case, thus allowing for a more straightforward view as a 
function of redshift and SFR.

As expected, the relative uncertainty on $M_{\rm dust}$ becomes worse, 
as redshift increases (due to Malmquist bias) and SFR decreases. 
Redshift plays a double role by dimming fluxes and thus raising the luminosity 
threshold; and by pushing the rest-frame SED coverage to shorter wavelengths 
(see also Sect. \ref{sect:restframe_simu}).
At the sensitivity of PEP and HerMES, dust masses are retrieved with a S/N$\ge$3 
for galaxies on the MS of star formation down to $M^\ast\sim10^{10}$ $[$M$_\odot]$
up to $z\sim 1$.  It is possible to obtain
an estimate of $M_{\rm dust}$ at comparable stellar masses up to $z\sim 2 $ for objects lying increasingly above the MS.

\subsection{Results of DL07: Systematics on $M_{\rm dust}$}\label{sect:NEW_syst}

\begin{figure*}[!ht]
\centering
\includegraphics[width=0.45\textwidth]{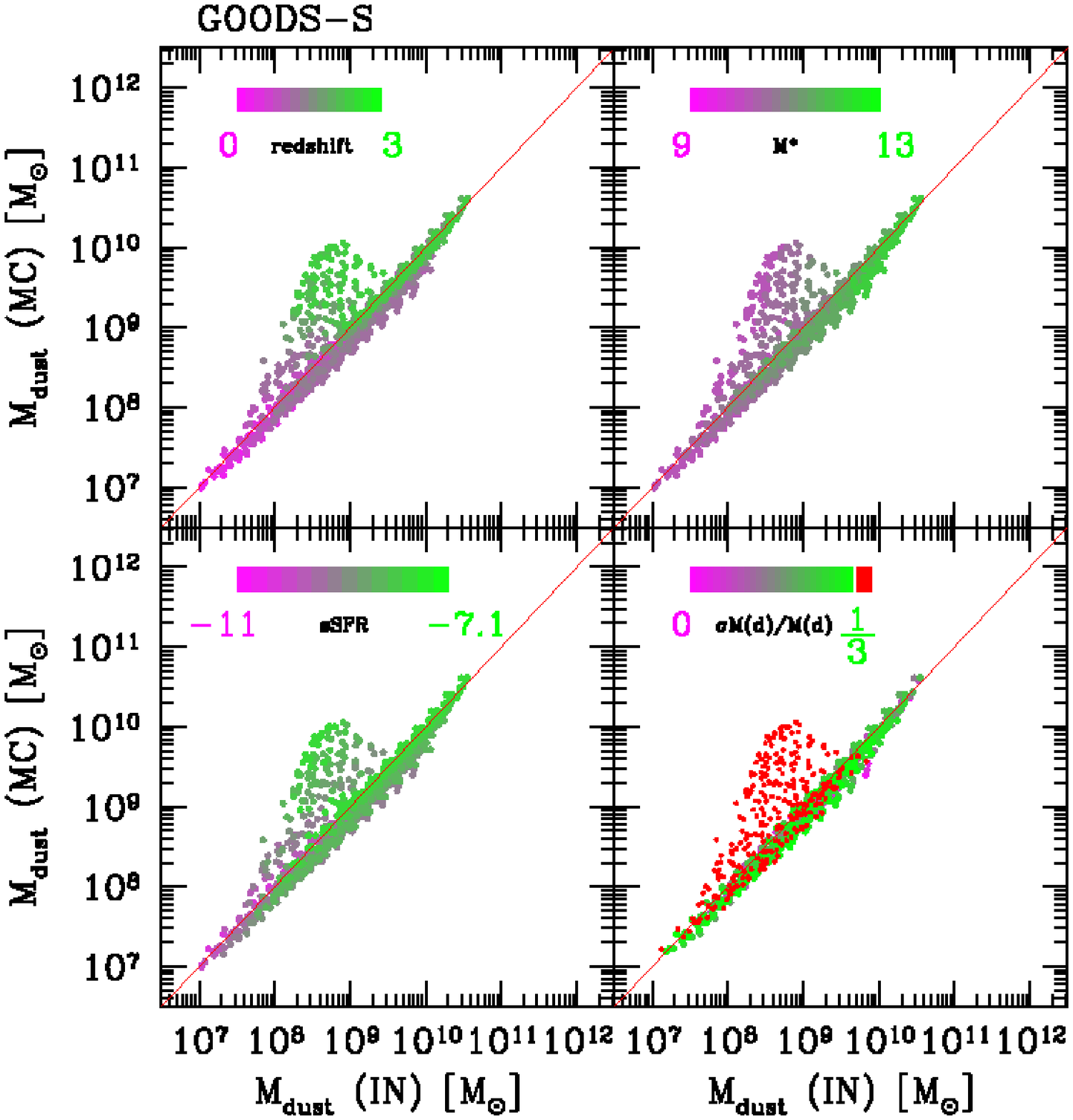}
\includegraphics[width=0.45\textwidth]{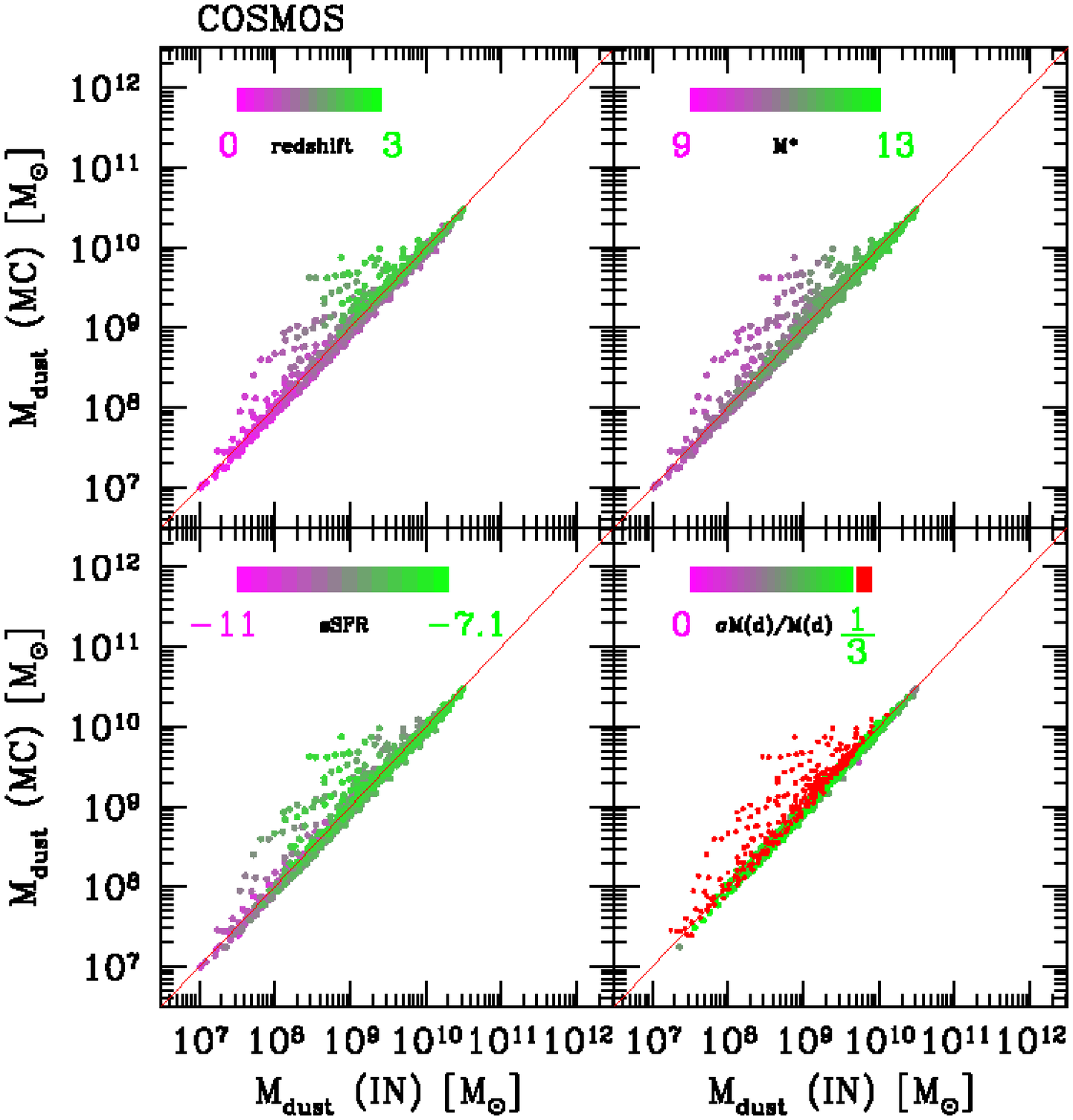}
\caption{Comparison of input and Monte Carlo (MC) output 
$M_{\rm dust}$ in DL07 simulations.
Color coding is based on redshift, stellar mass, specific SFR, and $M_{\rm dust}$ relative uncertainty.
As each dot belongs to an individual object, there is a general overlap of colors (green possibly hiding pink).
The two diagrams belong to two independent simulations obtained with GOODS-S and COSMOS depths.
Data points indicated in red have $M_{\rm dust}/\sigma(M_{\rm dust})<3$.}
\label{fig:dl07_in_mc_NEW}
\end{figure*}

It is now possible to study how well $M_{\rm dust}$ is recovered by comparing 
input and output dust masses.
Figure \ref{fig:dl07_in_mc_NEW} presents the mere comparison of input
and output $M_{\rm dust}$ for all bins in the considered $z$-$M^\ast$-SFR space.
Scatter increases and the incidence of outliers becomes larger as the depth of 
the available bands become less balanced, i.e., in the GOODS-S simulation, where 
MIPS and PACS are much deeper than SPIRE. 

The color coding in Fig. \ref{fig:dl07_in_mc_NEW} is based on redshift, stellar mass, specific SFR, and 
on $M_{\rm dust}$ relative uncertainty. 
Critical outliers lie at intermediate to high redshift and are characterized by 
low $M^\ast$. 
They turn out to 
be sources with few photometric points available and thus with limited 
wavelength coverage. At the adopted depths, it is not possible to 
derive $M_{\rm dust}$ with an accuracy better than 3$\sigma$ for them, therefore, 
they are indicated in red in Fig. \ref{fig:dl07_in_mc_NEW}.

Ignoring these critical cases, milder systematic offsets seem still to occur 
for those bins with larger relative 
uncertainties on $M_{\rm dust}$, which tend to lie preferentially 
below the 1:1 line. However, for these sources with $M_{\rm dust}$ measured at 
$>3\sigma$, such systematic offsets are well below a factor 2.

\begin{figure}[!ht]
\centering
\includegraphics[width=0.32\textwidth]{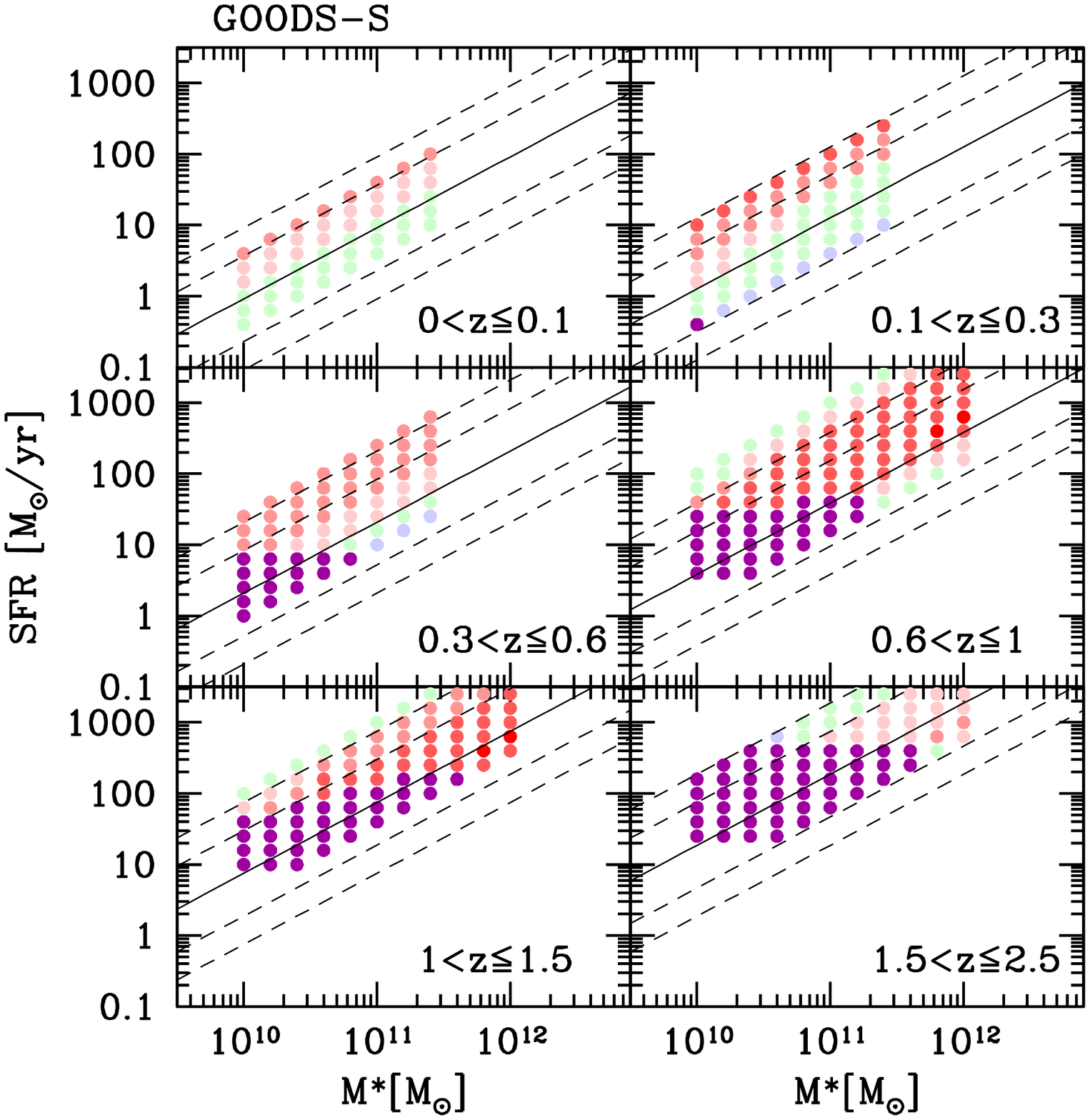}
\includegraphics[height=0.32\textwidth]{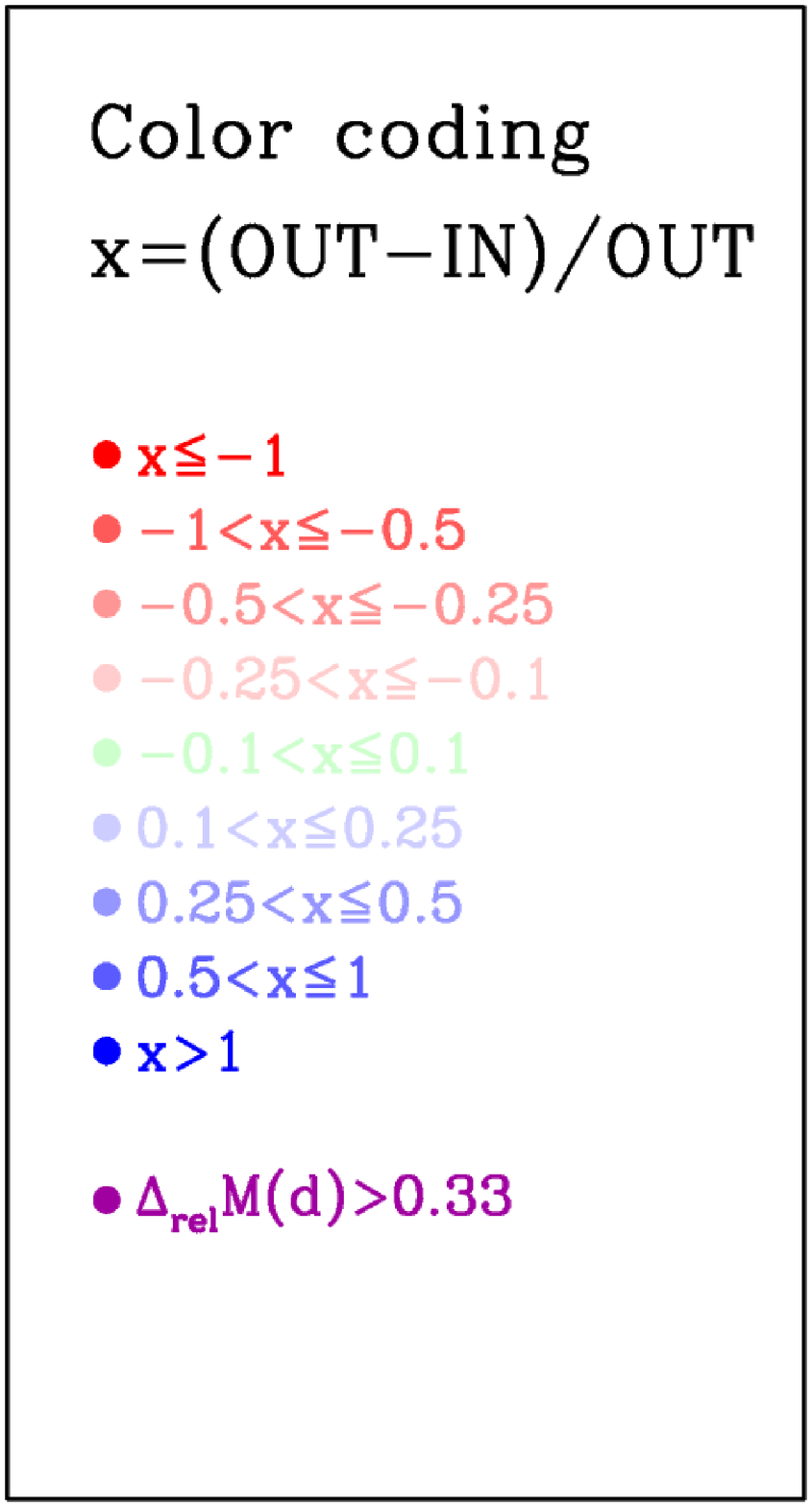}
\caption{Systematics on dust masses (defined as (OUT-IN)/OUT)
as a function of position in the $z$-$M^\ast$-SFR space, in the simulation with GOODS-S depth.
Black diagonal lines indicate the position of the MS \citep{elbaz2011} and 
$\pm4$, $\pm10$ MS levels (dashed).}
\label{fig:mstar_sfr_3b_NEW}
\end{figure}

Figure \ref{fig:mstar_sfr_3b_NEW} puts systematic offsets in the context 
of the $z$-$M^\ast$-SFR space in the case of GOODS-S.
With PACS and SPIRE depths strongly unbalanced, there is 
a trend to underestimate $M_{\rm dust}$ at intermediate sSFR. This effect 
is more critical at $z\simeq0.5-1.5$, where SPIRE bands are most important to constrain 
the peak of the far-IR SED. At lower redshift, deep PACS data provide a good 
constraint already, and at higher redshift the peak and $M_{\rm dust}$ are likely 
equally poorly constrained both in the ``IN'' and ``OUT'' cases.

Working on a small sample of high-redshift galaxies, \citet{magdis2012} showed that 
the presence of photometric data at rest-frame wavelengths larger that 
100-160 $\mu$m should avoid large systematics in the estimate of 
dust masses. Similar results were obtained locally by \citet{ciesla2014}.
Imposing the condition to have at least one 3$\sigma$ detection at 
$\lambda_{\rm rest}\ge160$ $\mu$m slightly mitigates these systematics
detected in our simulations, but does not solve the problem.

\subsubsection{The choice of templates and the role of $\beta$}\label{sect:role_beta}

The results presented above in Sections 
\ref{sect:NEW_rel_unc} and \ref{sect:NEW_syst} were obtained by fitting, with 
DL07 templates, the synthetic photometry computed by convolving 
the \citet{magnelli2014} SEDs with 
photometric filters. It is worth recalling that \citet{magnelli2014} assigned 
to each \citet[][DH02]{dale2002} template a value of dust temperature, $T_{\rm dust}$,
based on an MBB fit to the template itself. This $T_{\rm dust}$
was obtained fixing $\beta=1.5$. Then they used DH02 templates to fit 
the \textit{Herschel} stacked photometry of each $z$-$M^\ast$-SFR bin, 
thus linking each position in this space to $T_{\rm dust}$.
In short, we are fitting the DH02-based synthetic photometry 
of each $z$-$M^\ast$-SFR bin with DL07 templates.

The DL07 models were computed by adopting a fixed 
slope of the dust IR emissivity, $\beta=2.08$. On the other hand, in DH02 
models, $\beta$ varies as a function of the intensity of the radiation field, 
$U$, following the relation $\beta=2.5-0.4\log U$ at $\lambda>100$ $\mu$m.
This intensity varies in the range $0.3<U<1.e5$, thus one obtains $\beta=2.5,\ 2.1,\ 1.3$ for
$U=1,\ 100,\ 1000$.

We now  test whether the systematic trends 
seen in Fig. \ref{fig:mstar_sfr_3b_NEW} and described 
in Sect. \ref{sect:NEW_syst} could be, at least partially, driven by 
the different adopted models in input and output.
To this aim, the DL07 simulation (Sect. \ref{sect:simu_Mstar_SFR_z}) 
is now repeated with a new photometry based on DL07 models themselves.
This is carried out by convolving DL07 models with photometric filters at step number 
{\em 4a)} in the previous scheme, and adopting this new photometry during the SED 
fitting process instead of the photometry computed at step number {\em 3)}.

\begin{figure}[!ht]
\centering
\includegraphics[width=0.4\textwidth]{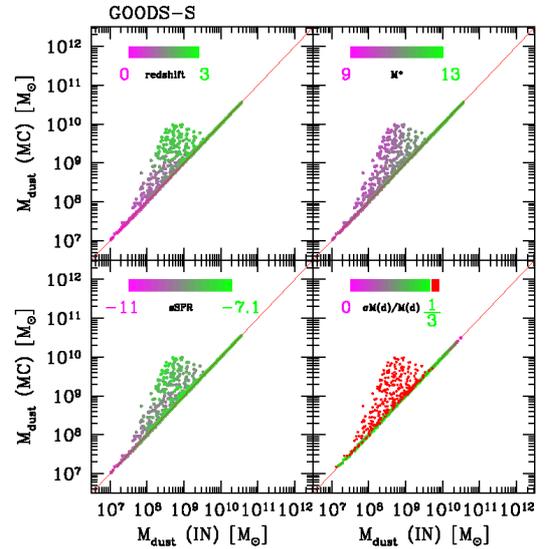}
\caption{Same as Fig. \ref{fig:dl07_in_mc_NEW}, for GOODS-S only 
and for MC results obtained using DL07-based photometry 
instead of DH02-based photometry.}
\label{fig:getting_rid_of_syst}
\end{figure}

Figure \ref{fig:getting_rid_of_syst} briefly shows the results for GOODS-S
(our worst case, see Sect. \ref{sect:NEW_syst}), i.e., reproduces Fig. \ref{fig:dl07_in_mc_NEW} 
with the new setup.
As expected, the majority of systematics are no longer found with the exception of 
the low S/N region of the parameters space, where dust masses tend to 
be overestimated. Their relative uncertainty on $M_{\rm dust}$ exceeds 33\% 
(red points). 
Similar results are obtained at the COSMOS depth.

The lesson to be learned is that the value of these kind of simulations 
and their ability to describe systematics is limited to the ideal case 
that the SED shape adopted in the simulation, and to fit the real sources
 is the same as that of real-life galaxies. In our case, this is exemplified
by the value of $\beta$.
Small differences can lead to misleading results or, more worrying, to physical overinterpretations of numerical effects.
Vice versa, the absence of systematics in these kind of simulations might not always 
be indicative of absence of systematics on real sources if the 
real SEDs differ from those adopted to model them.

\begin{figure}[!ht]
\centering
\includegraphics[width=0.4\textwidth]{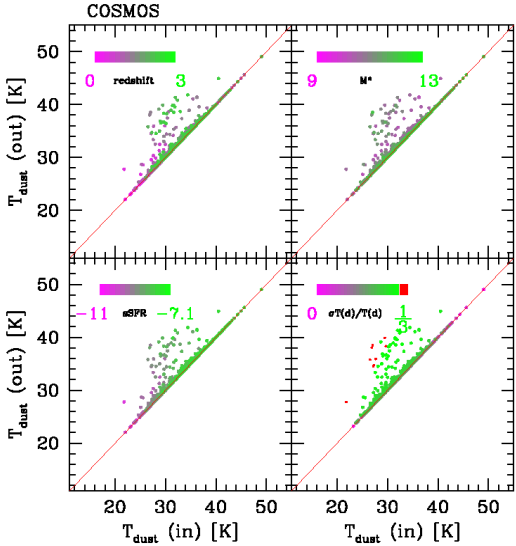}
\caption{Comparison of input and Monte Carlo (MC) output $T_{\rm dust}$ in MBB simulations
at the COSMOS depth, with fixed $\beta$. Color coding is based on redshift, stellar mass, 
specific SFR, and $M_{\rm dust}$ relative uncertainty.}
\label{fig:in_out_Td_cos}
\end{figure}

\subsection{Results of MBB: Constraints on $T_{\rm dust}$}

Following the scheme introduced for the DL07 simulations, 
we now analyze the results of MBB-based SED fitting 
of synthetic catalogs. The procedure followed is 
similar to the DL07 case and is described in Sect. \ref{sect:simu_Mstar_SFR_z}.
We fix the value of $\beta$ to ease the comparison of our simulations to 
recent literature works and later to the analysis of individual real sources for 
which few photometric points are available.

\begin{figure}[!ht]
\centering
\includegraphics[width=0.4\textwidth]{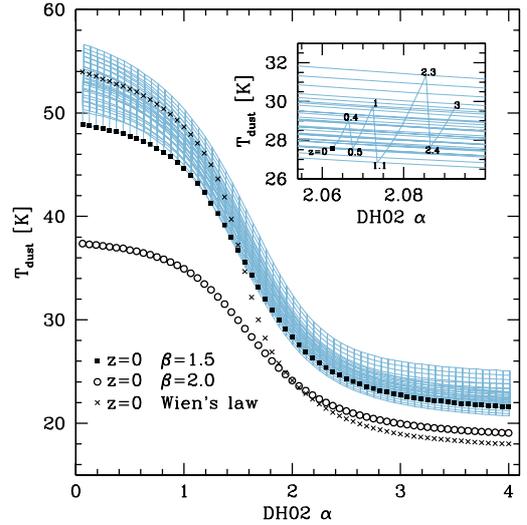}
\caption{Study of possible biases on $T_{\rm dust}$ in a MBB fit 
and dependence of DH02 templates \citep[e.g.,][]{magnelli2014}. DH02 templates
are characterized by their $\alpha$ parameter. Black filled squares and open circles
represent the values of $T_{\rm dust}$ associated with each DH02 template by 
fitting it with a MBB model with $\beta=1.5$ or 2.0, respectively.
The light blue grid maps the variation of $T_{\rm dust}$ as a function of redshift 
for each given template.
The inset includes a zoom on a single template, aimed at showing the 
details of the dependence of the $T_{\rm dust}$ derivation on redshift. Each redshift point 
was artificially shifted by a small amount along the x-axis  to avoid
overlapping. Black crosses refer to the result obtained by applying Wien's displacements 
law to DH02 templates.}
\label{fig:tdust_biases}
\end{figure}

We now deal with the case at COSMOS depth, which represents the 
shallowest and worst case scenario in this analysis. 
Figure \ref{fig:in_out_Td_cos} compares input and output 
dust temperatures.
Data points in the four panels are color coded as a function  
of redshift, $M^\ast$, sSFR, and $T_{\rm dust}$ relative error. 
Dust temperature is retrieved within few percents in most cases. 
Only few catastrophic failures are recorded: they are limited to 
bins with very poor S/N and, thus, with large uncertainties on $T_{\rm dust}$.
Only $\sim$4\% of the cases turn out to have $T_{\rm dust}$ 
overestimated by more than 10\%. These statistics become even better at
the GOODS-S depth.

When $T_{\rm dust}$ is not correct,  
$M_{\rm dust}$ can also be overestimated, up to a factor of $\sim$5.
Nevertheless, in these cases, the photometric S/N is not 
good enough to guarantee a relative error on $M_{\rm dust}$ smaller
than 3 $\sigma$.

\subsubsection{Biases in MBB $T_{\rm dust}$ determination}\label{sect:tdust_biases}

The SEDs of real galaxies are indeed not single-temperature modified 
blackbodies: as seen before, the FIR emission is produced by a mixture 
of dust grains, of different sizes and shapes, which turn into a 
nontrivial distribution of temperatures. Although convenient, the 
MBB approximation is an over-simplification in most of the cases.
More complex descriptions include using multitemperature blackbodies, 
DL07 modeling (see previous Sections), or mixed approaches as that by 
\citet[][described in Sect. \ref{sect:role_beta}]{magnelli2014}.
In this Section we study possible biases in MBB-based $T_{\rm dust}$
due to changes in $\beta$ and to redshift.

As in \citet{magnelli2014}, each template in the DH02 library is convolved 
with PACS/SPIRE passbands (70 to 500 $\mu$m) and the broadband photometry thus 
obtained is fitted with MBB models with varying dust temperature, $T_{\rm dust}$. 
The dust emissivity index, $\beta$, 
is fixed to 1.5 and 2.0 to test different scenarios.
In this way, a value of $T_{\rm dust}$ is associated with each DH02 
template. This is  the reference dust temperature.

Each template is then shifted to increasing redshifts, in the range $z=0$-$3$, 
and convolved again with filters. The $z>0$ synthetic photometry is then refit 
with a MBB, taking care to only use those bands with a rest-frame wavelength $\lambda_{\rm rest}>50$ $\mu$m.
        
For each template, characterized by the value of the parameters $\alpha$ \citep{dale2002}, Figure 
\ref{fig:tdust_biases} plots the values of $T_{\rm dust}$ obtained 
with $\beta=1.5$ or 2.0 at $z=0$ (black squares and open circles) and those 
derived at $z>0$ with $\beta=1.5$ (light blue grid). 
The inset includes a zoom on a single template, which is aimed at showing the 
details of the derivation of $T_{\rm dust}$ on redshift. Each redshift point 
was artificially shifted by a small amount along the x-axis to avoid
overlapping.

For reference, the results of applying  Wien's displacements law to 
DH02 templates, thus deriving dust temperatures from the wavelength of the 
SED's peak \citep{casey2012}, are
shown (black crosses).

Using different values of $\beta$ produces a systematic change in the inferred 
value of $T_{\rm dust}$, a smaller $\beta$ implying a larger $T_{\rm dust}$.
At higher redshift, $T_{\rm dust}$ further increases, but the change 
in temperature does not follow a linear trend, showing a saw-toothed 
pattern  instead. This reflects the fact that as redshift increases, the rest-frame 
wavelength sampled by each photometric band decreases, and the fitting procedure
tends to be biased toward higher dust temperatures because the SED is 
constrained at shorter wavelengths. When one band falls shorter than 50 $\mu$m
rest frame, it is discarded and $T_{\rm dust}$ shows a jump toward colder 
values because suddenly a short-$\lambda$ band is missing. This effect happens 
periodically, every time one band is discarded as a result of the $\lambda_{\rm rest}>50$ 
$\mu$m requirement.

In other words, for sources lying at different redshift,
the discreteness 
of the sampled wavelengths and their shift to the rest frame
cause a few K systematic difference in the determination 
of $T_{\rm dust}$.


\section{Rest-frame simulations}\label{sect:restframe_simu}

To understand the results 
presented in Sect. \ref{sect:simu_Mstar_SFR_z}, we ran
an additional simulation, which is limited to rest-frame SEDs, aimed at studying 
the effects of losing one or more photometric bands on the short- or long-wavelength side 
of the SED. At the same time, we also study the effects of having 
different amounts of noise in the data by randomly assigning 
S/N values to each band, independently. This means that, for each 
entry in the synthetic catalog, the distribution of 
relative photometric uncertainties among bands is neither flat nor 
based on the ratio of noise levels in selected \textit{Herschel} surveys, 
but is a random combination.

The new simulation is structured as follows:
\begin{enumerate}
\item First of all, a library of models (DL07 or MBB) is generated based on a grid of 
input parameters ($q_{\rm PAH}$, $U_{\rm min}$, 
and $\gamma$ in one case, and $T_{\rm dust}$ in the other). 
 We limit the analysis for DL07 models to the setup suggested by \citet{draine2007b}. 
In the MBB case,
the parameter $\beta$ is fixed to a value of 1.5 and $T_{\rm dust}$ spans the range 10-50 K. 
All models are renormalized to a fixed  
total dust mass of $10^8$ $[$M$_\odot]$ and then convolved with mid- and far-IR photometric bands.
We use ten bands for the DL07 models: \textit{Spitzer} IRAC 8 $\mu$m, IRS 16 $\mu$m, MIPS 24 $\mu$m, \textit{Herschel} PACS 70, 100, 160 $\mu$m, 
SPIRE 250, 350, 500 $\mu$m, and SCUBA 850 $\mu$m. We also performed an additional run, limited to six bands between 8 and 250 $\mu$m.
Only eight bands are used for MBB models: a square box filter centered at 40 $\mu$m and with a half width of 10 $\mu$m, 
the 70, 100, 160, 250, 350, 500 $\mu$m PACS bands, and the SCUBA 850 $\mu$m passband.
All entries in the catalog are at redshift $z=0$. This is the so-called input catalog.
\item The input catalog is degraded in two ways:
\begin{itemize}         
\item An increasing number of bands is removed, progressively
one by one, until only two are left. This procedure is repeated twice: 
first photometric bands are removed from the long-wavelength side of 
SEDs\footnote{So that the first realization has all eight bands available (40-850 $\mu$m);
the second has seven (40-500 $\mu$m); the third has six (40-350 $\mu$m); and so on until the last 
with only two bands (40-70 $\mu$m).}; then 
the number of bands is reset to eight (i.e., the full SED is re-installed) and bands 
are removed one by one from the short-wavelength side of SEDs\footnote{So that the first realization 
has all eight bands available (40-850 $\mu$m);
the second has seven (70-850 $\mu$m); the third has six (100-850 $\mu$m); and so on until the last 
with only two bands (500-850 $\mu$m).}.
\item The uncertainty on the photometry is increased. Relative uncertainties are increased randomly and independently for each band, 
spanning the range between 5\% and 50\% in relative errors.
\end{itemize}
\end{enumerate}
In practice, the degraded catalog contains all entries of the input catalog, 
each one modified eight times getting rid of bands on the 
short-$\lambda$ side and eight times getting rid of bands on the 
long-$\lambda$ side. Each of the $N\times16$ entries is modified ten times 
with random S/N values; each time, a random level of noise is assigned to each band, 
independently from the other bands.

By removing bands from the long-wavelength side 
of the SED, we are simulating the case of deep surveys, such as the deepest fields in PEP/HerMES \citep{lutz2011,oliver2012}, benefiting from deep PACS photometry for the majority of sources, and progressively missing 
SPIRE detections because of depth, confusion, and blending effects. Vice versa, when removing 
bands from the short-wavelength side, we are simulating the case of shallow surveys,
such as H-ATLAS \citep{eales2010}: 
SPIRE quickly reaches the relatively high confusion limit at these wavelengths, 
but a fast PACS observation cannot fully take advantage of the smaller beam and deeper confusion limit, 
and remains limited to bright and mostly lower redshift sources.

\subsection{Reading key}\label{sect:reading_key}

The results of the rest-frame simulations are presented with two main flavors of diagrams. 

First the relative uncertainty on parameters, as 
obtained with Monte Carlo runs, is shown. In these diagrams, 
each displayed dot belongs to a full MC run per one artificial object, i.e., the value and error associated with each 
dot are computed as the average and standard deviation of 1000 evaluations of 
the same entry of the synthetic catalog.
Color coding is based on the average relative photometric uncertainty of all available bands, if not otherwise 
specified. 
Darker colors refer to smaller average photometric uncertainties.
The $x$-axis shows the maximum or minimum wavelength covered by the data, $\lambda_{\rm max}$ (or $\lambda_{\rm min}$), 
depending on whether photometric bands were removed at the long- or short-wavelength side. 
In correspondence of the wavelength of each photometric band, a column of dots is plotted. 
These dots represent all those entries in the artificial catalog with that value $\lambda_{\rm max}$ (or $\lambda_{\rm min}$), 
therefore, the given object in the given column benefits from available photometry in all bands shortward (longward) 
of the $\lambda_{\rm max}$ ($\lambda_{\rm min}$) it belongs to. This is exemplified by the left arrows (right arrows) 
at the bottom of each diagram.

The second type of diagram (see also Appendix \ref{sect:app_syst}) probes 
for systematic effects. This diagram 
presents the comparison of output derived quantities to their known input values 
as a function of $\lambda_{\rm max,min}$. Each dot again belongs to a given 
object in the synthetic catalogs, i.e., the dot represents the average of 1000 MC evaluations.

\subsection{Results of DL07: Relative uncertainties}\label{sect:dl07_rel_err}

\begin{figure*}[!ht]
\centering
\includegraphics[width=0.38\textwidth]{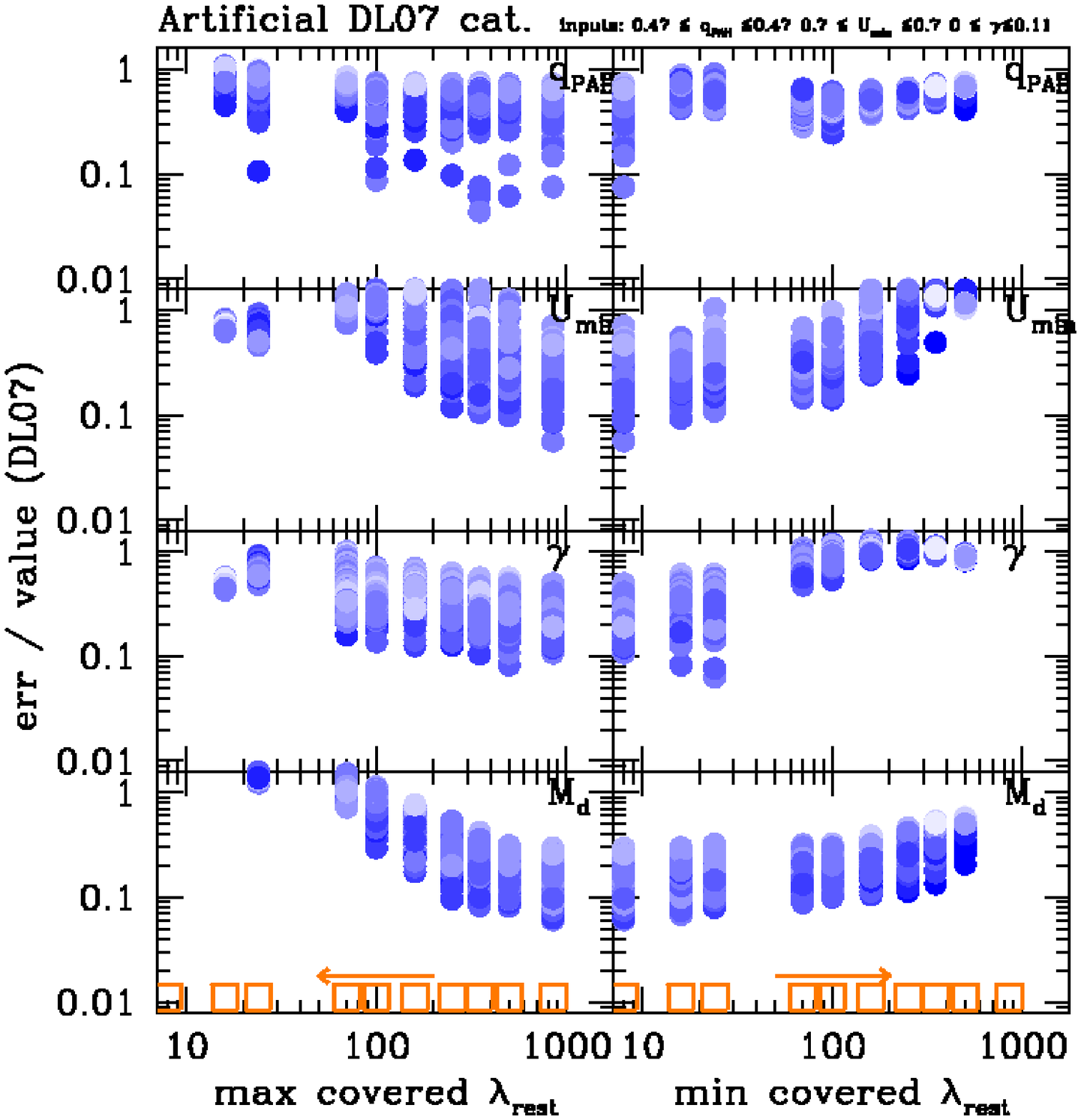}
\includegraphics[width=0.38\textwidth]{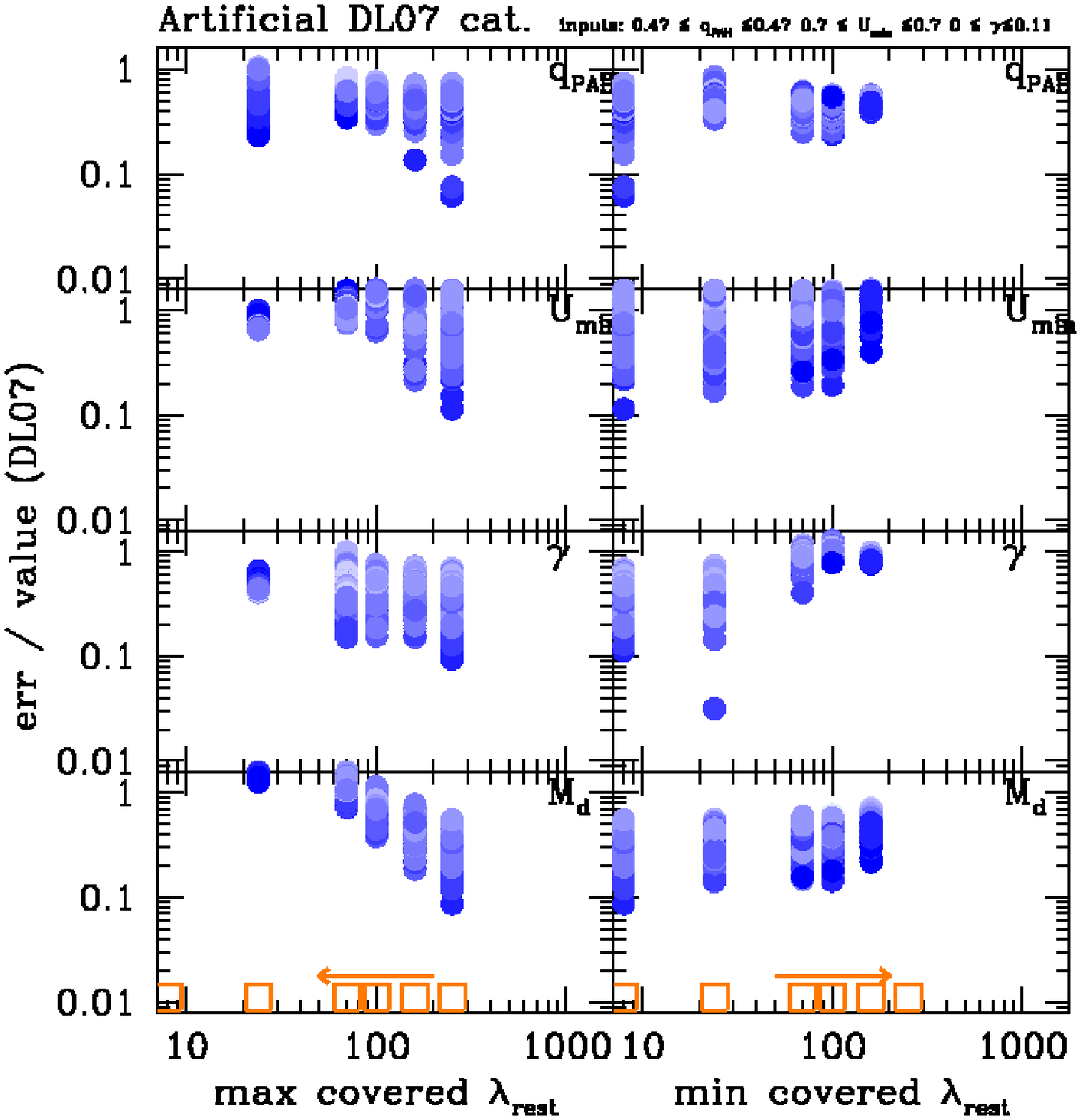}
\includegraphics[width=0.20\textwidth]{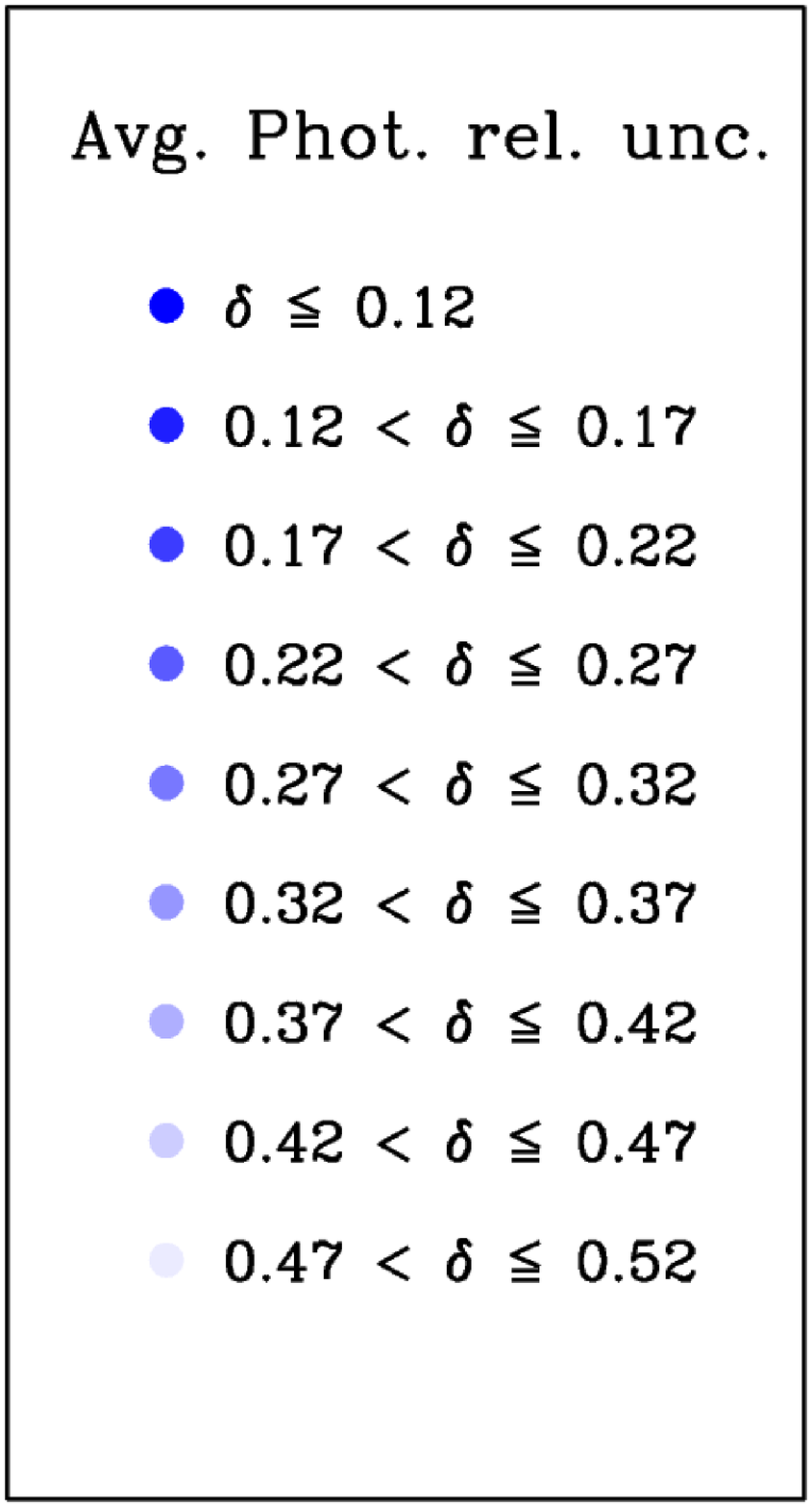}
\caption{Relative uncertainty on DL07 free parameters, as obtained with the 
rest-frame simulation. For simplicity, we show only the case 
with $\alpha=2$, $U_{\rm max}=10^6$, $U_{\rm min}=0.7$, $q_{\rm PAH}=0.47$, and $\gamma=0.01-0.11$. 
The {\em left} diagram is obtained using ten photometric bands from 8 to 850 $\mu$m; 
the {\rm right} diagrams uses only six bands between 8 and 250 $\mu$m.
{\em Left/right} columns belong to the case obtained by removing long- and short-wavelength bands.
Color coding is based on the average photometric uncertainty as computed over all available 
bands. See Sect. \ref{sect:reading_key} for more details.}
\label{fig:simu_dl07_1}
\end{figure*}

Figure \ref{fig:simu_dl07_1} presents the relative uncertainty on 
the free DL07 parameters and $M_{\rm dust}$ as a function of $\lambda_{\rm max,min}$ for 
cases obtained removing long-wavelength and short-wavelength bands. 
We limit the diagrams describing relative uncertainties 
to specific values of the
input parameters to avoid the piling up of too many 
cases.

Trivially, objects with larger photometric uncertainties have 
larger relative uncertainties on the derived quantities. 
However, since the photometric S/N ratio
of each band is independent from the others and the color coding based 
on an average photometric relative uncertainty is computed over all bands, 
in Fig. \ref{fig:simu_dl07_1} there is not a smooth gradient of colors, 
but a mixing of cases is present.

Unless the rest-frame mid-infrared SED is sampled by the available 
photometry (8-24 $\mu$m in the specific case), and with good S/N, constraints on the value $q_{\rm PAH}$ and $\gamma$ 
are weak. 
As long as the peak of the FIR SED is sampled, and if the average S/N ratio is good (darker dots and lower envelope
of the relative error distribution in Fig. \ref{fig:simu_dl07_1}), 
it is possible to have an estimate of $U_{\rm min}$ within a $\sim$30\% uncertainty.
A broader peak (induced by larger input values of $U_{\rm min}$) is more difficult to constrain
and the consequent uncertainty on $U_{\rm min}$ is larger.

The large uncertainties on the $\gamma$ and $U_{\rm min}$ parameters have 
strong consequences on the determination of $\langle U\rangle$. Unless the SED 
is robustly constrained (see below), the propagation of $\gamma$ and $U_{\rm min}$ 
errors onto Eq. \ref{eq:U_avg} significantly hinders the estimate of 
$\langle U\rangle$. On the other hand, $\langle U\rangle$ can also be derived 
from $L_{\rm dust}$ and $M_{\rm dust}$ (see Eq. \ref{eq:ldust}) under the 
assumption that the value of $P_0$ is known \citep[e.g.,][]{magdis2012} 
and that the two quantities can be measured independently enough.

The best constrained and most stable parameter turns out to be dust mass, $M_{\rm dust}$.
When several bands are available and the $\lambda>250$ $\mu$m SED is sampled, dust mass 
is constrained within a 30\% uncertainty even with poor photometry, but as soon as sub-mm data are missing, 
the maximal uncertainty (poor S/N, light dots) on $M_{\rm dust}$ immediately reaches 70-80\%, 
even with the full $8-250$ $\mu$m wavelength range covered.
When removing short-wavelength bands, the uncertainty on $M_{\rm dust}$ remains more stable than in the 
MBB case (see Sect. \ref{sect:mbb_rel_err}) on both the high and low S/N sides.
This might be because of the choices made on the basis of SINGS results \citep{draine2007b}, 
effectively limiting the freedom of choosing extreme models that would cause drastic changes of normalization.
In the best case of high quality photometry, the uncertainty on $M_{\rm dust}$ can be 
as low as 10-20\% even with only three FIR bands (rest-frame 100, 160, 250 $\mu$m) available.

Upper limits are simulated with relative photometric uncertainties larger
than 0.33 (see Fig. \ref{fig:simu_dl07_1}). If we compare the lightest blue dots in a given column to 
those in the subsequent column that has the given band fully removed, we find that 
the use of upper limits can produce an improvement in the determination of $M_{\rm dust}$, 
but this improvement is nevertheless modest.

Expectations for a typical PEP galaxy at $z=1.5-2.0$, and observed photometry covering 
the rest-frame 8$-$160 $\mu$m (i.e., $\lambda_{\rm obs}=24-350$ $\mu$m) can 
be read from the results of simulations carried out with six photometric bands
and removing long-wavelength data (Fig. \ref{fig:simu_dl07_1}, right). The column of dots 
at $\lambda=160$ $\mu$m represents results for sources with rest-frame 8$-$160 $\mu$m
photometry available.
In such cases, the relative 
uncertainty on $M_{\rm dust}$ can be as low as $\sim$10-20\% for good S/N data
(darker dots and lower envelope of the dots distribution), the exact value depending on the value of input parameters.  
Nevertheless, it easily grows beyond 50\% for 3$\sigma$-only detections. 
Results only get worse by a factor $\sim2$ 
for $\lambda_{\rm rest}$ up to 100 $\mu$m.

\subsection{Results of DL07: Systematics}\label{sect:dl07_syst}

Possible systematic offsets are studied by comparing 
output and input values. We focus on dust mass estimates as the other 
parameters are much more poorly constrained (see Sect. \ref{sect:dl07_rel_err}).
Moreover, only a short compendium of results is presented here, while 
a thorough description of details is given in Appendix \ref{sect:app_syst}.

\begin{figure*}[!ht]
\centering
\includegraphics[width=0.38\textwidth]{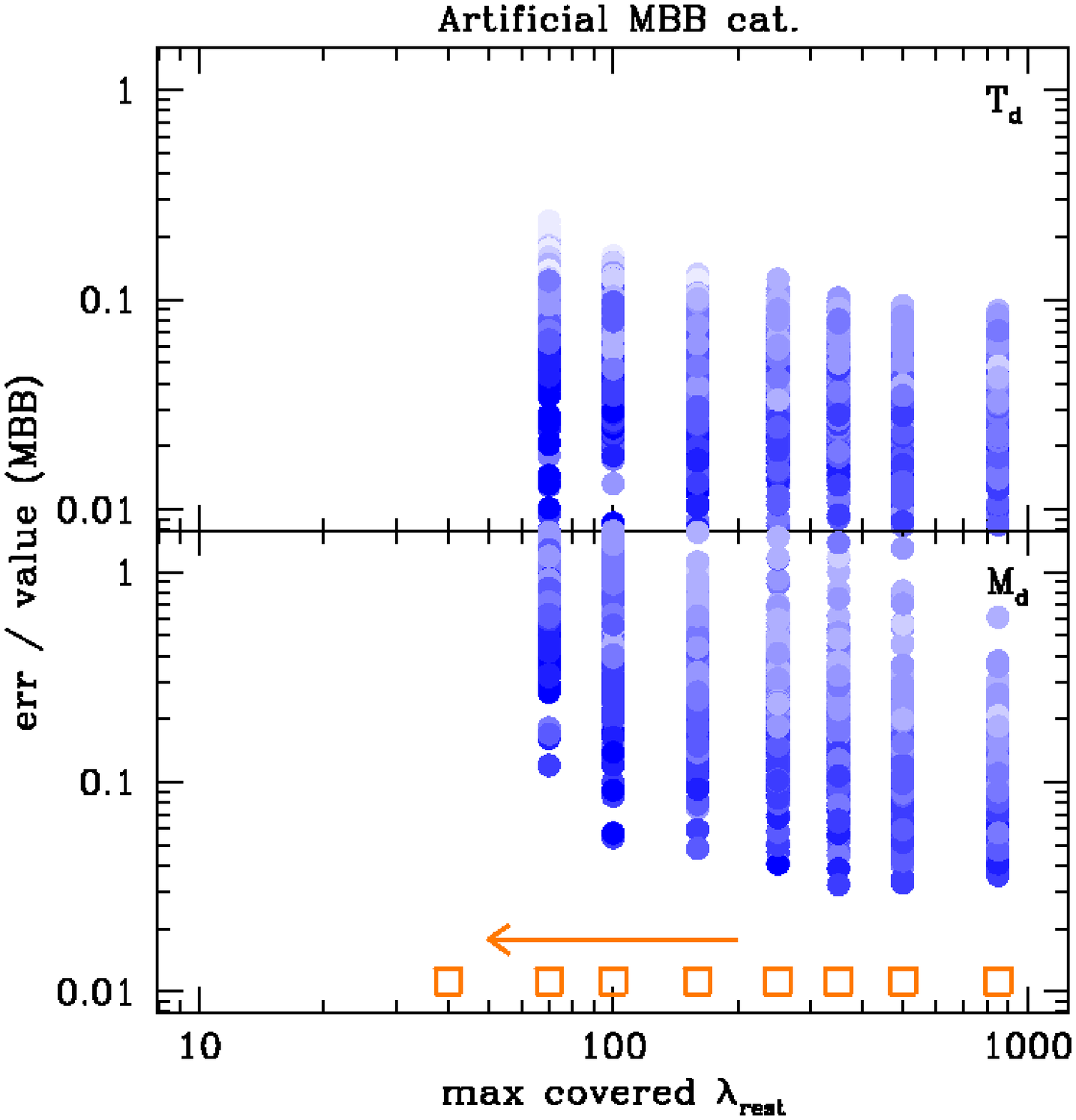}
\includegraphics[width=0.38\textwidth]{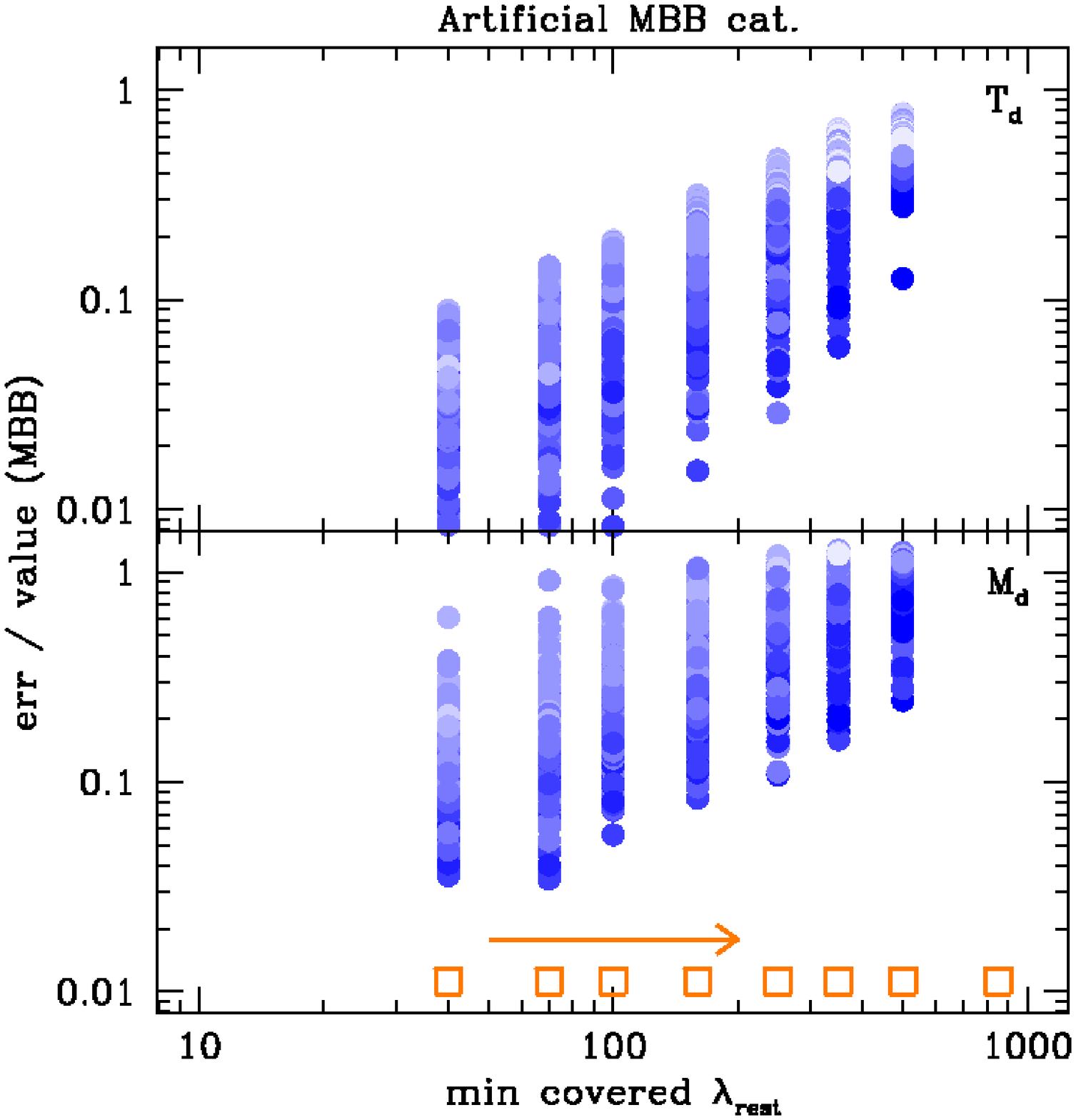}
\includegraphics[width=0.20\textwidth]{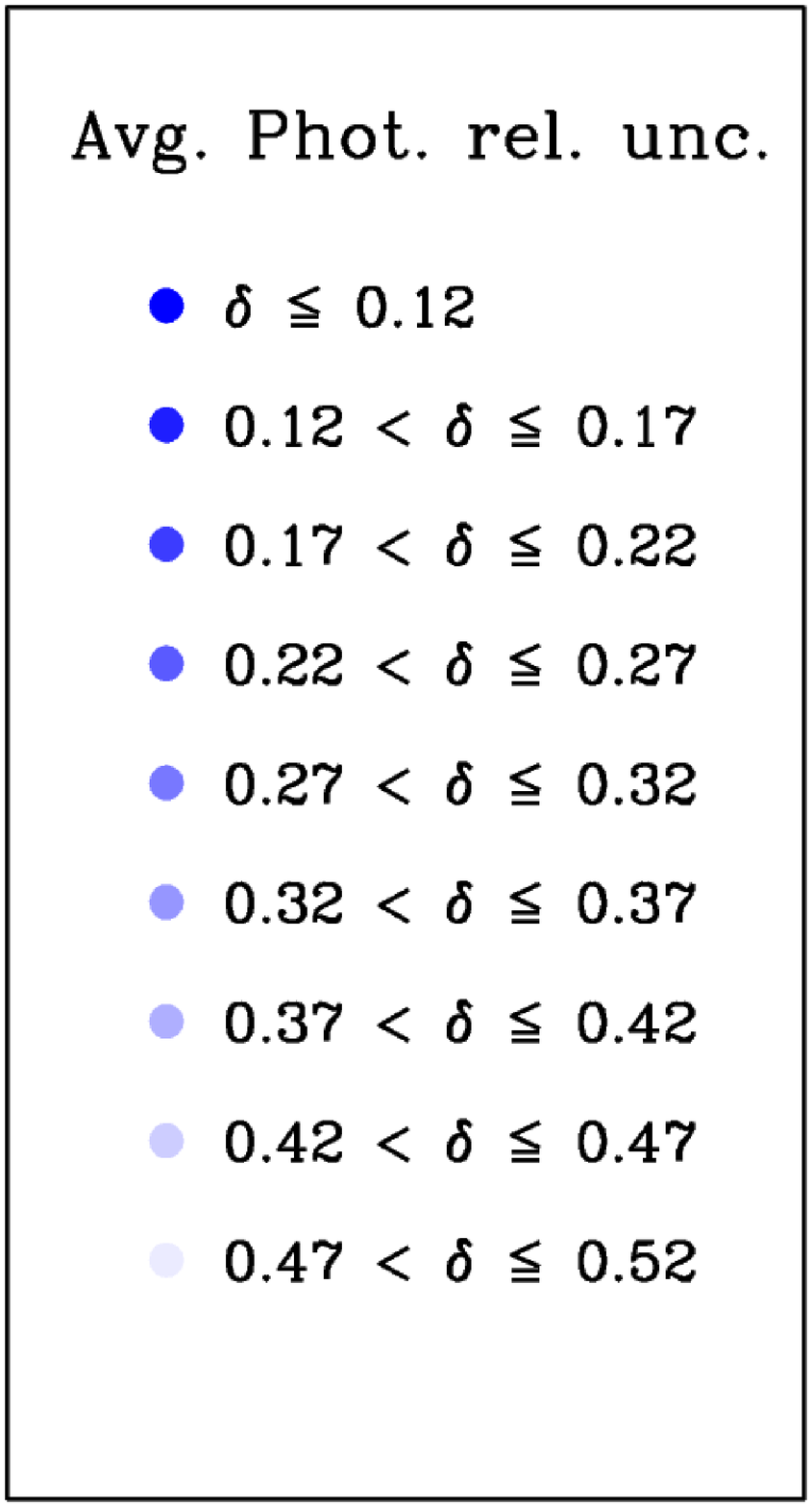}
\caption{Relative error on dust temperature ({\em top}) and dust mass ({\em bottom})
obtained with MBB simulations, when erasing long-wavelength bands ({\em left}) and 
short-wavelength bands ({\em right}). Color coding is based on the average relative 
photometric uncertainty computed on the available bands, and is explained in the legend.
See text for more details on how to read these diagrams.}
\label{fig:mbb_rel_err_1}
\end{figure*}

When removing short wavelength bands, $M_{\rm dust}$ is usually 
easily retrieved and there are no very significant trends related to $\gamma_{\rm in}$ 
in over- or underestimating $M_{\rm dust}$. This also holds in varying $q_{\rm PAH,in}$ and $U_{\rm min,in}$. 
When dealing with a limited number of bands, i.e., with MC runs using only six filters in total,
systematics as function of $U_{\rm min,in}$ can be triggered. This is because 
not only bands at the short wavelength side are progressively missing, but also 
the long wavelength SED is sampled only up to 250 $\mu$m rest frame.

When removing long-wavelength photometric bands, 
systematics show trends as a function of the position in the 
$q_{\rm PAH,in}$, $U_{\rm min,in}$, and $\gamma_{\rm in}$
parameter space. 
At low $U_{\rm min,in}$ there is a tendency to underestimate $M_{\rm dust}$ if the 
band coverage is poor, while at high $U_{\rm min,in}$ there is a tendency 
to overestimate it.
A larger value of $\gamma_{\rm in}$ produces an increased
chance to overestimate $M_{\rm dust}$. Similarly, 
the larger $q_{\rm PAH,in}$, the more $M_{\rm dust}$ can be overestimated.

The three effects can add up or compensate each other.
If there is a general tendency to under-estimate $M_{\rm dust}$ 
(e.g., because $U_{\rm min,in}$ is small), then 
the larger values of $\gamma_{\rm in}$ or $q_{\rm PAH,in}$ mitigate it. 
In contrast, large values of $U_{\rm min,in}$ combined to larger 
$\gamma_{\rm in}$ or $q_{\rm PAH,in}$ can induce a tendency to significantly 
overestimate $M_{\rm dust}$ when the SED coverage is poor at long wavelengths.
          
Generally speaking, if the maximum covered rest-frame wavelength is 
$\lambda_{\rm max}<160-200$ $\mu$m systematics on $M_{\rm dust}$ can become significant, especially 
when the S/N of the available FIR photometry is poor.

\subsection{Results of MBB: Relative uncertainties}\label{sect:mbb_rel_err}

We now analyze the case of MBB models with fixed $\beta$ and how relative errors on $T_{\rm dust}$ and $M_{\rm dust}$ depend on 
the available photometry and on its S/N ratio.
Figure \ref{fig:mbb_rel_err_1} shows the behavior of relative uncertainties on $T_{\rm dust}$ and $M_{\rm dust}$, 
for all entries in the artificial catalog built on MBB models without distinction of input dust temperature.

The left panel shows that,
as the long-wavelength part of the SED is progressively less 
sampled, the uncertainty on $T_{\rm dust}$ and $M_{\rm dust}$ increases, as expected. The parameter
$T_{\rm dust}$ is always constrained relatively well within a 25\% uncertainty. 
As the average photometric uncertainty increases (brighter color tones), 
naturally the uncertainty on the derived quantities increases. This is 
particularly true for $M_{\rm dust}$.
When rest-frame sub-mm data are available, with the exception 
of a few outliers dust masses can  generally be retrieved with 
relative uncertainties, within 20-30\% in case of poor photometry. As the maximum covered wavelength decreases, 
the uncertainty 
on $M_{\rm dust}$ explodes when the available bands do not sample the SED 
at $\lambda\ge350$ $\mu$m (restframe) anymore and the photometry is poor.
This turns out to be slightly worse than DL07 modeling, which produces 
more stable $M_{\rm dust}$ uncertainties even with SEDs only limited to 
$\sim$160 $\mu$m rest frame (see Sect. \ref{sect:dl07_rel_err}).

The right-hand panel shows instead what happens when short-wavelength bands 
are progressively removed. The uncertainty on $T_{\rm dust}$ is now 
larger than in the previous case, reflecting the fact 
that while the long-wavelength side of the MBB has a weak dependence on 
$T_{\rm dust}$, there is a stronger dependence on the short-wavelength side. 
Uncertainties on $M_{\rm dust}$ are not 
significantly enhanced with respect to the previous case.

As the temperature increases the 
minimum uncertainty on $T_{\rm dust}$ tends to become larger,
reflecting the fact that the peak of the MBB moves to 
shorter wavelengths.
A similar trend is also detected  for $M_{\rm dust}$, although with smaller
amplitude.

The case of a PEP $z=1.5-2.0$ galaxy can be studied in 
Fig. \ref{fig:mbb_rel_err_1}, focusing on the left panel, 
at $\lambda=40-160$ $\mu$m. The observed photometry 
(e.g., at 100, 160, 250, 350 $\mu$m) samples the rest-frame 
40-200 $\mu$m wavelength range.
The simulation shows that it is possible to constrain $T_{\rm dust}$ within a $<20\%$ accuracy, but 
the uncertainty on $M_{\rm dust}$ could easily grow up to 80\% with poor S/N. 
In the best case, the smallest uncertainty possible ($\sim15$\%) is 
reached if photometric accuracies of $\sim15-20$\% are available 
in almost all bands (350 $\mu$m included).

\subsection{Results of MBB: Systematics}\label{sect:mbb_syst}

Here the possible systematics on dust mass are tested for the MBB 
modeling along with $T_{\rm dust}$ reliability.
As in the DL07 case, we only summarize the main results, while we defer to 
Appendix \ref{sect:app_mbb_syst} for fine details.

When the photometric accuracy is poor, $M_{\rm dust}$ can be 
systematically overestimated if bands are missing
on the long-wavelength side of the SED; 
the problem is relatively milder when removing bands
on the short-wavelength side.

\begin{figure*}[!ht]
\centering
\rotatebox{-90}{\includegraphics[height=0.45\textwidth]{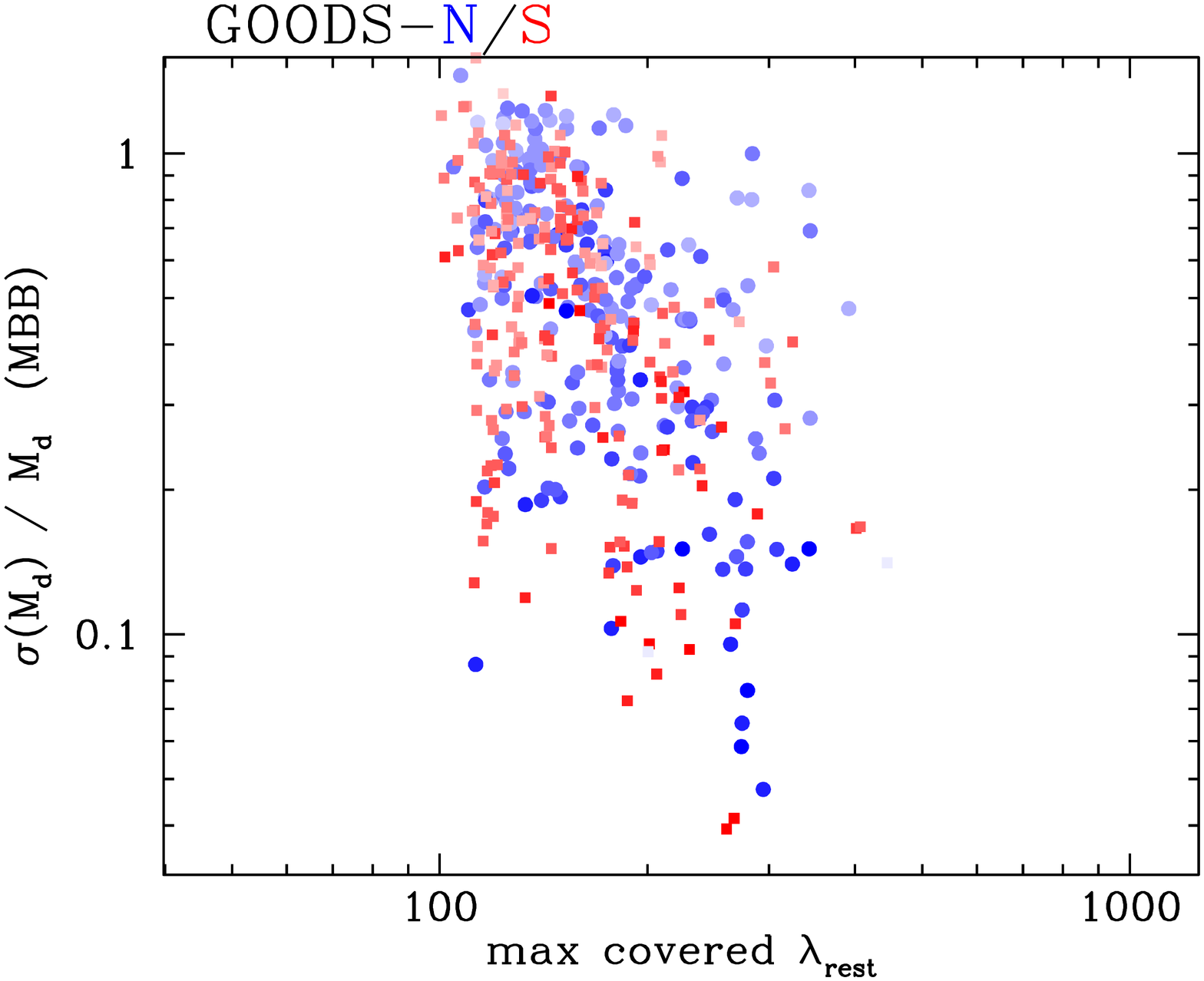}}
\rotatebox{-90}{\includegraphics[height=0.45\textwidth]{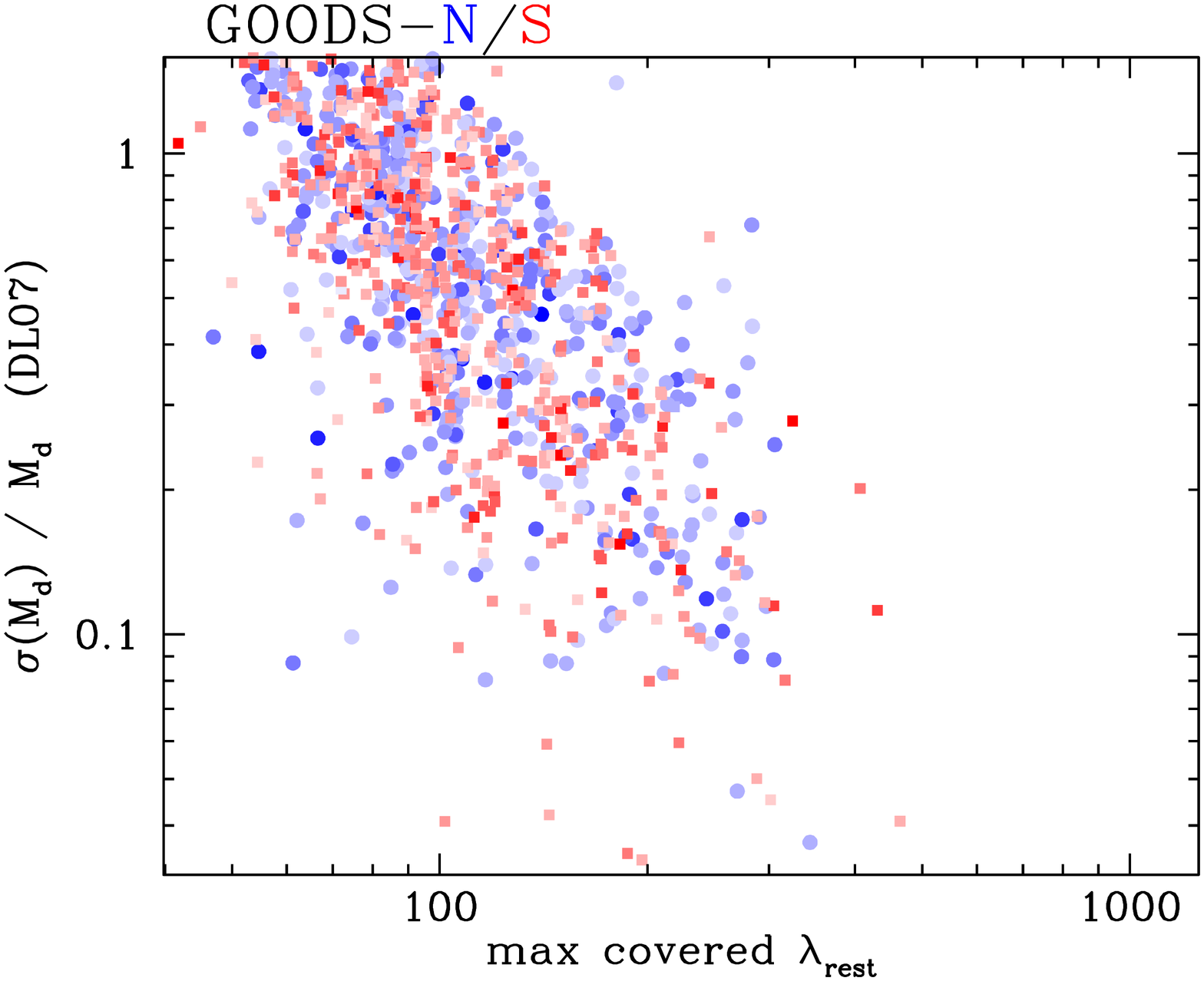}}
\caption{Relative uncertainty on $M_{\rm dust}$ for MBB ({\em left}) and DL07 ({\em right})  
fits to the FIR SEDs of GOODS-N (blue) and GOODS-S (red) sources. Color coding is based on the maximum 
S/N ratio of the available photometry (at $\lambda_{\rm obs}\ge 100$ $\mu$m) for each object,
ranging from a value of 3.0 to 1000.}
\label{fig:mdust_gn}
\end{figure*}

The estimate of $T_{\rm dust}$ is not very sensitive 
to the lack of long wavelength datapoints. 
On the contrary, it can be 
 systematically underestimated or overestimated when removing 
data on the blue side of the SED.
This happens because the 
peak of the MBB emission shifts to longer wavelengths as dust temperature 
decreases, and the effect of poor SED sampling (due to the removed datapoints)
is thus amplified.
Overall, only $\sim10$\% of cases turn out to have 
$M_{\rm dust,out}>2\times M_{\rm dust,in}$.


\section{Discussion}\label{sect:discussion}

In the previous Sections, we built expectations on DL07 and MBB 
SED fitting applied to \textit{Herschel}-detected sources. 
It is now time to derive dust and gas masses for real sources
and stacked data (see Sect. \ref{sect:data}) and to study their actual 
uncertainties and the properties of \textit{Herschel} galaxies 
in this context.

\subsection{The real world and its limitations}\label{sect:real}

We apply the methods described in Sect. \ref{sect:sed_fitting} to the GOODS-N and GOODS-S data, 
taking care to limit the MBB fit to $\lambda_{\rm rest}\ge50$ $\mu$m to avoid contamination from warm dust. We adopted a value of $\beta=2.08$  to simplify the comparison of MBB results to DL07 
dust masses, and we used the \citet{draine2003} revisitation of $\kappa_\nu$  \citep[see also][]{ld01}.

The errors analysis is only focused on $M_{\rm dust}$. In fact, 
simulations have already shown that via MBB fitting, $T_{\rm dust}$ can be 
determined within a 30\% uncertainty as long as the short-wavelength side 
of the SED is constrained; the values of the $\gamma$, $q_{\rm PAH}$, 
$U_{\rm min}$, and DL07 parameters can hardly be constrained by the available photometry. 
These results are also shared in the SED fitting of real SEDs discussed here and therefore 
are not covered further in this Section.

The left panel of Fig. \ref{fig:mdust_gn} shows the relative uncertainty
on $M_{\rm dust}$ for a MBB fit to 
all 160$\mu$m-selected GOODS-N/S sources as a function of the maximum available rest-frame 
wavelength, $\lambda_{\rm max}$, of each object. The right panel of Fig. \ref{fig:mdust_gn}
deals with DL07 models. In both cases, color coding is based on the 
maximum S/N ratio of the available photometry for each object with darker 
symbols indicating a higher S/N. 

At larger rest-frame $\lambda_{\rm max}$,
the number of available bands is also larger, mostly because of redshift effects.
In the case of the MBB fit, we emphasize that sources with smaller $\sigma\left(M_{\rm dust}\right)/M_{\rm dust}$
tend to benefit from a higher quality photometry, i.e., a higher (maximum) photometric 
S/N at $\lambda_{\rm obs}\ge 100$ $\mu$m. On the other hand, no such trend is seen in the DL07 case. 

The relative uncertainty on $M_{\rm dust}$ increases as the maximum 
covered rest-frame wavelength decreases, thus confirming the findings of Monte Carlo simulations 
(see Sect. \ref{sect:restframe_simu}). In the case of GOODS fields, the use of 
photometric upper limits on the long-wavelength side does not provide a significant advantage in SED fitting 
because SPIRE data are much shallower than PACS data (up to a factor of $\sim$10; 
see Table \ref{tab:depths}).

The performance turns out to be better in the DL07 case:
The trend of $\sigma\left(M_{\rm dust}\right)/M_{\rm dust}$ vs. $\lambda_{\rm max}$ 
is flatter, and the  distribution is characterized by a smaller scatter at a given 
value of  $\lambda_{\rm max,rest}$. For example, at a rest-frame $\lambda_{\rm max}$=200 $\mu$m 
with DL07 modeling, it is always possible to constrain $M_{\rm dust}$ to a $\sim$40\% relative 
uncertainty, while the relative error on the MBB-based $M_{\rm dust}$ 
can reach values as large as 70\%.

In the best case scenario of five to six bands available (covering up to 500 $\mu$m
in the observed frame), the uncertainty on dust mass is typically on the order of 20\% 
but can reach up to 30-40\%, depending on the S/N of the available photometry and 
on the adopted modeling.

\subsection{Comparison of DL07 and MBB $M_{\rm dust}$ estimates}\label{sect:compa_dl07_mbb}

\begin{figure*}[!ht]
\centering
\includegraphics[width=0.45\textwidth]{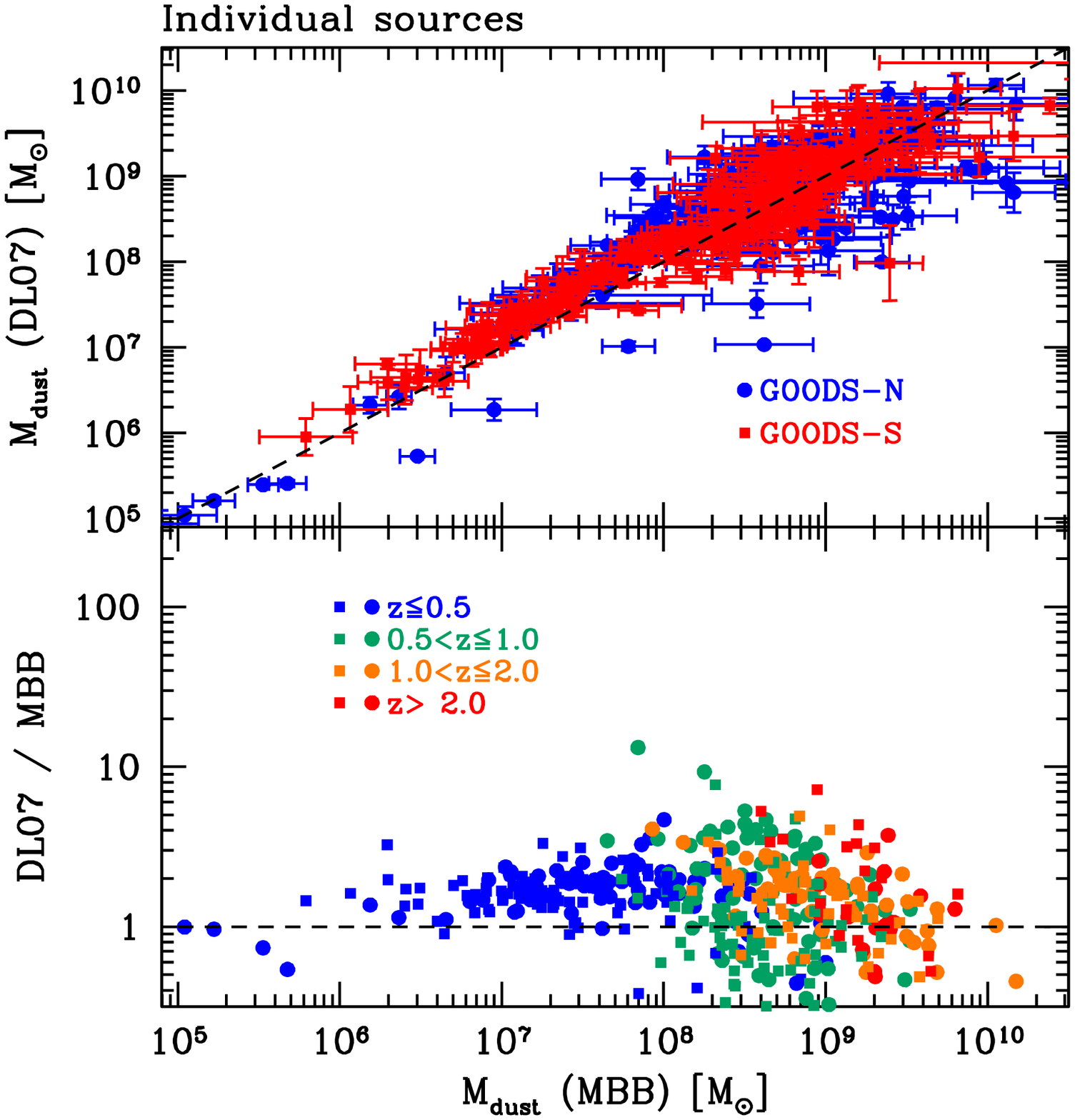}
\includegraphics[width=0.45\textwidth]{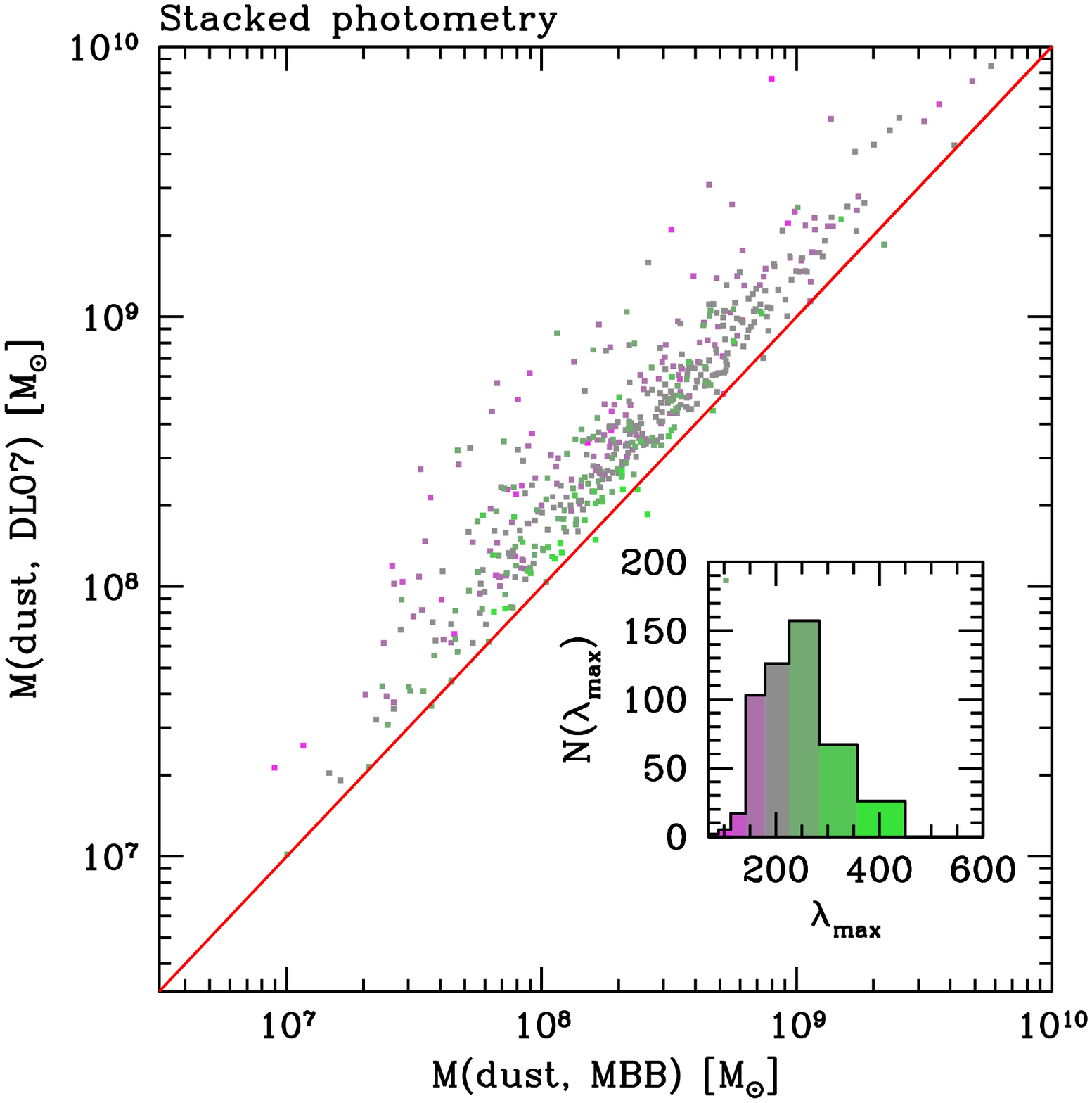}
\caption{Comparison of MBB and DL07 dust estimates. The MBB fitting adopts $\beta=2.08$ and 
a minimum rest-frame wavelength of 50 $\mu$m in the fit. The DL07 estimates are obtained 
with the whole 8-500 $\mu$m photometric set for a maximum of nine bands, and 
DL07 parameters are limited as prescribed by \citet{draine2007b}. {\em Left}: Individual GOODS-N/S sources detected by \textit{Herschel}.
{\em Right}: Stacked photometry by \citet{magnelli2014} color coded on the basis of the 
maximum available rest-frame wavelength (see also inset).}
\label{fig:cfr_mbb_dl07_mdust}
\end{figure*}

Figure \ref{fig:cfr_mbb_dl07_mdust} reports on the comparison 
of $M_{\rm dust}$ estimated with MBB and DL07 SED fitting.
A median systematic offset of $\le50$\% between the two estimates is found. 
Studying objects with fully sampled SEDs (up to sub-mm wavelengths), \citet{magdis2012,magdis2013} 
report a systematic offset of a factor $\sim2$. \citet{magnelli2012a} report a factor $\sim$ 3 
discrepancy. Comparing these different results is not as straightforward as it might seem.
In fact, the underlying assumptions for the MBB modeling differ  significantly. We have already 
mentioned in Sect. \ref{sect:role_beta} there is a discrepancy between the 
$\beta=1.5$ emissivity adopted by \citet{magnelli2012b} and the 
$\beta$ used by the DH02 models, which depends on the intensity of the radiation field.
Finally, the $\kappa_\nu$ \citep{ld01,draine2003} adopted by most authors has a frequency 
dependency on the power of $\beta=2.08$. 
\citet{bianchi2013} shows that when fitting local NGC 
galaxies and using $\beta=2.08,$ any offset should be washed away (see Sect. \ref{sect:kappa}). 
Although we are using a consistent $(\beta,\kappa_\nu)$ set for the two approaches, this 
incongruence with respect to \citet{bianchi2013} results might come from possible mismatched 
temperatures due to less well-sampled SEDs of high-$z$ objects.
It is therefore always
important to  report on the adopted setup when referring to 
this delicate comparison.

\begin{figure*}[!ht]
\centering
\rotatebox{-90}{
\includegraphics[height=0.7\textwidth]{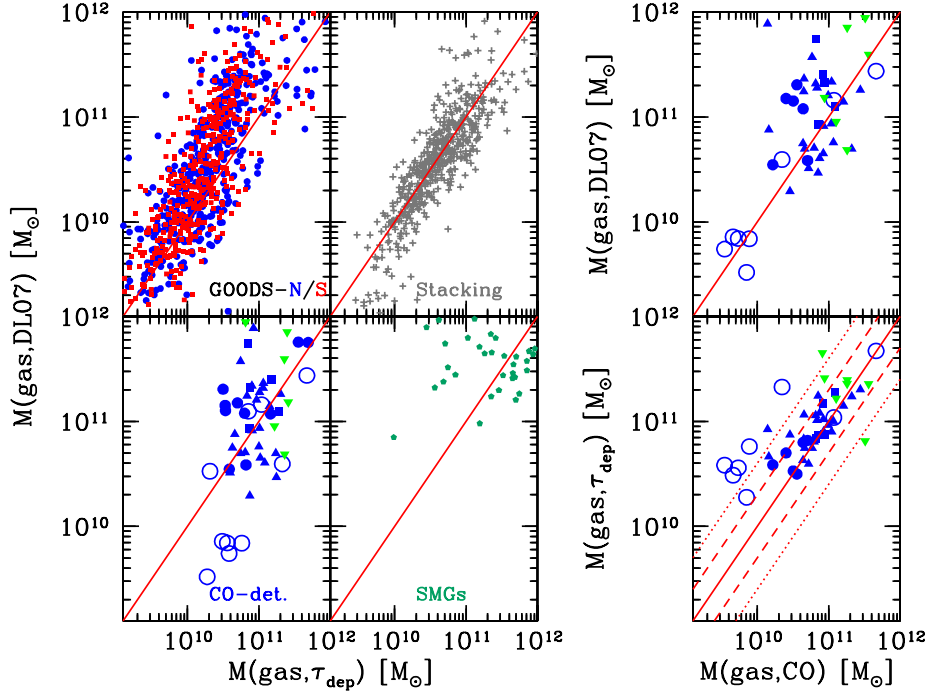}
}
\caption{Comparison of three different $M_{\rm gas}$ estimates: The first
is based on CO (rescaled to a common $\alpha_{\rm CO,MW}=4.36,$  including 
a metallicity correction; Genzel et al. \citeyear{genzel2012,bolatto2013,genzel2015});
the second based on the scaling of depletion times, $\tau_{\rm dep}$; and the third derived from 
$M_{\rm dust}$, using the $\delta_{\rm GDR}$-$Z$ and $M^\ast$-$Z$ 
relations in the PP04 metallicity scale.
{\em Left}: Datasets include GOODS-N (blue filled circles) and GOODS-S (red filled squares) 
160$\mu$m-detected sources; \textit{Herschel}-detected SMGs \citep[green filled pentagons;][]{magnelli2012a};
CO-detected galaxies (different blue symbols, see below); and the \citet{magnelli2014} stacked 
points (gray crosses). 
The {\rm right}-hand panels only include CO-detected sources, namely:
PHIBSS galaxies (blue triangles, Tacconi et al. \citeyear{tacconi2013});
BzK galaxies (blue squares, Daddi et al. \citeyear{daddi2010a});
other star-forming galaxies (blue filled circles, Magnelli et al. \citeyear{magnelli2012b}); 
lensed galaxies (empty blue circles, Saintonge et al. \citeyear{saintonge2013});
and SMGs (green upside down triangles, Bothwell et al. \citeyear{bothwell2013}).
The dashed and dotted lines in the lower-right panel mark 
the $\pm0.3$ and $\pm0.6$ dex deviations from the 1:1 
locus (solid line).}
\label{fig:cfr_mgas_DL_KS_CO}
\end{figure*}

\begin{figure}[!ht]
\centering
\rotatebox{-90}{
\includegraphics[height=0.45\textwidth]{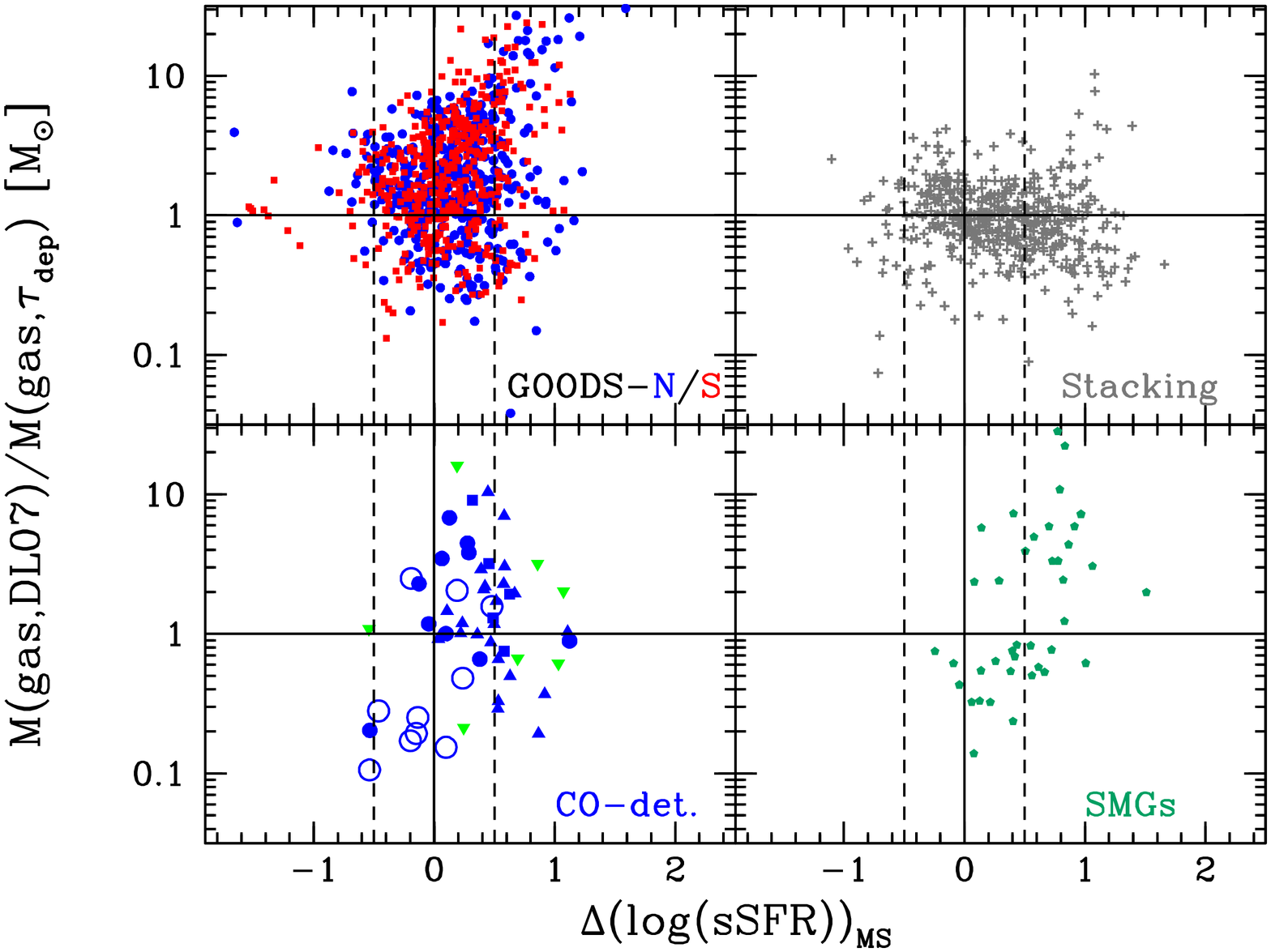}
}
\caption{Comparison of $M_{\rm gas}$ as based on dust 
masses and on the scaling of $\tau_{\rm dep}$, as a function of distance from the main sequence of star formation. 
The \citet{whitaker2014} definition of the MS has been adopted. 
Symbols are as in Fig. \ref{fig:cfr_mgas_DL_KS_CO}.}
\label{fig:cfr_mgas_DL_KS_3}
\end{figure}

Color coding the ratio of the two estimates by redshift, we see that the scatter 
in the distribution of points becomes very large above $z\sim1$. At this redshift, one begins to 
lose PACS bands, on the blue side of the SED, because of $k$ correction and sensitivity issues; at the same time, 
 the SPIRE photometry also becomes poorer and more affected by confusion noise. 

Similar results are obtained fitting the stacked photometry by \citet{magnelli2014}, 
which has the advantage of a more extensive SED coverage on the 
long-wavelength side (Fig. \ref{fig:cfr_mbb_dl07_mdust}, right). In this case, it is seen that objects 
with longer $\lambda_{\rm rest,max}$ lie closer to the 1:1 locus. 
On the contrary, for poorly sampled SEDs, the discrepancy becomes larger (see also Sect. 
\ref{sect:restframe_simu}).
We conclude that two concomitant effects contribute 
to the difference in MBB/DL07 mass ratios obtained by different authors: 
the underlying MBB model assumptions and the available spectral coverage.

\subsection{Dust-based $M_{\rm gas}$}

Following Sect. \ref{sect:mgas_mdust}, the DL07-based dust masses are converted 
into the molecular gas content of galaxies following $\delta_{\rm GDR}$-$Z$ scaling of \citet{magdis2012}. Metallicities are computed from stellar masses with the parameterization of the $M^\ast$-$Z$ relation by \citet{genzel2015}.
Both of these relations are calibrated to the PP04 metallicity scale.

Figure \ref{fig:cfr_mgas_DL_KS_CO} compares different $M_{\rm gas}$ estimates 
for the datasets in hand, all inclusive of the helium contribution. 
CO-based masses have been rescaled to a common Milky Way conversion 
factor $\alpha_{\rm CO,MW}=4.36$, and we have taken a metallicity correction 
 into account as well. The adopted correction
is the geometric mean of the \citet{bolatto2013} and \citet{genzel2012} 
dependencies of $\alpha_{CO}$ on metallicity \citep[see][]{genzel2015}.

Gas masses based on the scaling of depletion times
have been obtained adopting Eq. \ref{eq:genzel2015_tau_dep} in Sect.
\ref{sect:ks} \citep[][and priv. comm.]{genzel2015} 
and the \citet{whitaker2014} definition of star formation main sequence.

\begin{figure}[!h]
\centering
\rotatebox{-90}{
\includegraphics[height=0.48\textwidth,trim=0cm 0cm 6cm 0cm,clip=true]{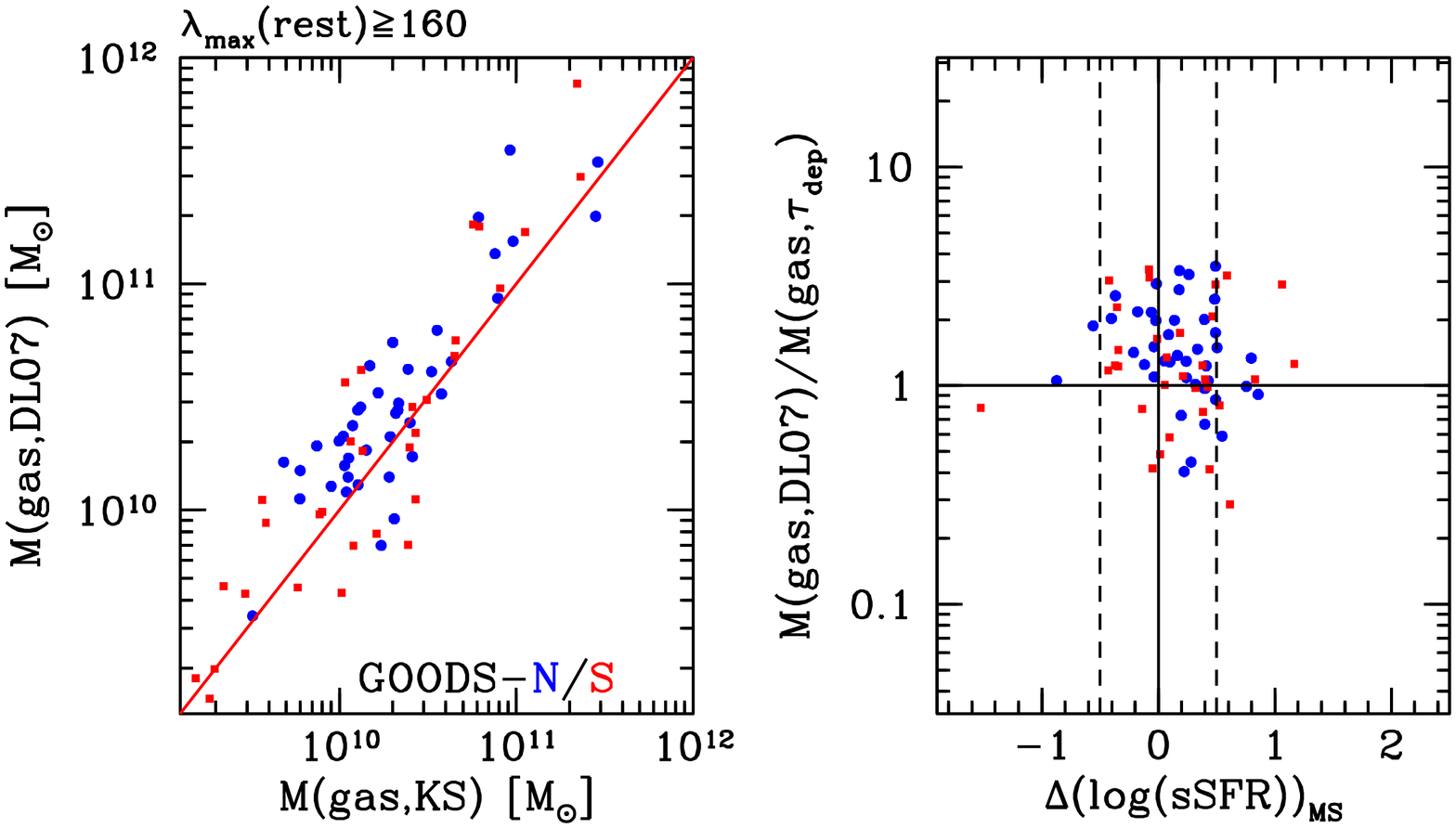}
}
\caption{Same as Figs. \ref{fig:cfr_mgas_DL_KS_CO} and \ref{fig:cfr_mgas_DL_KS_3} 
for GOODS-S/N sources only, limiting the maximum available rest-frame wavelength to 
$\lambda_{\rm max}\left(\textrm{rest}\right)\ge160$ $\mu$m.}
\label{fig:cfr_mgas_DL_KS_CO_160}
\end{figure}

In the case of CO-detected galaxies, the CO-based and $\tau_{\rm dep}$-based $M_{\rm gas}$ 
estimates are  consistent overall within a factor of $\sim2$ with only few exceptions.
For the low-mass lensed galaxies, \citet{saintonge2013} adopted 
the metallicity-dependent values of 
$\alpha_{CO}$ derived using the \citet{genzel2012} relation, nevertheless, the metallicities 
of most of their sources are out of the range where the \citet{genzel2015} $\tau_{\rm dep}$
scaling was calibrated (and holds). 
 It is thus no surprise that these sources show an offset between 
CO- and $\tau_{\rm dep}$-based determinations of M$_{\rm gas}$ (bottom right panel of Fig. \ref{fig:cfr_mgas_DL_KS_CO}).

As far as CO- and dust-based estimates are compared (top-right panel 
of Fig. \ref{fig:cfr_mgas_DL_KS_CO}), for roughly one third of the sources 
the two $M_{\rm gas}$ estimates differ by more than 0.3 dex with dust systematically 
providing a higher $M_{\rm gas}$ than CO. The SEDs of these sources are 
missing some long-wavelength bands and, therefore, the $M_{\rm dust}$ estimate
is affected by systematics, as seen in  previous Sections.

Dust-based $M_{\rm gas}$ of the stacked photometry by \citet{magnelli2014} are in good 
agreement with $\tau_{\rm dep}$-based estimates with a relatively small scatter of $\sim$0.23 dex, which is 
on the same order of the scatter in the adopted 
$\tau_{\rm dep}\left(z,M^\ast,\textrm{sSFR}\right)$ scaling relation \citep{genzel2015}.
 The situation for individual objects is more complex: 
at the low-mass end, the two estimates of $M_{\rm gas}$ are in fair agreement, 
but, at high masses, the dust-based estimate is significantly larger
than $\tau_{\rm dep}$ results. 
The mismatch between the two $M_{\rm gas}$ determinations becomes 
larger as the number of available bands decreases,  
 mainly driven by limited wavelength coverage. This holds both for PACS-selected sources 
and for CO-detected objects. 

For the general GOODS-N/S far-IR population, $\lambda_{\rm max}(\textrm{rest})\le200$ $\mu$m
and can be as low as $\le$100 $\mu$m even at intermediate-low redshift because of the large
noise in SPIRE bands (see Table \ref{tab:depths}). According to 
our simulations, in this case dust masses can be overestimated (see Sect. \ref{sect:dl07_syst}), 
thus explaining the difference in $M_{\rm gas}$ in these cases. At higher redshift, 
$\lambda_{\rm max}(\textrm{rest})$ becomes even shorter if no sub-mm detections are available, 
and the overestimate of masses becomes more critical (see Sect. \ref{sect:NEW_syst}).
Figure \ref{fig:cfr_mgas_DL_KS_CO_160} limits the results on GOODS-N/S sources to 
$\lambda_{\rm max}(\textrm{rest})\ge160$ $\mu$m.

When an object benefits from sub-mm observations (e.g., 
as is the case for most CO-detected galaxies and SMGs),
then the maximum rest-frame wavelength available is on average longer than for 
other PACS-selected galaxies. If a sufficient number of bands is available, 
and the SED coverage is fine enough, then $\tau_{\rm dep}$- and DL07-based masses turn  out 
to be in good agreement. Outliers with $M_{\rm gas}(DL07)\gg M_{\rm gas}(\tau_{\rm dep})$
suffer from as poor SED coverage as analogous GOODS-N/S cases.

It is worthwhile recalling that the adopted $\tau_{\rm dep}$ scaling was 
calibrated for MS galaxies of nearly solar metallicity \citep{genzel2015}. 
Therefore, we also expect a contribution to the
$M_{\rm gas}(DL07)/M_{\rm gas}(\tau_{\rm dep})$ mismatch from the adopted 
$\tau_{\rm dep}$ scaling.
Figure \ref{fig:cfr_mgas_DL_KS_3} exemplifies the possible trends of 
this ratio as a function of distance from the MS of star formation 
in terms of $\Delta\left(\log\left(\textrm{sSFR}\right)\right)_{MS}$.
The \citet{whitaker2014} definition of MS has been adopted.
The dependence of $M_{\rm gas}$ on the distance from the 
star-forming MS found by \citet{genzel2015} does not play a role in this case 
because it is factorized out by taking the $M_{\rm gas,DL07}/M_{\rm gas,\tau_{\rm dep}}$ ratio.

\begin{figure*}[!ht]
\centering
\rotatebox{-90}{
\includegraphics[height=0.45\textwidth]{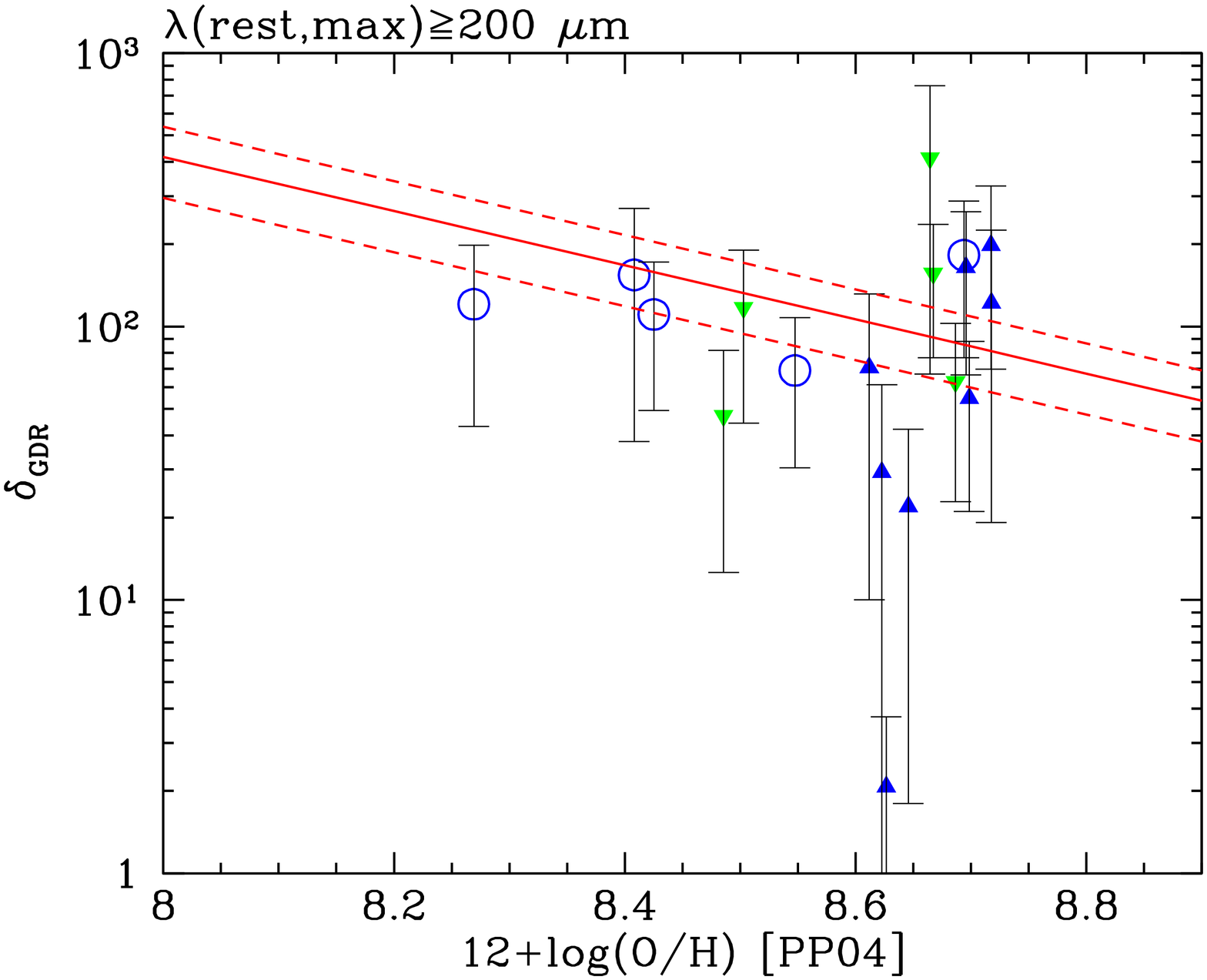}
}
\rotatebox{-90}{
\includegraphics[height=0.45\textwidth]{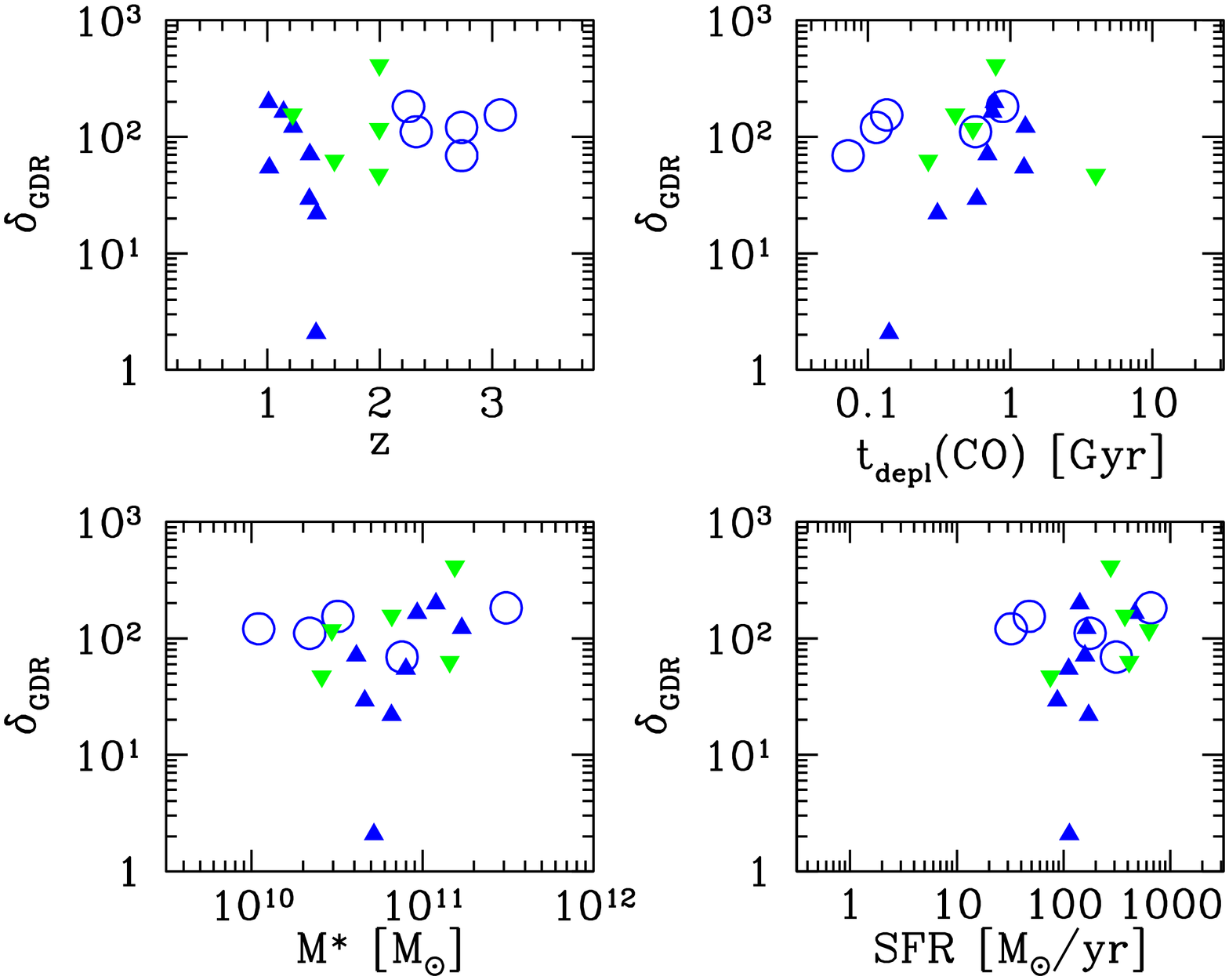}
}
\caption{Gas-to-dust mass ratio, $\delta_{\rm GDR}$, of CO-detected galaxies
with rest-frame $\lambda_{\rm max}\ge200$ $\mu$m, computed by combining CO-based 
gas masses and DL07-based dust masses. Symbols are as in Fig. \ref{fig:cfr_mgas_DL_KS_CO}. 
{\em Left}: $\delta_{\rm GDR}$ as a 
function of metallicity. The red lines represent the local relation by \citet{magdis2012} 
and its 0.15 dex scatter. {\em Right}: $\delta_{\rm GDR}$ as a function of 
other relevant parameters: redshift $z$; depletion times based on CO; 
$M^\ast$ \citep{wuyts2011a}; and $\textrm{SFR}\left(\textrm{IR}\right)$. }
\label{fig:delta_gdr}
\end{figure*}

In the case of the \citet{magnelli2014} stacked photometry, the data show 
a trend of the $\tau_{\rm dep}$-based 
$M_{\rm gas}$ estimates of bins above the MS to be larger than those based on dust.
Limiting to $\Delta\left(\log\left(\textrm{sSFR}\right)\right)_{MS}\le\pm 0.5$ dex
(vertical dashed lines), we restrict the analysis to the MS proper locus. The scatter of the 
$M_{\rm gas}(DL07)/M_{\rm gas}(\tau_{\rm dep})$ decreases to 0.19 dex for the stacked photometry. 
As a result of stacking, the SED is covered up to
$\lambda_{\rm max}(\textrm{rest})\sim160$ $\mu$m and up to $z=2$. Our simulations
(Sect. \ref{sect:dl07_syst}) prove that this is enough to avoid underestimates 
of $M_{\rm dust}$ and keep overestimates to less than a factor of 2 (details depending on the 
actual values of $q_{\rm pah}$, $U_{\rm min}$, and $\gamma$). 

As for individual sources, the majority of catastrophic failures lying at $>3\sigma$ 
(measured above using the result for stacking)
above $M_{\rm gas}(DL07)/M_{\rm gas}(\tau_{\rm dep})=1$ turn out to be 
on the high-$z$ tail of the \textit{Herschel} sample. The SEDs of these galaxies 
are poorly sampled and are only covered  up to $\lambda_{\rm max}(\textrm{rest}) \le 100$ $\mu$m.

In summary, the analysis of the SEDs of real sources and stacked photometry
by means of DL07 fitting confirms the expectations from the MC 
exercise discussed before. When the mid- and far-IR rest-frame SEDs are covered
up to at least 160 $\mu$m, the DL07 dust-based estimate of $M_{\rm gas}$ is reliable, 
not affected by systematics, and consistent with independent estimates based on
$\tau_{\rm dep}$ scalings or CO observations. If the SED wavelength coverage 
is poorer, i.e., limited to shorter wavelengths, dust masses can be overestimated.
This might happen because the SED is redshifted and therefore the available bands 
sample shorter wavelengths and also because of a decreased S/N ratio and increased
confusion in the case of faint, high-$z$ galaxies. 
If the maximum rest-frame wavelength available falls shorter than 100 $\mu$m, DL07-based 
dust masses are severely overestimated and should be not recommended anymore.


\subsection{The dust-to-gas ratio}

 One assumes a value of the gas-to-dust mass ratio, $\delta_{\rm GDR}$ to compute the molecular gas content of galaxies from their dust masses. 
As seen in Sect. \ref{sect:mgas_mdust}, it is common to adopt the local 
scaling of $\delta_{\rm GDR}$ with the metallicity derived by \citet{leroy2011} 
or some a variant of it \citep[e.g.,][]{magdis2012}. This procedure implicitly 
assumes that the relation holds regardless of redshift.

The sample of CO-detected sources in hand benefits from fully independent 
measurements of gas and dust masses.
We combine CO-based $M_{\rm gas}$ with the DL07-based determinations of 
$M_{\rm dust}$ to derive $\delta_{\rm GDR}$. All gas masses have been renormalized to the 
Galactic $\alpha_{\rm CO,MW}=4.36$ and a metallicity dependence of $\alpha_{\rm CO}$ 
has been included as well \citep{genzel2015}.
The $M^\ast$-$Z$ relation 
described in Eq. \ref{eq:mass_metall} is used with stellar masses 
from BC03 SED fitting \citep{wuyts2011a} to produce metallicities on the PP04 scale.

We only consider sources with $\lambda_{\rm max}\left(\textrm{rest}\right)\ge200$ $\mu$m
 hereto minimize the uncertainties on $M_{\rm dust}$ 
and avoid systematics (see Sects. \ref{sect:simu_Mstar_SFR_z} and \ref{sect:restframe_simu}).
The left panel in Fig. \ref{fig:delta_gdr} shows the resulting trend
of $\delta_{\rm GDR}$ vs. $12+\log\left(O/H\right)$
compared to the \citet{magdis2012} local relation and its 0.15 dex scatter. 
A $\pm$1.0 dex systematic uncertainty on its zero point is also 
reported by \citet{magdis2012}. 

The data lie close to the local relation and are consistent with it within
the uncertainties and possible systematics. At the high metallicity end, 
the scatter is very large and there exist cases with very small $\delta_{\rm GDR}$,
which are significantly below the locus occupied by local galaxies. The right-hand panel of 
Fig. \ref{fig:delta_gdr} seeks  possible dependencies 
of $\delta_{\rm GDR}$ on other derived quantities. No significant trend is 
found as a function of $M^\ast$, SFR, $\tau_{\rm dep}$, or redshift. 
The most critical outlier turns out to be a $z\simeq1.4$ galaxy with a very 
high $M_{\rm dust}\sim 8\times10^9$ $[$M$_\odot]$ and an SED with 10-205 $\mu$m 
rest-frame coverage.

\subsection{Synergies between \textit{Herschel} and ALMA}

The photometric determination of dust masses through 
SED fitting allows us, in principle, to obtain an estimate of
gas masses for a very large number of objects via the 
$\delta_{\rm GDR}$ scaling. This approach offers the advantage 
that photometric observations are still significantly faster  
than sub-mm spectroscopy, even with the last 
generation of far-IR or sub-mm facilities, and especially 
for galaxies at $z>1$.

\begin{figure}[!ht]
\centering
\rotatebox{-90}{
\includegraphics[height=0.45\textwidth]{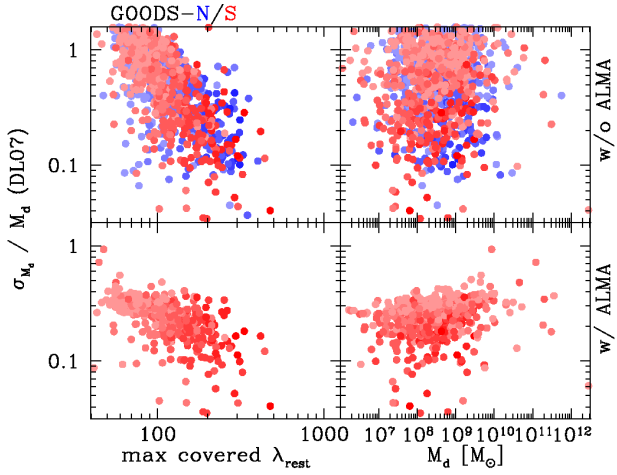}
}
\caption{Results of DL07 SED fitting to galaxies detected by 
\textit{Herschel} in the GOODS fields, including artificial ALMA band 6 
photometry. Color coding is based 
on the number of available photometric bands, ranging from only four bands  
(lighter colors) to eight bands (darker colors). This number reflects the 
maximum rest-frame wavelength available as well as redshift dependencies. 
{\em Top}: Relative uncertainties on dust mass, 
obtained only with the real photometry (see also Fig. \ref{fig:mdust_gn}).
{\em Bottom}: The same quantity obtained when adding artificial photometry in 
the ALMA band 6 ($\sim$1100 $\mu$m).}
\label{fig:alma_simu_1}
\end{figure}

Far-infrared photometry also has the advantage of 
sampling the dust emission of galaxies near the SED peak, 
providing a calorimetric measurement of their SFR \citep{elbaz2011,nordon2010}
for a relatively cheap time investment. Nevertheless, the analysis has shown 
the effects of the limited wavelength coverage:
When the SED extends only up to $\lambda_{\rm rest,max}\le200$ $\mu$m,
the uncertainty on the derived dust mass can be very large
and systematics might affect this measurement in a fraction of cases
(see Sects. \ref{sect:real}, \ref{sect:restframe_simu}). 

On the sub-mm side, 
assuming the Rayleigh-Jeans (RJ) tail of SEDs is observed, \citet{scoville2014} 
developed a strategy aimed at deriving gas (dust) masses on the basis 
of a one-band sub-mm continuum measurement. If confirmed reliable, 
this approach would be extremely competitive in terms of 
exposure time with respect to CO spectroscopy. 
\citet{genzel2015} showed that 
this method can lead to incorrect (up to a factor >3) gas masses, even  when applying corrections 
to take  the RJ approximation into account. In fact, the proposed
scaling of sub-mm luminosity holds for a specific dust temperature 
$T_{\rm dust}=25$ K, which is a condition that is only met  in a few cases. \citet{magnelli2014} have 
shown that $T_{\rm dust}$ varies across the MS of star formation, and 
increases as a function of $\Delta\left(\log\left(\textrm{sSFR}\right)\right)_{MS}$.
A dust temperature measurement is hence necessary to reach a 
correct estimate of $M_{\rm dust}$ from the sub-mm emission, even in the
simple MBB, RJ approximation.

\begin{figure}[!ht]
\centering
\rotatebox{-90}{
\includegraphics[height=0.45\textwidth]{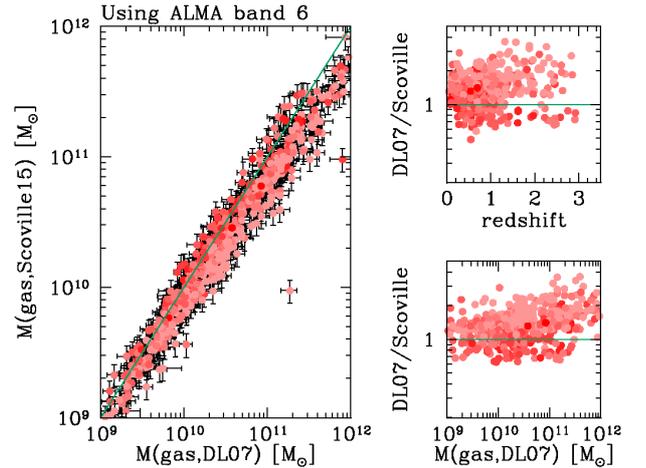}
}
\caption{Comparison of the DL07-based $M_{\rm gas}$ estimate obtained, including artificial ALMA band 6 
photometry to expectations, based on the \citet{scoville2015} recipe, and applied 
to the same artificial ALMA band 6 data. Color coding is based 
on the number of available photometric bands (see Fig. \ref{fig:alma_simu_1}).
{\em Left}: Direct comparison: error bars only include  statistic noise. 
{\em Right}: Ratio of the two estimates as
a function of redshift and DL07-based gas mass.}
\label{fig:alma_simu_2}
\end{figure}

If not using the available 
$T_{\rm dust}\left(z,M^\ast,\textrm{sSFR}\right)$ scaling \citep[e.g.,][]{magnelli2014}, 
\citet{genzel2015} suggest a strategy based on continuum observations in two 
distinct sub-mm bands. 
In this context, using ALMA bands 6 and 7 (centered at 1100 and 850 $\mu$m),
the best relative uncertainty of $M_{\rm dust}$ that can be reached is 
on the order of 60\%, for galaxies at $z\sim2$ with 10$\sigma$ detections 
in the two sub-mm bands. The uncertainty on dust mass becomes worse at lower redshift. In practice,
the SED coverage is too small and too far off the SED peak to allow for a robust determination of
$T_{\rm dust}$ and $M_{\rm dust}$. Adopting bands 7 and 9 (at 850 and 450 $\mu$m)
lowers $\sigma\left(M_{\rm dust}\right)/M_{\rm dust}$ to 30-40\% at best 
and using bands 6 and 9 finally brings it to 20-30\%, i.e., a better than $3\sigma$
estimate of $M_{\rm dust}$. However,  band 9 is particularly demanding in terms
of atmospheric conditions. At $z<1,$ this approach is not 
necessarily faster than ALMA CO spectroscopy.
We defer to \citet{genzel2015} for further details.

It is interesting to study how FIR and sub-mm observations complement
each other and how combining them improves $M_{\rm dust}$
measurements. We simulate the ALMA performance in conjunction with 
\textit{Herschel} data in the following way. 
For each galaxy detected by PACS in the GOODS-S field, 
we assume that the best-fit DL07 model obtained in Sect. \ref{sect:real}
represents the actual emission of the galaxy. This best-fit 
model is then convolved with a box ALMA passband, centered at $\sim$1100 $\mu$m 
(i.e., in ALMA's band 6). The artificial photometry thus obtained is added to the 
real \textit{Spitzer} and \textit{Herschel} data and the extended SED is 
fitted again following the usual procedure. 
We assume that a S/N ratio of 5 holds in ALMA band 6.

Figure \ref{fig:alma_simu_1} compares the uncertainties on $M_{\rm dust}$ 
obtained with and without the artificial band  6 photometric point 
 as a function of the 
maximum rest-frame wavelength covered by real data, $\lambda_{\rm rest,max}$, 
and dust mass (see also Fig. \ref{fig:mdust_gn}). 
The availability of good quality (S/N=5 in this case) photometry at 
$\lambda_{\rm obs}$=1100 $\mu$m reduces the uncertainty on 
$M_{\rm dust}$ to $\le$33\% for $>$85\% of sources. For comparison, 
only $\sim$20\% of the sample have  similarly good quality $M_{\rm dust}$ estimates 
when no ALMA data are available.
Similar results are obtained for the \citet{magnelli2014} stacked data,
adding the ALMA band 6 
filter to the simulation described in Sect. \ref{sect:simu_Mstar_SFR_z}.

We then use the artificial ALMA band 6 flux density obtained above to 
derive a gas mass expectation applying the recipe by Scoville et al. 
(\citeyear{scoville2015}; see their Appendix A and also Scoville et al. 
\citeyear{scoville2014}). As usual, DL07-based dust masses are transformed into 
$M_{\rm gas}$ adopting the \citet{genzel2015} $M^\ast$-$Z$-$z$ and
the \citet{magdis2012} $\delta_{\rm GDR}$-$Z$ relations 
(see Sect. \ref{sect:mgas_mdust}). The $M_{\rm gas}$ expectation based on
band 6 flux only \citep{scoville2015} includes the RJ correction prescribed 
by the authors. Figure \ref{fig:alma_simu_2} compares the two.

\begin{figure*}[!ht]
\centering
\rotatebox{-90}{
\includegraphics[height=0.45\textwidth]{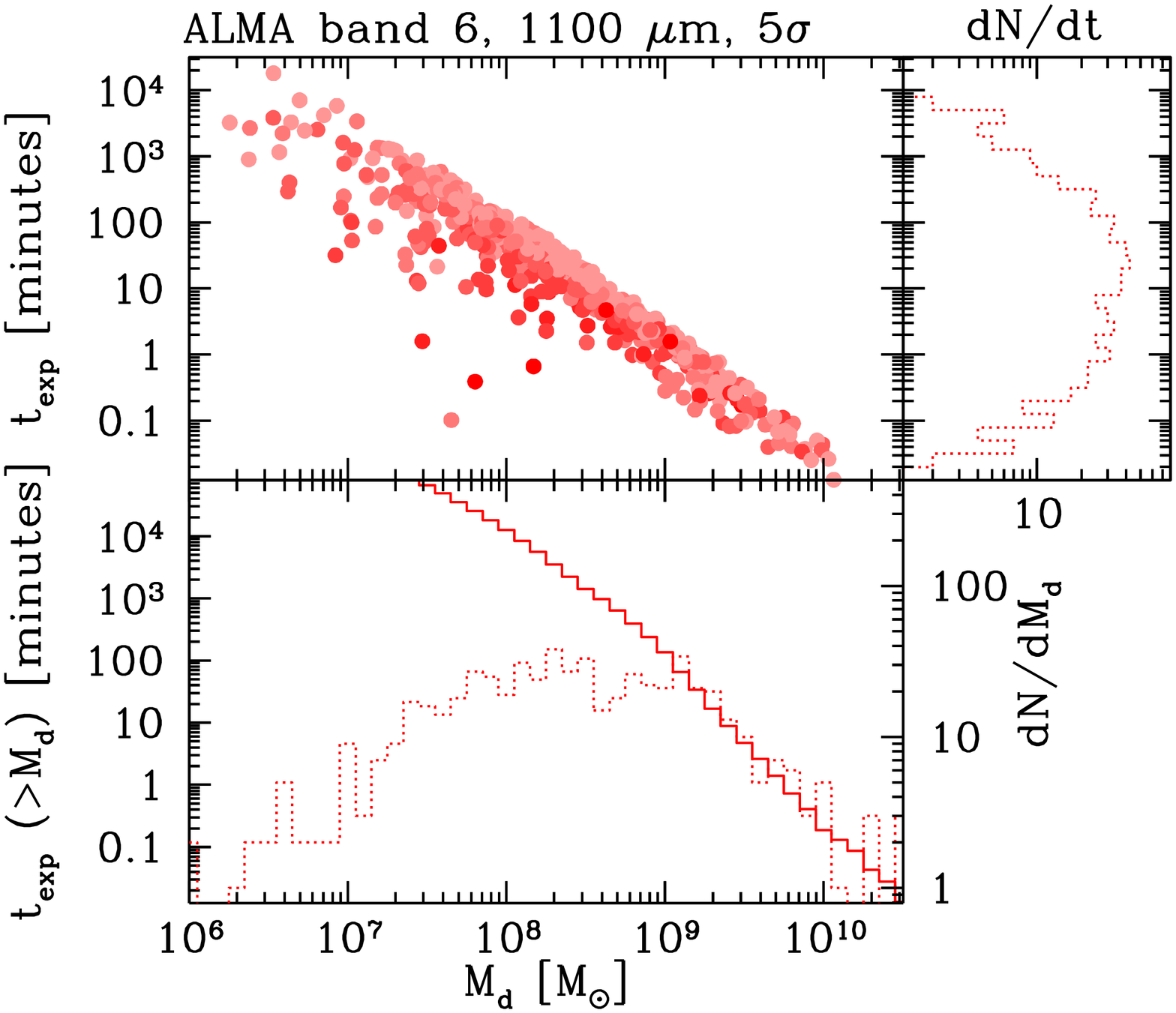}
}
\rotatebox{-90}{
\includegraphics[height=0.45\textwidth]{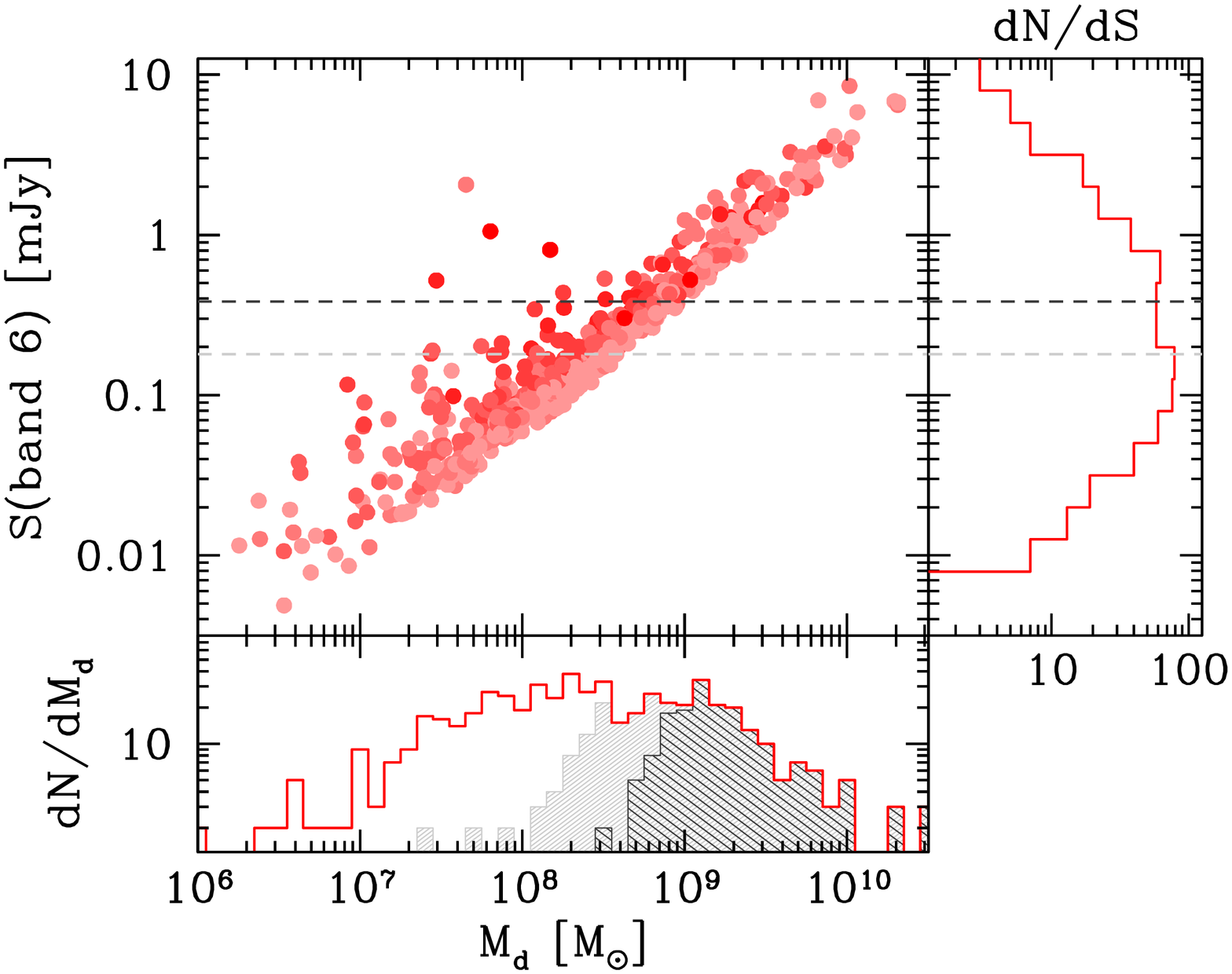}
}
\caption{Further results of DL07 SED fitting including artificial ALMA band 6 
photometry. Color coding is based 
on the number of available photometric bands (see Fig. \ref{fig:alma_simu_1}). 
{\em Left}: 
Exposure time estimate (without overheads) for ALMA band 6 observations of 
\textit{Herschel}-detected galaxies. The $t_{\rm exp}$ estimate for 
individual sources (upper bigger panel, dots) are shown and summed up
to compute the cumulative $t_{\exp}$ needed to observe all sources above a 
given value of $M_{\rm dust}$ (solid histograms). Dotted histograms represent 
the number distribution of sources. {\em Right}: 
Distribution of ALMA band 6 expected fluxes, as a function of 
dust mass, for galaxies detected by \textit{Herschel} in the GOODS-S field. The horizontal 
dashed lines correspond to 0.38 and 0.18 $[$mJy$]$ flux limits (dark and light gray, respectively).
The right side and bottom panel show the projected 1-D distributions. The hatched dark and light
gray histograms only include  sources above the flux limits indicated.}
\label{fig:band6_alma_fluxes}
\end{figure*}

The DL07-based $M_{\rm gas}$ estimate tends to be systematically larger than 
the expectation based on the recipe by \citet{scoville2015}. 
For a small fraction of the sample the opposite trend holds.
The overestimate
becomes larger at the high-mass end. The trend seen here is the combination of 
the systematic higher $M_{\rm dust}$ obtained with DL07 modeling with respect to 
MBB (see Sect. \ref{sect:compa_dl07_mbb}), the fact that dust temperature 
is varying as a function of the position in the $M^\ast$-SFR-$z$ space 
\citep[see][]{genzel2015,magnelli2014}, a slight difference in the $\beta$ values 
underlying the two methods ($\beta=2.08$ vs. $1.8$), and possibly other systematics 
(e.g., related to metallicity dependencies). This is different from the 
problem encountered in Fig. \ref{fig:cfr_mgas_DL_KS_CO} at large $M_{\rm gas}$ because now 
we are reasoning in relative terms (DL07 vs. Scoville methods), and the long 
wavelength side of the SED is constrained by the artificial ALMA photometry.

Following a similar path to \citet{genzel2015}, 
we now estimate the {\em on source} exposure time, $t_{\rm exp}$, 
needed to reach S/N=5 in band 6 for our \textit{Herschel}-detected star-forming galaxies
based on DL07 models.

To this aim, the ALMA sensitivity 
calculator\footnote{http://almascience.eso.org/proposing/sensitivity-calculator}
is used with the standard Cycle 3 configuration via an array of 36 12m antennas,
requesting an angular resolution of 1 arcsec.
The left-hand panels of Fig. \ref{fig:band6_alma_fluxes} show 
$t_{\rm exp}$ for individual sources and the cumulative 
$t_{\rm exp}$ for sources below a given dust mass (solid histograms) 
along with the number distribution of objects (dotted histograms).
The dispersion of the points in the upper panel reflects the spread
in redshift of the sample.  However, the sub-mm 
negative $k$ correction implies that exposure times 
for a given mass are of the same order of magnitude at all redshifts.

Targeting all \textit{Herschel} galaxies detected by the PEP/GOODS-H
survey in the GOODS-S field down to $M_{\rm dust}=10^9$ $[$M$_\odot]$
would require a few minutes of on-source exposure, i.e., without accounting for 
overheads. Reaching down to $M_{\rm dust}=10^{8.5}$ $[$M$_\odot]$
requires roughly 20-25 hours on source with ALMA band 6 in the 
above mentioned configuration. 
Using ALMA band 7 would need a similar amount of time.

The right-hand panel of Fig. \ref{fig:band6_alma_fluxes} includes the distribution of 
the expected ALMA band 6 fluxes as a function of dust mass. Instead of 
requesting a fixed S/N ratio, a 0.38 mJy and 0.18 mJy depths are shown, 
similar to recent ALMA Cycle-3 approved surveys\footnote{See list of ALMA Cycle-3 high priority projects:
http://almascience.eso.org/observing/highest-priority-projects.}. At these depths,
ALMA  detects the majority of \textit{Herschel} sources above 
$M_{\rm dust}=10^{8.5}$ $[$M$_\odot]$ with a fainter tail 
extending down to $\sim10^8$ $[$M$_\odot]$, and with the added value 
of possible undetected PACS objects, which were not included in the current analysis.

Assuming an average dust/gas ratio of 0.01, the limits mentioned above correspond to 
$M_{\rm gas}=10^{10.5-11}$ $[$M$_\odot]$. \citet{berta2013b} showed that 
the Schechter characteristic mass, $M_{\rm gas}^\ast$ of the 
molecular gas mass function lies between $M_{\rm gas}=10^{10.3-10.9}$ 
$[$M$_\odot]$ at $z=0.2-2.0$. 
A measurement of the molecular gas mass function based on dust observations, modulo the 
$\delta_{\rm GDR}$ scaling, and on the dust mass function itself 
down to $M_{\rm gas}^\ast$ and up to $z\sim2$, is 
within reach of ALMA in 20-25 hours spent on source.


\section{Conclusions}\label{sect:summary}

We have exploited the deepest FIR blank field maps available to date from  the PEP, GOODS-H, and HerMES surveys to study the feasibility of deriving dust and gas masses 
via SED fitting for individually detected sources. 
In parallel, we  built extensive Monte Carlo simulations to study 
the limitations of real data, and to understand how they influence the uncertainties 
and systematics on such dust mass determinations. 
We  focused the analysis on two popular modeling approaches: SED 
fitting by means of a single-temperature modified blackbody and 
by means of the \citet{dl07} model.
The main results of the analysis based on artificial sources and MC 
sampling are:
\begin{itemize}
\item FIR SED fitting recovers dust mass consistently as long as 
the wavelength coverage offered by the data extends at least up to 
160-200 $\mu$m (rest frame). In this case, no systematics are expected on $M_{\rm dust}$.
In contrast, if the  $>3\sigma$ detections fall shorter than this 
wavelength limit, $M_{\rm dust}$ can be severely overestimated with 
the amplitude of the systematic effect depending on the details 
of the model.
\item The uncertainty of $M_{\rm dust}$ also strongly depends on 
wavelength coverage. As a rule of thumb, it is not possible to reach 
a $\ge3\sigma$ determination of $M_{\rm dust}$ if the available 
photometry does not extend at least up to $\sim$200 $\mu$m in the rest frame.
\item The determination of dust temperatures based on MBB fitting is rather stable,
even if long-wavelength bands are missing. In addition,  $T_{\rm dust}$ is always 
constrained to better than a 10-20\% uncertainty as long as the 
blue side of the FIR SED is constrained.
Nevertheless, a small offset 
on $T_{\rm dust}$ can induce a large systematic error on 
$M_{\rm dust}$.
\item The discreteness of the sampled wavelengths (photometric bands), 
combined with redshift, can cause significant residual systematics
in the determination of $T_{\rm dust}$, up to few K.
\item Reliability tests based on artificial catalogs tend to minimize 
systematics on $M_{\rm dust}$ if the SED shape adopted for SED fitting 
is consistent with the one used to produce the mock catalog. 
In other words, these kinds of simulations assume 
that the adopted SED library 
is consistent with real-world' SEDs. On the other hand, small differences in 
SED shape (e.g., DH02 vs. DL07 models) can induce strong and not easily predictable 
systematic effects. 
\end{itemize}

The GOODS fields benefit from an extensive multiwavelength coverage
from the UV to far-IR and sub-mm wavelengths. The SEDs of galaxies 
detected by \textit{Herschel} can reach up to 200 $\mu$m rest frame and beyond.
The SEDs of individual sources have been fitted with DL07 and MBB 
models. At the same time,  stacked photometry of near-IR selected 
galaxies has also undergone the same analysis, binned in $z$-$M^\ast$-SFR space.
The main results of these pieces of analysis are:
\begin{itemize}
\item At the depth of the deepest \textit{Herschel} extragalactic surveys 
(GOODS-S as observed by PEP, GOODS-H, and HerMES), it is possible to 
retrieve dust masses with a S/N$\ge$3 for galaxies on the 
main sequence of star formation down to $M^\ast\sim10^{10}$ $[$M$_\odot]$
up to $z\sim 1$. At higher redshift ($z\le2$), the same goal is 
achieved  for objects only lying increasingly above the MS 
at similar stellar masses or for galaxies at the tip of the MS (i.e., with higher $M^\ast$).
At shallower depths (e.g., in the case of the COSMOS field), this reasoning
shifts to even higher values of SFR.
Dust temperatures (based on MBB fit) can be constrained within a 10\% accuracy
in most of the cases across the $z$-$M^\ast$-SFR space, modulo residual systematics
due to the discreteness of SED sampling (see above).
\item As in the case of simulated data, \textit{Spitzer} and \textit{Herschel} data alone are not sufficient to produce 
an estimate of $M_{\rm dust}$ to better than 30\% uncertainty if the 
maximum rest-frame wavelength covered by the data is shorter than $\sim$160-200 $\mu$m.
\item Comparing MBB- and DL07-based masses, 
the average offset between the two, regardless 
of redshift, is a factor $\sim1.5$.
At $z>1,$ the scatter of the MBB/DL07 mass ratio becomes very
large, mainly because photometric points are progressively missed.
 We stress that to allow a direct and meaningful 
comparison, it is paramount to 
adopt a consistent set of parameters, taking special care 
in the value of $\beta$, i.e., the slope of the dependence of 
dust emissivity on frequency.
\item Dust masses estimated with DL07 modeling are more robust than 
those based on MBB: Relative errors are more mildly dependent on 
the maximum covered rest-frame wavelength and less scattered.
\item Dust mass estimates, based on DL07 SED modeling and on 
scaling of depletion times, $\tau_{\rm dep}$, are consistent with each other 
as long as the data guarantee a sufficient wavelength coverage.
Applying the local dependence of $\delta_{\rm gdr}$ on metallicity to transform 
$M_{\rm dust}$ into $M_{\rm gas}$, these estimates 
are overall consistent with CO-based estimates for a small sample of 
star-forming galaxies. While comparing $M_{\rm gas}$ estimates 
based on different methods, it is important to adopt a 
consistent set of relations, calibrated to a common metallicity scale.
\item Using CO-based $M_{\rm gas}$ renormalized to $\alpha_{\rm CO,MW}=4.36$, 
$M_{\rm dust}$ obtained through DL07 modeling,
and metallicities computed with the $M^\ast$-$Z$-$z$ relation \citep{genzel2015},
the $\delta_{\rm GDR}$ of $z>1$ galaxies depends on metallicity in 
a similar manner as for local galaxies within uncertainties and systematics.
\end{itemize}

CO-based $M_{\rm gas}$ estimates, which also represent an important anchor for validation of indirect methods despite 
$\alpha_{\rm CO}$ uncertainties, are still limited at $z>1$ \citep[see also][]{genzel2015}. 
More CO observations of individual
galaxies covering a wide range of parameters are highly desirable and are becoming
more easily accessible with ALMA and NOEMA. In parallel,
also the determination of gas-phase abundances and, therefore, of the
redshift evolution of ISM physical properties \citep[e.g.,][]{kewley2013,
shapley2015}, will undergo significant progress over the next few years thanks to 
NIR multiobject spectroscopy (e.g., with KMOS, MOSFIRE).

Finally, as $M_{\rm dust}$ estimates based on \textit{Spitzer} and 
\textit{Herschel} photometry are limited to cases with high quality SEDs available, 
we recalled the advantages and 
limitations of estimates including only sub-mm data \citep[e.g.,][]{genzel2015}. 
For example, a scaling of sub-mm fluxes into $M_{\rm dust}$ can 
be affected by strong systematics if the characteristic dust temperature of 
the SED is not known. Continuum observations in two sub-mm continuum bands might eventually help, but they are time consuming and the results are still limited by 
large uncertainties because the wavelength range covered is relatively small.

We therefore explored a combined IR plus sub-mm approach, combining existing \textit{Herschel}
data to expected ALMA 850 or 1100 $\mu$m continuum fluxes.
These single band observations 
allow one to reduce the uncertainties on $M_{\rm dust}$ down to $<30$\% 
for virtually all \textit{Herschel}-detected galaxies in the GOODS-S field.
A direct measurement of the 
molecular gas mass function based on dust observations up to $z\sim2$ will be soon
within reach.



\begin{acknowledgements}  
The authors would like to thank the referee, Dr. Georgios Magdis, for his useful 
and effective comments that improved the quality of this manuscript, 
as well as for his positive and fast response.
PACS has been developed by a consortium of institutes led by MPE (Germany) and 
including UVIE (Austria); KU Leuven, CSL, IMEC (Belgium); CEA, LAM (France); 
MPIA (Germany); INAF-IFSI/OAA/OAP/OAT, LENS, SISSA (Italy); IAC (Spain). 
This development has been supported by the funding agencies BMVIT (Austria), 
ESA-PRODEX (Belgium), CEA/CNES (France), DLR (Germany), ASI/INAF (Italy), 
and CICYT/MCYT (Spain). SPIRE has been developed by a consortium of institutes
led by Cardiff University (UK) and including University of Lethbridge (Canada),
NAOC (China), CEA, LAM (France), IFSI, University of Padua (Italy), IAC
(Spain), Stockholm Observatory (Sweden), Imperial College London, RAL,
UCL-MSSL, UKATC, University of Sussex (UK), Caltech, JPL, NHSC,
University of Colorado (USA). This development has been supported by national
funding agencies: CSA (Canada); NAOC (China); CEA, CNES, CNRS (France);
ASI (Italy); MCINN (Spain); SNSB (Sweden); STFC, UKSA (UK) and NASA
(USA).
\end{acknowledgements}




\bibliographystyle{aa}
\bibliography{27746}




\begin{appendix}

\section{Ancillary data for CO samples}\label{sect:app_CO_data}

The PHIBSS survey \citep{tacconi2013} performed CO spectroscopy 
of galaxies at $z=1$-$2,$ and included additional data from 
past work, for a total of 73 objects. The survey observed 38 sources in the 
Extended Groth Strip (EGS) field, also targeted by PEP and HerMES. 
Another 18 objects belong to the ``Q-fields'' \citep{steidel2004}; \citet{tacconi2013}
also included six BzK sources by \citet{daddi2010a} and six sources by \citet{magnelli2012b}, 
plus a few other additional objects.

Far-IR photometry is sought in the EGS from the PEP and HerMES data \citep{lutz2011,oliver2012}, using 
a closest-neighbor algorithm and visual inspection of multiwavelength maps 
(IRAC, MIPS 24$\mu$m, PACS, and SPIRE). The sources by 
Magnelli and Daddi, and few other isolated objects  studied
by \citet{magnelli2012b} and \citet{saintonge2013}, and their photometry 
can be found in their works.

Out of the 38 PHIBSS sources in EGS, we have the following 
detection statistics: 37, 18, 15, 14, 15, and 13 are detected at 
24, 100, 160, 250, 350, and 500 $\mu$m, respectively. 
Five out of the six BzK sources  by \citet{daddi2010a} included in PHIBSS 
have 24 $\mu$m to 500 $\mu$m photometry by PEP and HerMES 
\citep{lutz2011,oliver2012}. 
The object BzK21000 benefits from 
millimeter photometry (1.3, 2.2, 3.3 mm) by Dannerbauer et al. (quoted as in prep. by Magdis et al. 2011),
\citet{carilli2010}, and \citet{daddi2009a} as collected by \citet{magdis2011}.
CB58 and the Cosmic Eye have photometry collected by \citet{saintonge2013}.
For the other sources in the PHIBSS sample (mainly the Q-fields), no 
FIR photometry has been retrieved from the literature yet.
        
\citet{magnelli2012b} included in their analysis 
six PEP sources, two HDF sources, and the well-studied object GN20.
They actually included nine sources in their observations, but 
only six are used by \citet{tacconi2013}.
For the six PEP objects, 24-to-500 $\mu$m photometry has been retrieved 
from PEP and HerMES catalogs \citep{lutz2011,oliver2012}.
The two HDF169 and HDF242 objects also have 24-to-500 $\mu$m photometry \citet{magnelli2012b}.
In addition to the 24-to-500 $\mu$m imaging, GN20 also has 850, 1100, 2200, 3300 6600 $\mu$m 
photometry collected by \citet{magdis2011} and obtained by 
\citet{pope2006}, \citet{perera2008}, \citet{dannerbauer2009}, \citet{daddi2009a},
\citet{carilli2011}. For this source,  
a 1160 $\mu$m photometry is also available, obtained by \citet{penner2011}
from the combination of 1100 $\mu$m Atzec and 1200 $\mu$m MAMBO maps.

\citet{saintonge2013} have observed 17 lensed galaxies, out of which ten have CO
detections (and derived physical quantities).
The available photometry includes 100-to-500 $\mu$m PACS and SPIRE data,
and 1200 $\mu$m IRAM/MAMBO measurements for five of them.

\citet{bothwell2013} observed 40 SMGs in CO and detected 32 of them.

They also included some additional data taken from the literature.
Their sample is spread over several different sky areas:
three sources in the Subaru deep field, or UDS; three sources in smaller fields; 
five sources in the Lockman Hole East (LH-East); 16 sources in the HDF;
three sources in SSA-13; seven sources in ELAIS-N2 (seven sources); and three sources in 
SSA-22.
The Subaru Deep field (SXDF or UDS) is part of HerMES; 
the LH-East is part of PEP and HerMES; 
the HDF is part of PEP and HerMES (GOODS-N);
ELAIS-N2 is part of HerMES, but has not been released yet 
(at the time of writing, in DR2 and DR3).
In the UDS field, only 250, 350, 500 $\mu$m bands from HerMES DR2, plus the 
850 $\mu$m by \citet{bothwell2013} are available. 

In the HDF (GOODS-N) there are the 24, 100, 160, 250, 350, 500 $\mu$m data
by PEP, GOODS-\textit{Herschel} and HerMES, plus the 850 $\mu$m fluxes by Bothwell.
In the LH-East  there are the 24, 100, 160, 250, 350, 500 $\mu$m data
by PEP + HerMES, plus the 850 $\mu$m fluxes by Bothwell.
None of the \citet{bothwell2013} sources is detected in the AzTEC 1.1 mm 
maps by \citet{michalowski2012}.

In synthesis, out of the 40 SMGs in the sample by \citet{bothwell2013},
16 have a secure CO detection and six have a candidate detection.
Out of these, 16 detected objects lie in UDF, LH, or HDF, and two candidates
lie in HDF.
Out of these 16+2, 13 have enough mid- and far-IR photometric detections to ensure 
an estimate of dust mass based on DL07 SED fitting.
Out of these 13, seven sources have enough data to be included in 
our analysis (i.e., $M_{\rm gas,CO}$, $M_{\rm dust}$, $M_{\rm stars}$).

\section{Detailed analysis of systematics}

We dissect the fine details of possible systematic effects 
in the derivation of $M_{\rm dust}$, as emerging from 
the MC runs with rest-frame photometry  
(see Sects. \ref{sect:restframe_simu}, \ref{sect:dl07_syst}, 
and \ref{sect:mbb_syst}).

\subsection{DL07 systematics}\label{sect:app_syst}

The DL07 case is analyzed first.
We only focus on dust mass estimates  because the other parameters are much more poorly 
constrained (see Sect. \ref{sect:dl07_rel_err}).

\begin{figure*}[!ht]
\centering
\includegraphics[width=0.37\textwidth]{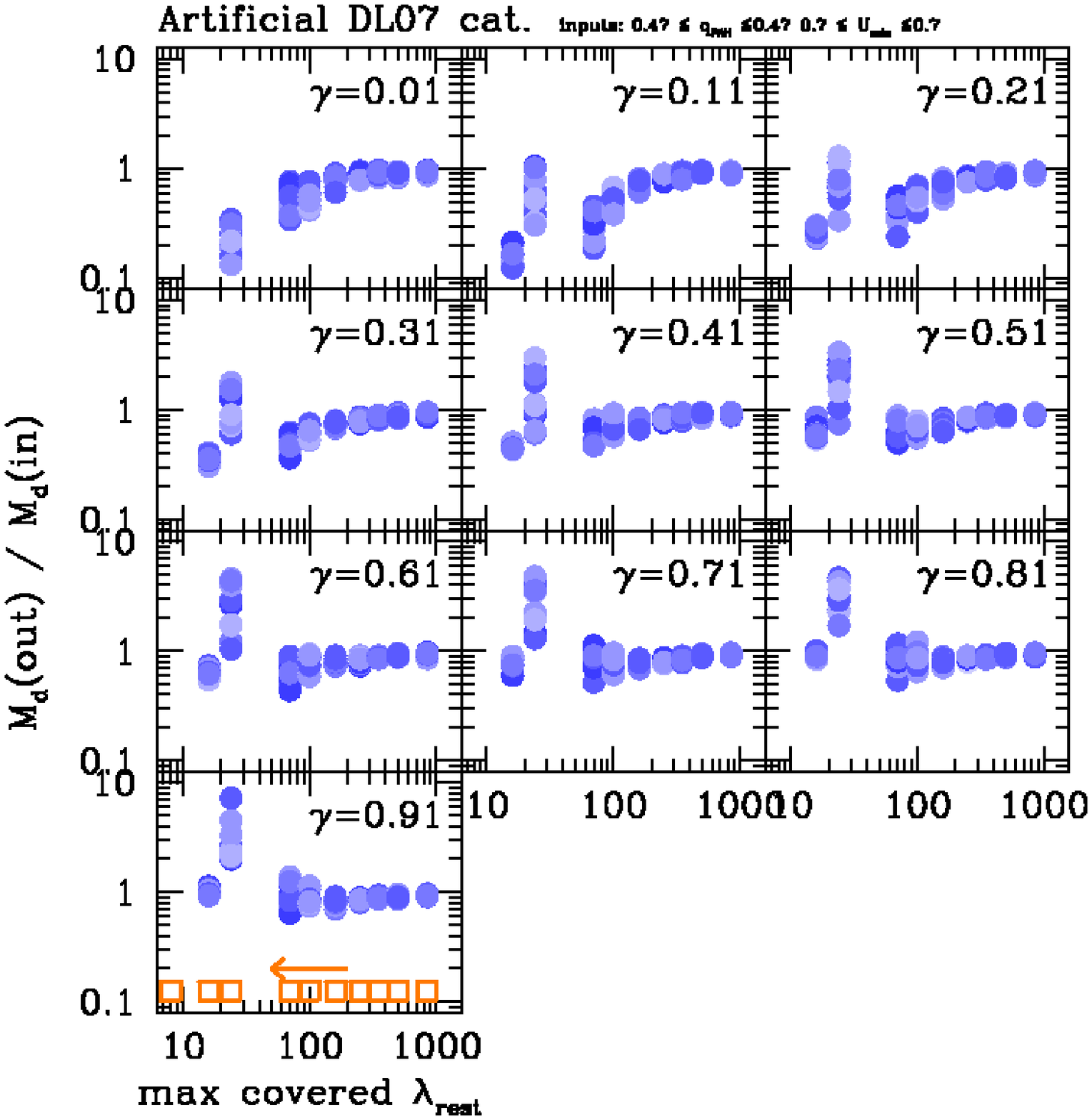}
\includegraphics[width=0.37\textwidth]{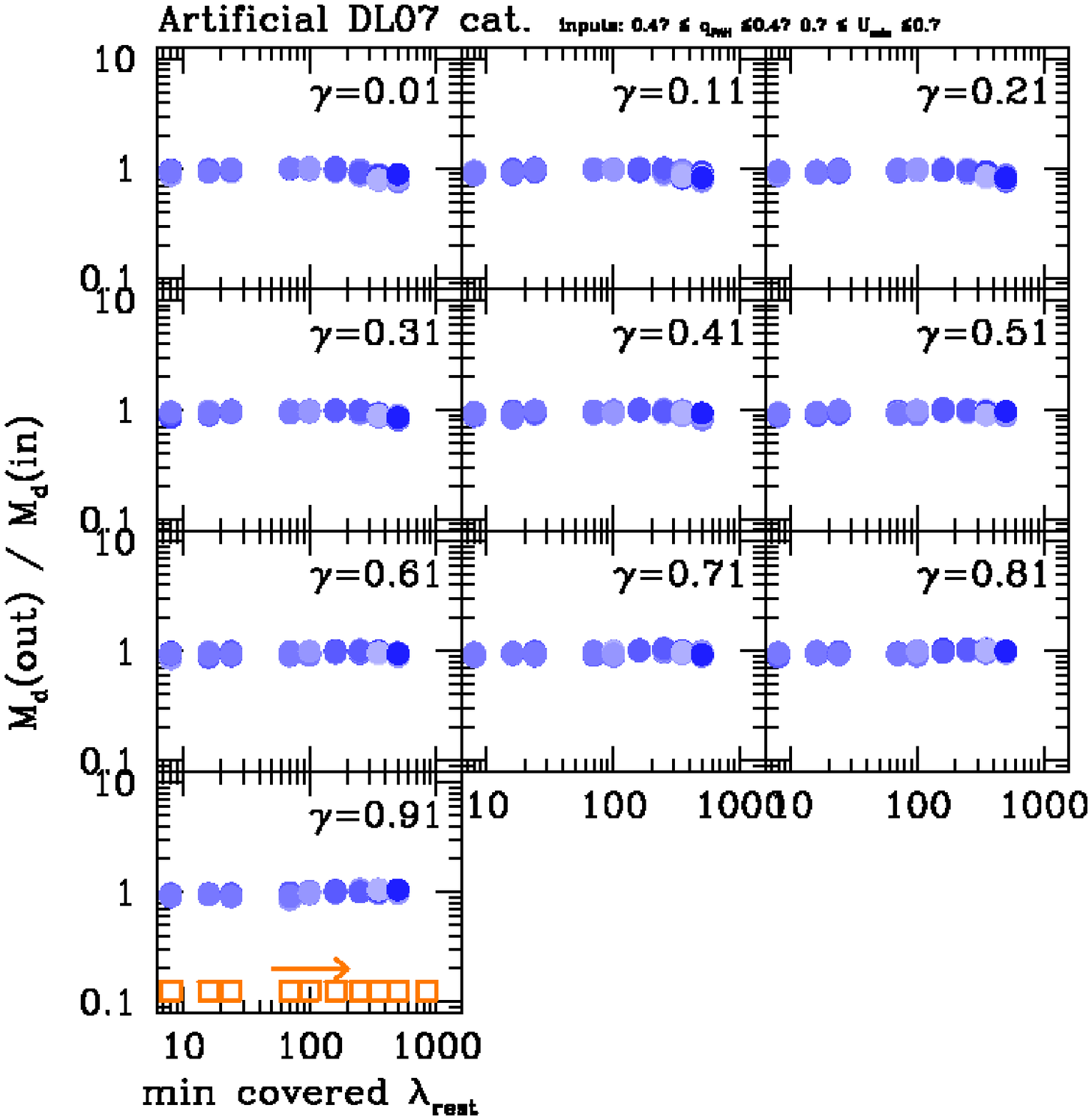}
\includegraphics[width=0.37\textwidth]{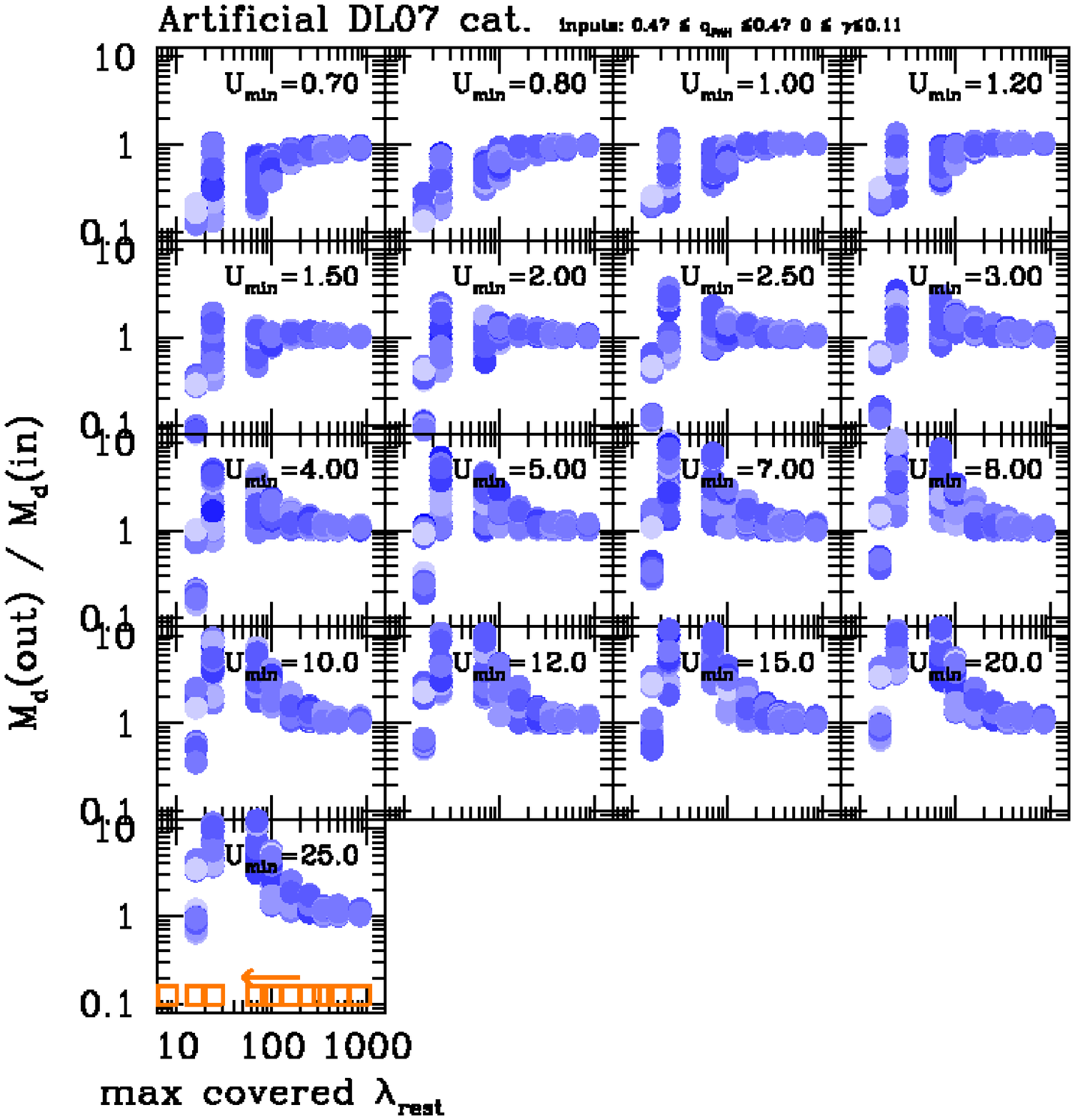}
\includegraphics[width=0.37\textwidth]{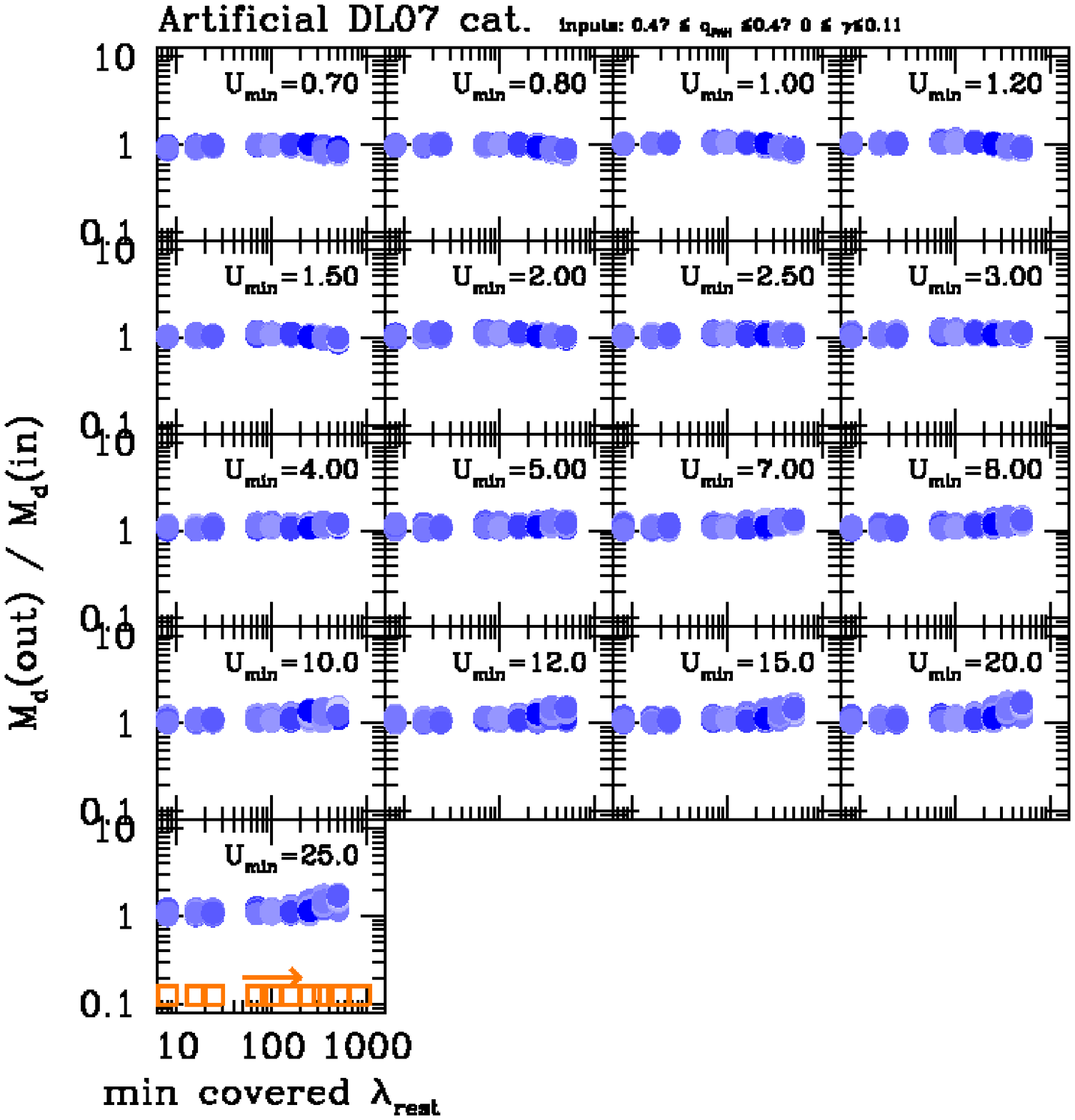}
\includegraphics[width=0.37\textwidth]{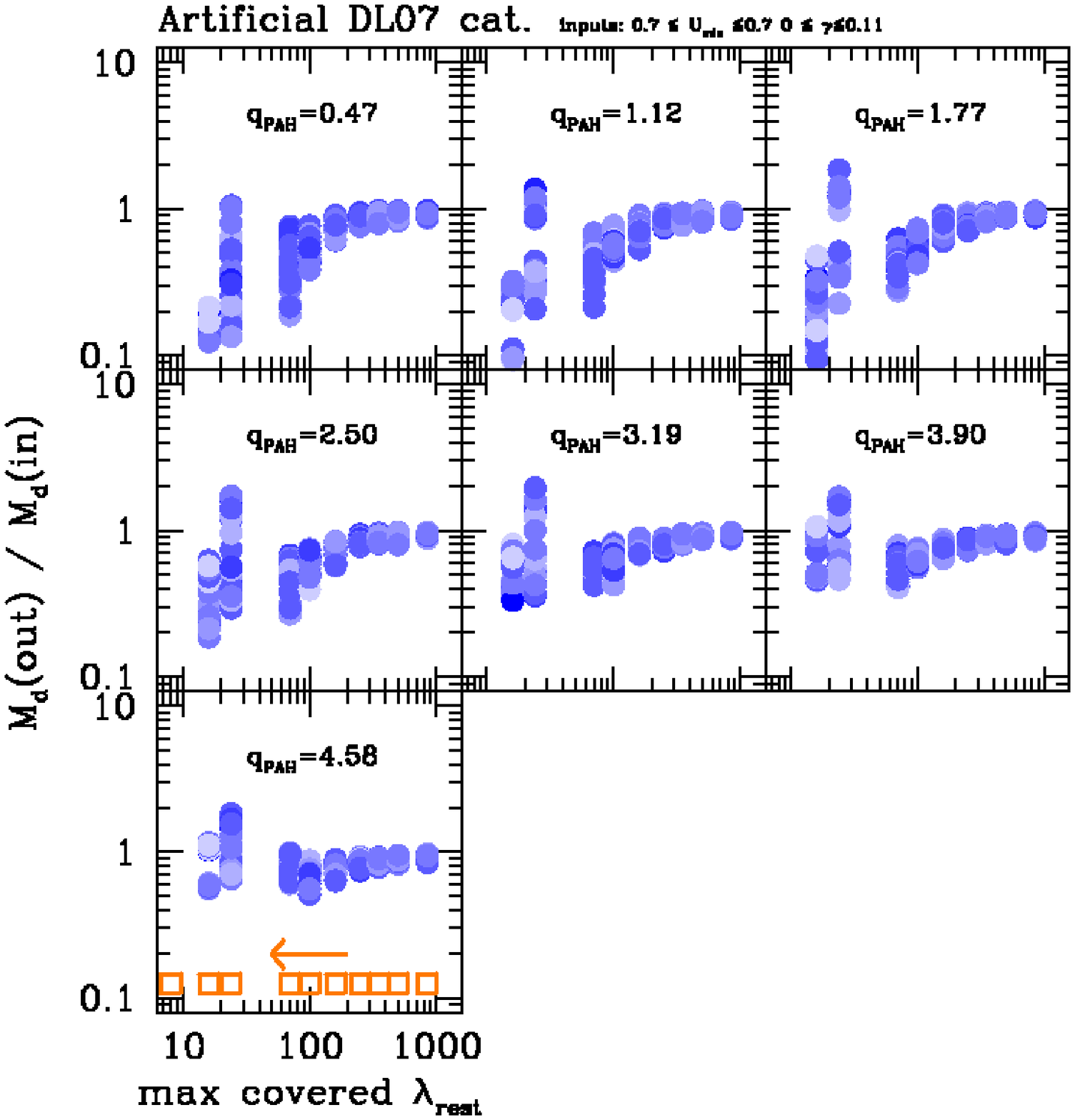}
\includegraphics[width=0.37\textwidth]{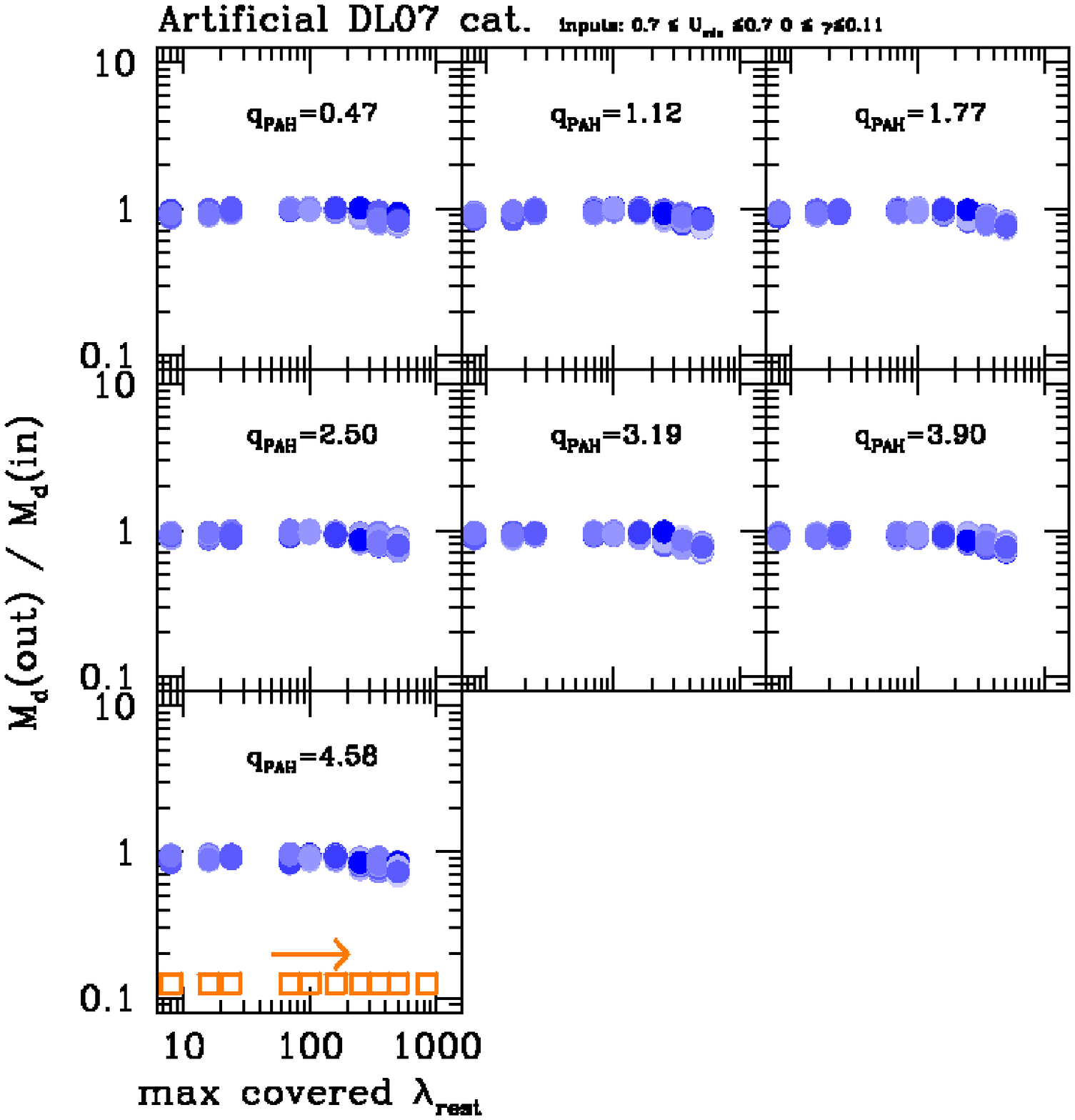}
\caption{Comparison of input and output $M_{\rm dust}$ in DL07 simulations obtained with ten photometric bands.
{\em Left/right} columns present catalog entries obtained removing bands from the long-wavelength 
and short-wavelength side of SEDs, respectively. {\em Top/middel/bottom} diagrams
show results obtained with specific values of two parameters and sampling the 3rd vary
within the ranges recommended by \citet[][see also Sect. \ref{sect:dl07}]{draine2007b}. 
When fixed, the adopted values are: $q_{\rm PAH}=0.47$, $U_{\rm min}=0.7$, and $\gamma=0.11$.}
\label{fig:dl07_in_out_1}
\end{figure*}

Figure \ref{fig:dl07_in_out_1} presents the comparison 
of $M_{\rm dust,out}$ and $M_{\rm dust,in}$ as a function of 
$\lambda_{\rm max,min}$, i.e., for cases obtained 
removing long-wavelength bands (left hand diagrams) and 
short-wavelength bands (right). Simulations including ten photometric bands are shown. 
Possible dependencies on $q_{\rm PAH,in}$, $U_{\rm min,in}$, and $\gamma_{\rm in}$ are studied by fixing two
parameters and splitting the analysis in bins of the third.

First of all, we focus on the top row of diagrams, 
showing cases at specific values of 
$q_{\rm PAH,in}=0.47$ and $U_{\rm min,in}=0.70$ and sampling 
$\gamma_{\rm in}$ in the allowed range.  
When removing long-wavelength bands, there is a tendency to underestimate
$M_{\rm dust}$, if the bands coverage and the S/N are not adequate.
This tendency becomes smaller for larger values of $\gamma_{\rm in}$.
          
This is still a special case, and it is necessary to disentangle the effect of the other two 
parameters $q_{\rm PAH}$ and $U_{\rm min}$ (see middle and bottom panels in Fig \ref{fig:dl07_in_out_1}).
For example, moving to $U_{\rm min,in}=2.0$, the tendency is to overestimate  $M_{\rm dust}$, 
rather than to underestimate it.
This difference is indeed mainly driven by $U_{\rm min,in}$ (see middle panels), but
the overestimate becomes larger for larger $\gamma_{\rm in}$.
A similar effect is also shared by $q_{\rm PAH}$ (bottom panels).
A larger value of $\gamma_{\rm in}$ produces an increased
chance to overestimate $M_{\rm dust}$. Nevertheless, if there is a general tendency to underestimate $M_{\rm dust}$ 
(e.g., because $U_{\rm min,in}$ is small), then this underestimate becomes milder 
for the larger values of $\gamma_{\rm in}$. If, instead, there is the tendency to 
overestimate $M_{\rm dust}$ (e.g., because of the large value of $U_{\rm min,in}$), then this tendency 
becomes even worse with larger $\gamma_{\rm in}$. 
          
However,  here we are speaking of large values of 
$\gamma_{\rm in}$ ($>$0.4), which are rarely seen in real 
galaxies \citep[see, e.g., Tab. 4 in][]{draine2007b}.
Keeping $\gamma_{\rm in}<0.2$, differences in trends are more 
difficult to appreciate.
          
When removing short-wavelength bands, $M_{\rm dust}$ is usually 
easily retrieved and there are not very significant trends related to $\gamma_{\rm in}$ 
in over- and underestimating $M_{\rm dust}$. This holds also varying $q_{\rm PAH,in}$ and $U_{\rm min,in}$.

We now examine the middle panels in Fig. \ref{fig:dl07_in_out_1}, 
highlighting what happens when sampling $U_{\rm min,in}$ variations at 
$q_{\rm PAH,in}=0.47$ and $\gamma_{\rm in}=0.11$.
When removing long-wavelength bands, there are now
big differences in the in/out trends as a function of $\lambda_{\rm max}$, 
depending on the actual value of $U_{\rm min,in}$.
At low $U_{\rm min,in}$ there is a tendency to underestimate $M_{\rm dust}$ if the 
band coverage is poor, while at high $U_{\rm min,in}$ there is a tendency 
to overestimate it.
          
This recalls what is happening with the MBB (see Sect. \ref{sect:mbb_syst}). Changing 
the input $U_{\rm min,in}$, in fact, produces a change of the position of 
the FIR peak (see Fig. \ref{fig:dl07} and Sect. \ref{sect:dl07}).
Consequently, there is a tendency to overestimate $M_{\rm dust}$ when the combination 
of bands coverage and ``equivalent temperature'' is  produces the effects 
described for the MBB case study (see Sect. \ref{sect:mbb_syst}). At lower $U_{\rm min,in}$, the SED peaks 
at longer wavelengths and at some point it exits the range 
where overestimate of fluxes is ``preferred'', causing underestimates
of $M_{\rm dust}$ to dominate.  
          
When removing short-wavelength bands the same effects are seen, although with a much smaller amplitude.

Finally, in the bottom panels of Fig. \ref{fig:dl07_in_out_1}, we focus on $U_{\rm min,in}=0.7$ and 
$\gamma_{\rm in}=0.11$,   
and follow the variation of $q_{\rm PAH,in}$.
In this case, as in the case of variable $\gamma_{\rm in}$, the trend to over- or underestimate $M_{\rm dust}$ changes
as a function of the value of $q_{\rm PAH,in}$.
          
When removing long-wavelength bands, 
for the quoted choice of $U_{\rm min,in}$ and $\gamma_{\rm in}$, dust masses can be 
underestimated for small values of $q_{\rm PAH,in}$, but 
this trend becomes much milder for large values.
On the other hand, remember that for $U_{\rm min,in}=2.0$ and same range of $\gamma_{\rm in}$, the tendency
is to overestimate $M_{\rm dust}$ (see middle panels of Fig. \ref{fig:dl07_in_out_1}), 
and the tendency is to have a larger overestimate when $q_{\rm PAH,in}$ becomes larger.

\begin{figure*}[!ht]
\centering
\includegraphics[width=0.45\textwidth]{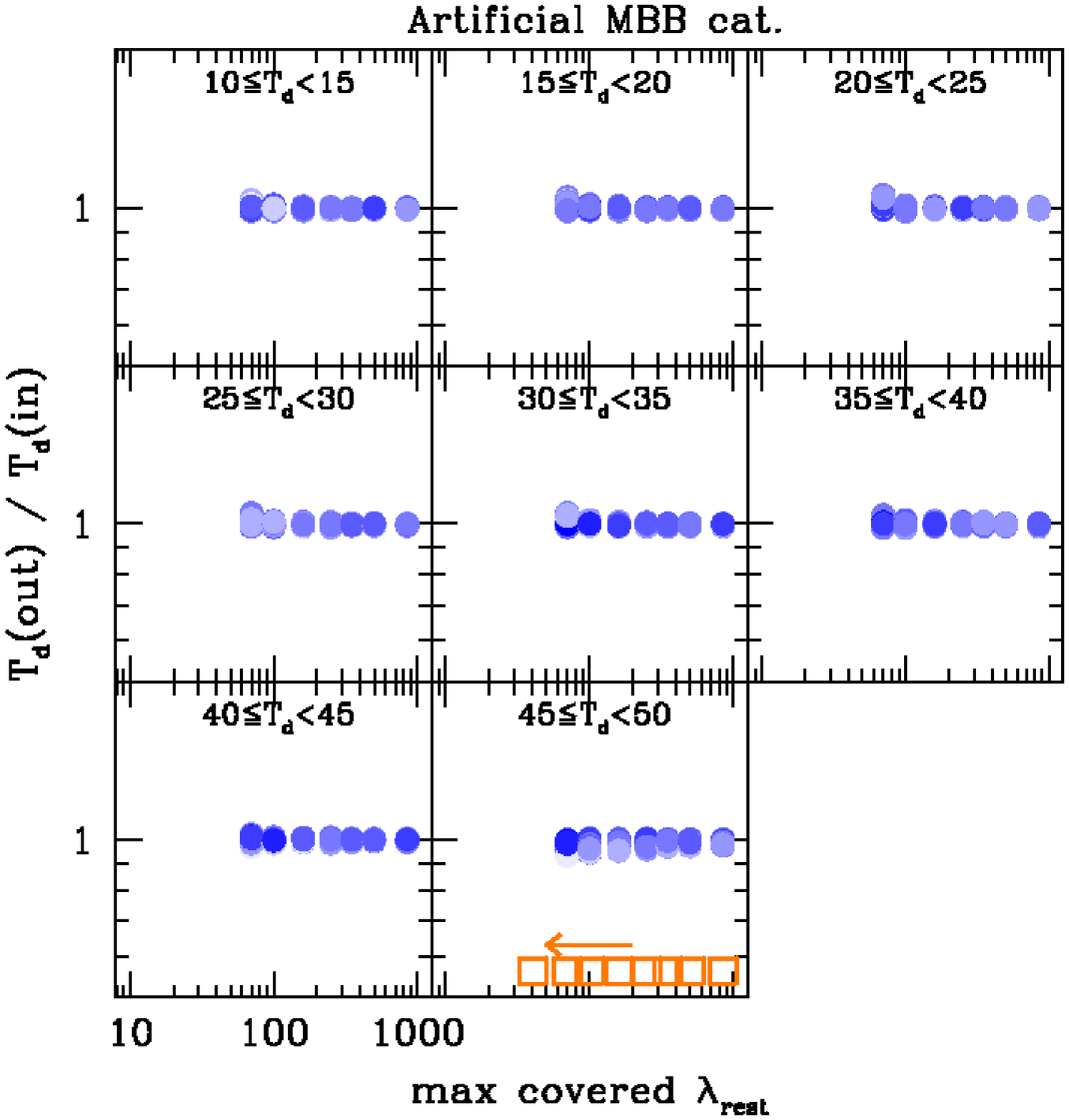}
\includegraphics[width=0.45\textwidth]{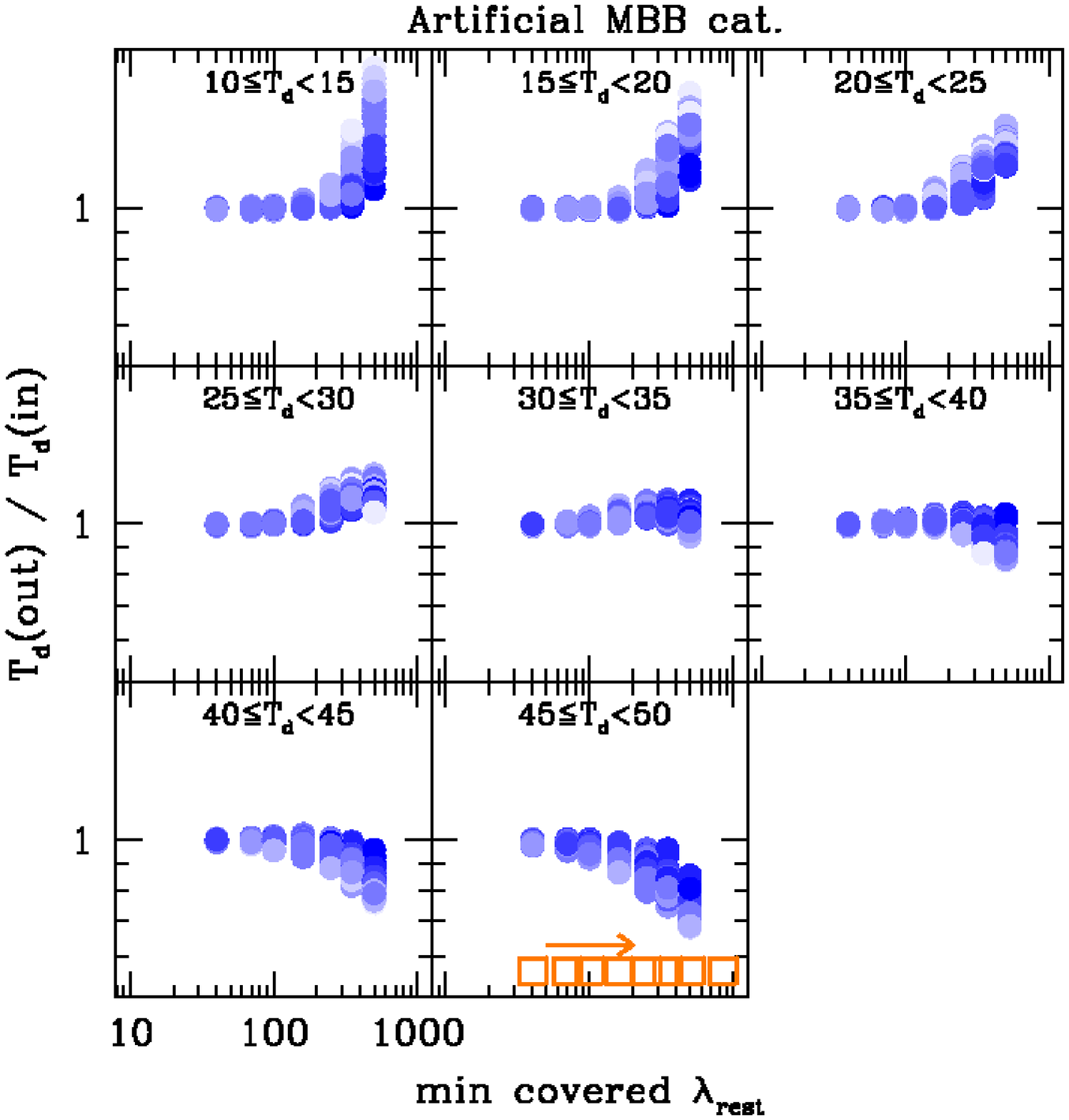}
\includegraphics[width=0.45\textwidth]{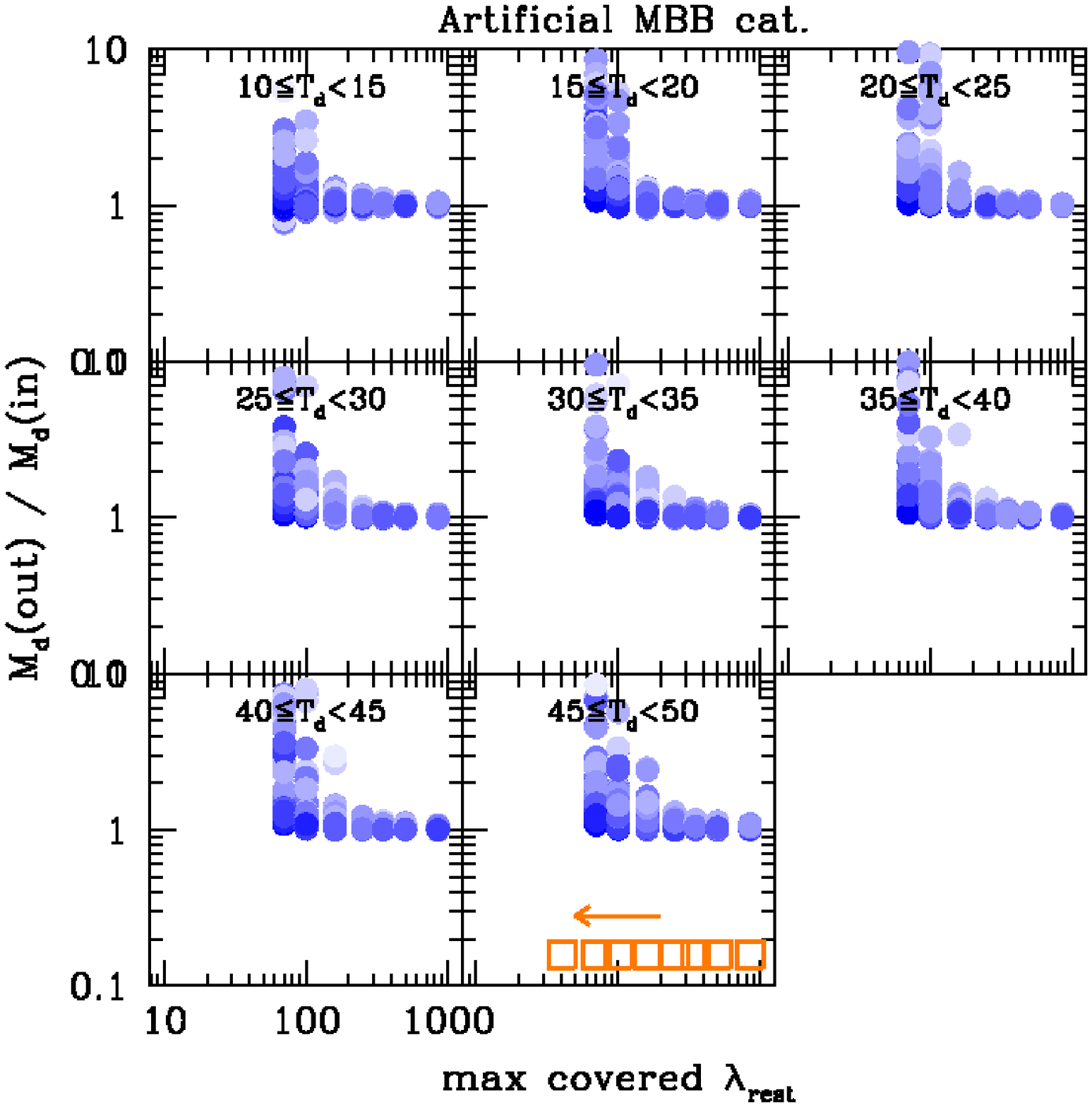}
\includegraphics[width=0.45\textwidth]{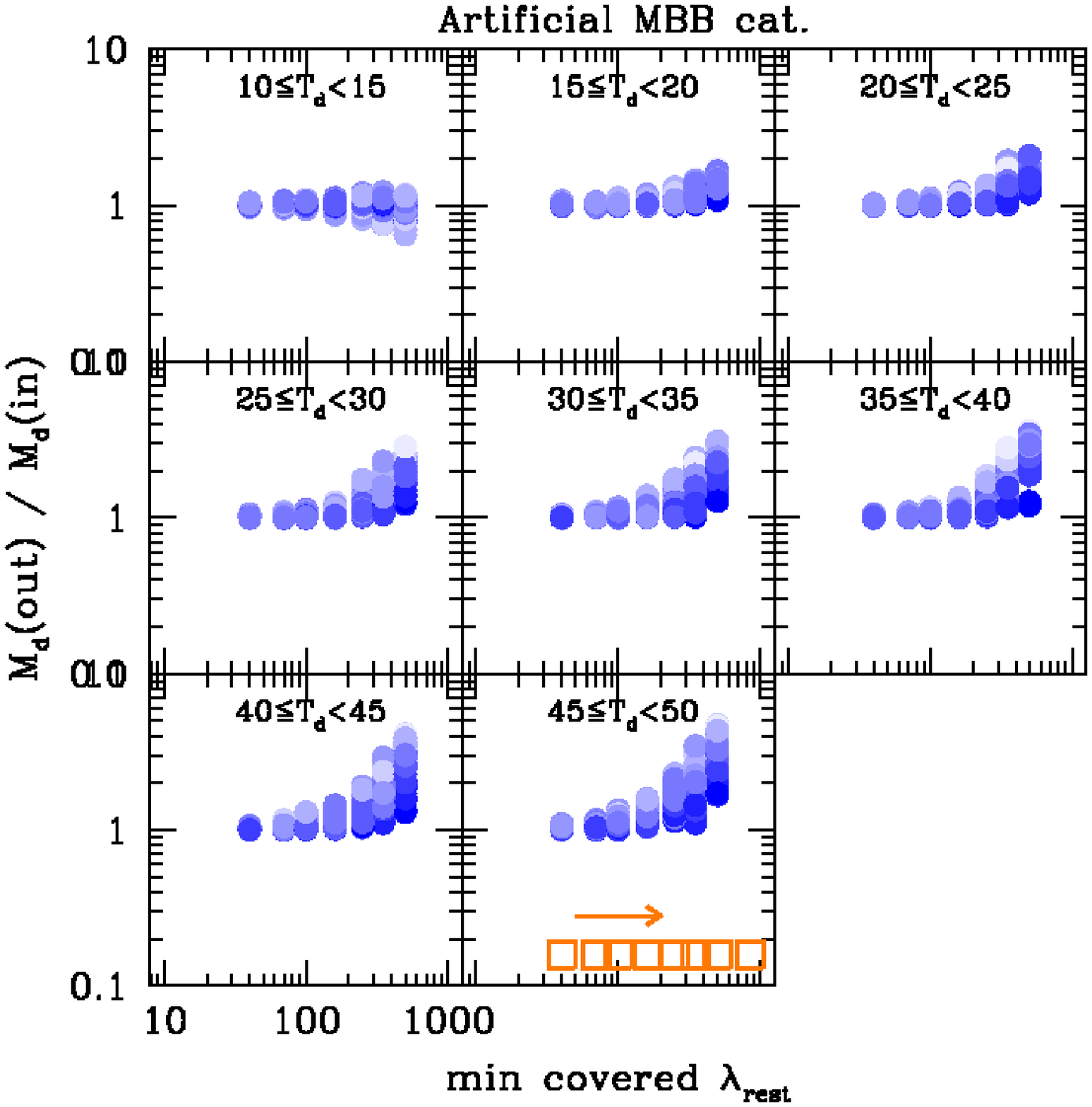}
\caption{Comparison of input and output $T_{\rm dust}$ ({\em top}) and $M_{\rm dust}$ ({\em bottom}) in MBB simulations, 
split in bins of input dust temperature. {\em Left/right} panels again refer to cases with long- or short-wavelength bands removed, 
respectively. Color coding is the same as in Fig. \ref{fig:mbb_rel_err_1}}
\label{fig:mbb_in_out_1}
\end{figure*}

In summary, the larger $q_{\rm PAH,in}$, the more $M_{\rm dust}$ can be overestimated.
If there is a tendency of underestimating $M_{\rm dust}$ (e.g., driven 
by a small value of $U_{\rm min,in}$), then a large $q_{\rm PAH,in}$ makes this underestimate
milder. On the other hand, if $U_{\rm min,in}$ is large and thus there is already 
the tendency to overestimate $M_{\rm dust}$, then a large $q_{\rm PAH,in}$ makes things 
worse.

Using Eq. \ref{eq:U_avg} to derive the mean radiation field $\langle U\rangle$, 
a similar behavior of the ratio $\langle U\rangle_{\rm out}/\langle U\rangle_{\rm in}$
is observed. Depending on the actual values of $U_{\rm min}$, $\gamma$ and 
$q_{\rm PAH}$, the chance of overstimate or underestimate $\langle U\rangle$ 
can be amplified or suppressed due to summing or compensating effects of the 
three parameters. Since $\langle U\rangle$ and $M_{\rm dust}$ is linked 
by Eq. \ref{eq:ldust}, this explains the relative stability of the 
$M_{\rm dust}$ estimate on a more fundamental level.

When removing short-wavelength bands, as usual there are very small
tendencies for under- or overestimates of $M_{\rm dust}$. 
In this case, 
by varying $q_{\rm PAH,in}$ these trends do not seem to change significantly,  
or at least not significantly enough to be appreciated by this analysis.

Simulations limited to a maximum wavelength of 250 $\mu$m, 
obviously behave equivalently when removing long-wavelength bands.
On the other hand, when removing short-wave bands, things are different because the three longest wavelength bands are missing. 
Trends as a function of $\gamma_{\rm in}$ are still very small and are only slightly 
enhanced with respect to the ten bands case, but still much smaller than 
when removing long-wavelength bands. Trends as a function of $U_{\rm min,in}$ also start having a 
larger amplitude  when removing short-wavelength bands, 
while trends with ten bands continued to be marginal. Finally $q_{\rm PAH,in}$ trends are marginal,
as in the case of ten bands.

\subsection{MBB systematics}\label{sect:app_mbb_syst}

It is worthwhile to verify how reliably the MBB fit to the artificial 
photometry can retrieve the input values of dust temperature and mass. 
Figure \ref{fig:mbb_in_out_1} shows the comparison of input and output quantities, 
as a function of $\lambda_{\rm max,min}$, split in bins of input dust temperature, $T_{\rm dust,in}$.

\begin{figure*}[!ht]
\centering
\includegraphics[width=0.45\textwidth]{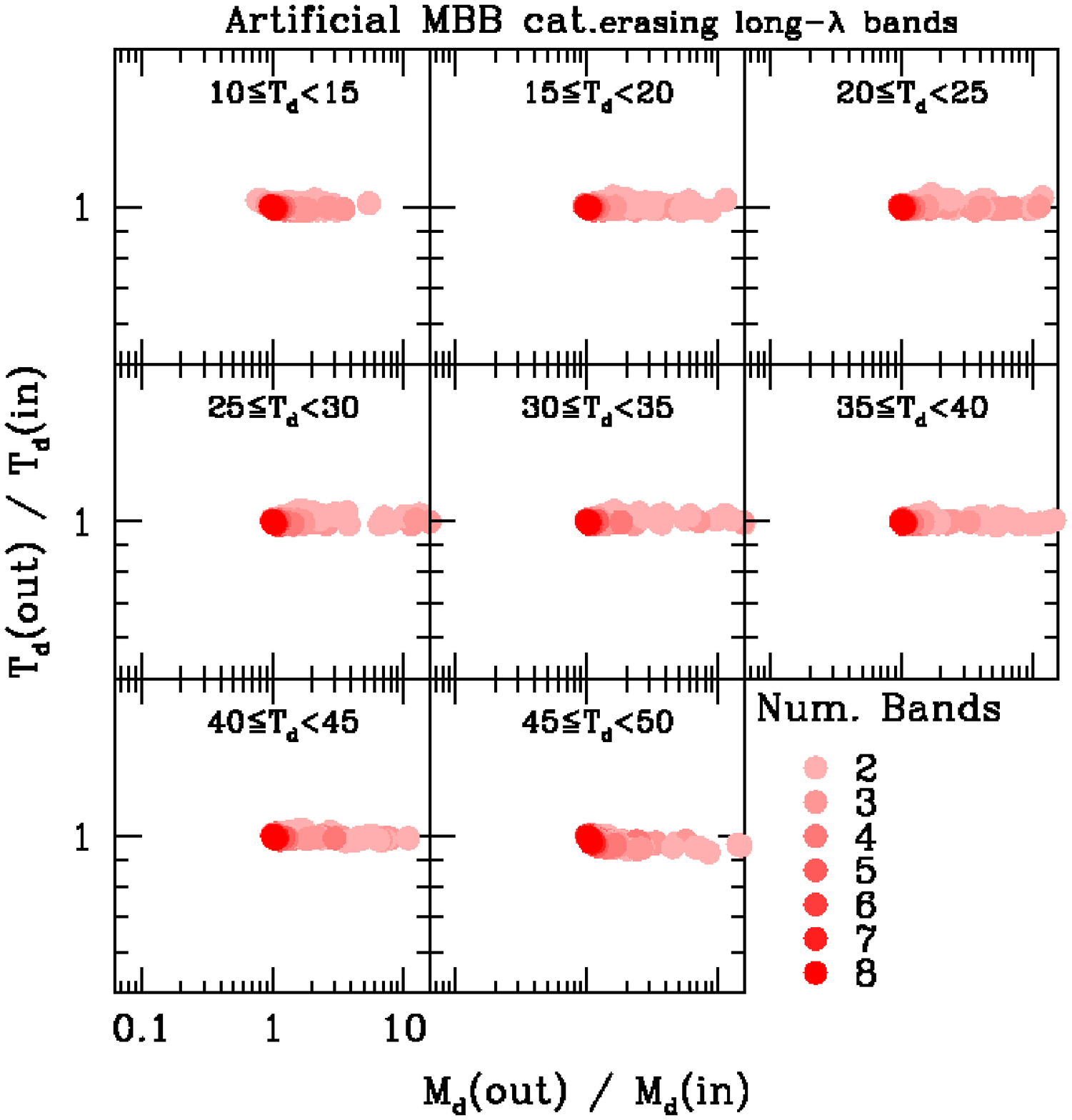}
\includegraphics[width=0.45\textwidth]{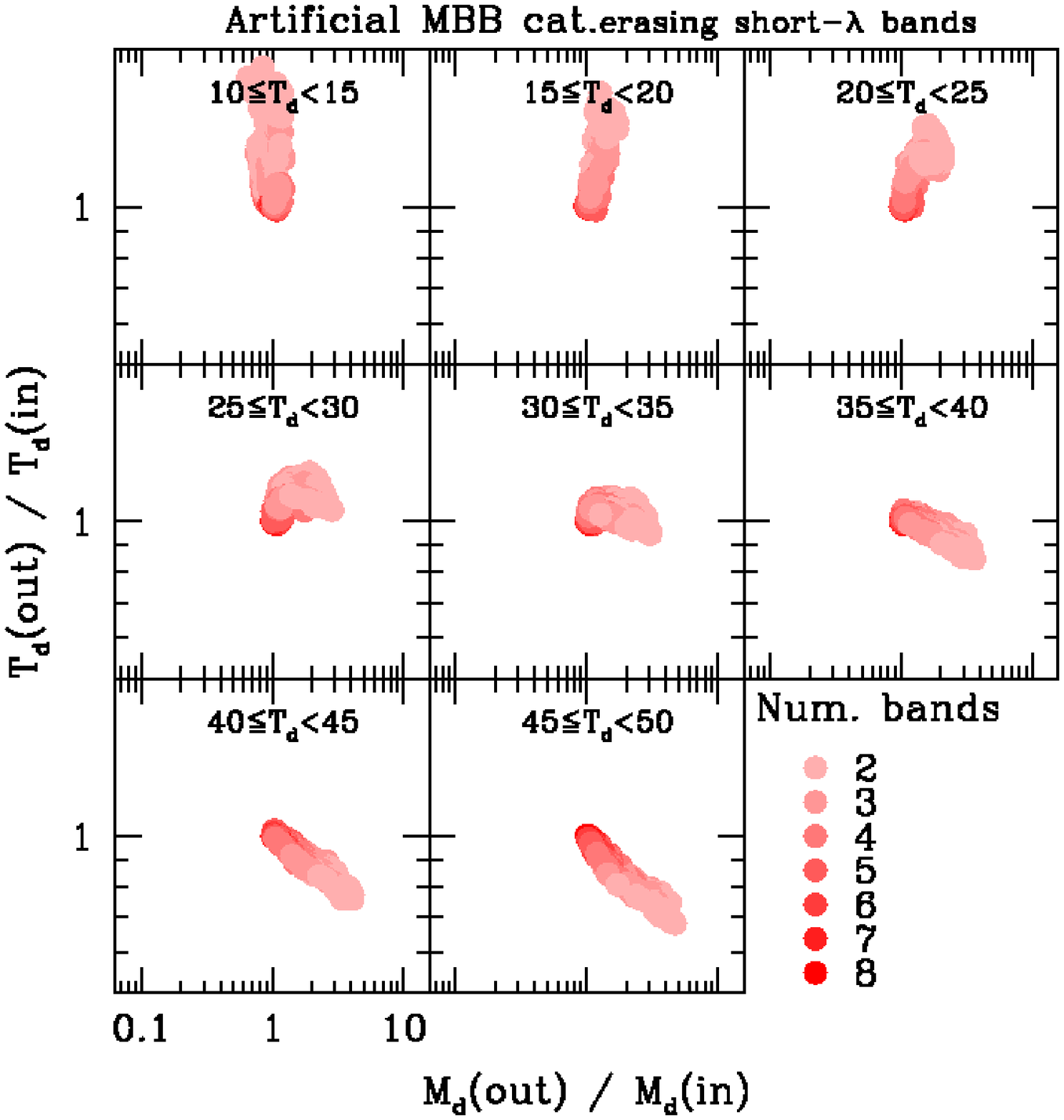}
\caption{Direct comparison of input/output $T_{\rm dust}$ vs. $M_{\rm dust}$ in MBB simulations, 
split in bins of input dust temperature. {\em Left/right} panels again refer to cases with long- or short-wavelength bands removed, 
respectively. Color coding is based on the number of available bands; from two (light pink) to eight (red).}
\label{fig:mbb_in_out_2}
\end{figure*}

As the average photometric uncertainty grows (lighter blue dots), there is the chance to 
systematically overestimate $M_{\rm dust}$. The overestimate worsens
when removing datapoints at the long-wavelength side, while the 
problem is relatively milder (but still significant) when removing bands
at the short-wavelength side. 

When removing datapoints at the long-wavelength side, 
$T_{\rm dust}$ is always well recovered, but
when removing datapoints at the short-wavelength side 
it can be systematically underestimated or overestimated.
This effect reflects the fact that the 
peak of the MBB shifts to longer wavelengths as dust temperature 
decreases, combined with the poor sampling of the SED obtained 
by removing datapoints.

Nevertheless, the incidence of 
catastrophic $M_{\rm dust}$ overestimates is small: 
Only $\sim10$\% of cases turn out to have $M_{\rm dust,out}>2\times M_{\rm dust,in}$. 
Finally, there exists a 0.1\% of cases with the tendency 
to underestimate $M_{\rm dust}$ for the lowest $T_{\rm dust,in}$ considered, when removing 
datapoints at the short-wavelength.

Although affecting a limited fraction of cases, it is interesting to study in detail the nature 
of these systematic trends to understand what causes them.
Figure \ref{fig:mbb_in_out_2} directly compares the ratio of input/output 
$T_{\rm dust}$ to the same quantity for $M_{\rm dust}$. Color coding is now based on the 
number of available bands. 

As long as short-$\lambda$ bands are progressively removed, , it is possible 
to overestimate $T_{\rm dust}$ at low input
temperatures 
($T_{\rm dust,in}\le20$ K, for example).
Consequently, the shape of the MBB SED changes, but 
the long-$\lambda$ bands still give a good constraint on the 
model normalization, hence, on $M_{\rm dust}$. 
Something different happens at higher input dust temperatures. 
At $T_{\rm dust,in}=$25-35 K, the output $T_{\rm dust}$ is more 
stable but there are still variations, which start driving an 
overestimate of dust mass. The peak of the MBB starts moving out 
of the covered wavelength range and the shape of the model in 
the covered range of wavelengths is roughly constant (Rayleigh-Jeans regime; RJ). 
Nevertheless, an underestimate of $T_{\rm dust}$ implies a lower emissivity, 
and thus a larger normalization (i.e., a larger dust mass) is 
needed to reproduce the ``observed'' fluxes. This causes an 
overestimate of dust mass.

When removing long-$\lambda$ bands, the trend is again to 
overestimate $M_{\rm dust}$, but the cause is more subtle. In fact, 
in this case, $T_{\rm dust,out}$ is rather consistent with $T_{\rm dust,in}$.
This is because the short-wavelength side of the SED is always constrained by the 
available photometry. There is still a possibility of slightly 
underestimating dust temperatures, however, because by anchoring the 
model at the short-wavelength side, it is still possible 
to reproduce the short-$\lambda$ colors with a lower temperature, 
which allows for a different normalization (i.e., a different $M_{\rm dust}$). The smaller 
the number of long-wave bands, the larger is the freedom 
in renormalization. 
However,  this time a small effect on $T_{\rm dust}$ translates into 
a big effect on $M_{\rm dust}$. A small 
variation of $T_{\rm dust}$ implies a significant change in the shape of the 
MBB SED blueward of the peak. Now the short-wavelength side of the SED is constrained by the data because we are 
removing long-$\lambda$ bands. Therefore a small underestimate of $T_{\rm dust}$
induces a significant overestimate of $M_{\rm dust}$.
          
In the previous case (i.e., removing short-$\lambda$ bands), the 
opposite was happening: A big variation on $T_{\rm dust}$ implied a 
relatively smaller variation of $M_{\rm dust}$ because we were 
sampling the RJ side of the SED, where shape variations 
due to $T_{\rm dust}$ changes are less prominent than at the short-wavelength
side. 
For example, focusing on input temperature in the range $T_{\rm dust,in}=$45-50 K
one can note that:
\begin{itemize}
\item When removing long-wavelength bands, a variation of 10\% in $T_{\rm dust}$ turns out to 
produce a change of normalization of up to a factor of 10.
\item When removing short-wavelength bands, a variation of 30\% in $T_{\rm dust}$ induces
a change in normalization of up to a factor of 4-5 ``only''.
\end{itemize}

\end{appendix}


\end{document}